\documentclass[12pt,preprint]{aastex}
\pdfoutput=1

\usepackage{epsfig}

\def\rhoIn{\rho_{_{\rm in}}}
\def\vIn{v_{_{\rm in}}}

\def\vZ{v_{_0}}
\def\aZ{a_{_0}}
\def\PZ{P_{_0}}
\def\RhoZ{\rho_{_0}}
\def\v1{v_{_1}}
\def\P1{P_{_1}}
\def\Rho1{\rho_{_1}}
\def\w0{w_{_0}}
\def\y0{y_{_0}}
\def\z0{z_{_0}}
\def\vstar1{v_{_{*1}}}
\def\a1{a_{_1}}
\def\el0{\ell_{_0}}
\def\rhoMinus{\rho_{_{-}}}
\def\uMinus{u_{_{-}}}

\newcommand{\etar}{\eta_{_r}}
\newcommand{\etai}{\eta_{_i}}
\newcommand{\zetar}{\zeta_{_r}}
\newcommand{\zetai}{\zeta_{_i}}
\newcommand{\pir}{\pi_{_r}}
\newcommand{\pii}{\pi_{_i}}
\newcommand{\delr}{\delta_{_r}}
\newcommand{\deli}{\delta_{_i}}

\newcommand{\xstar}{x_{_*}}
\newcommand{\xstarZ}{x_{_{*0}}}
\newcommand{\xstarO}{x_{_{*1}}}

\newcommand{\rstar}{r_{_*}}
\newcommand{\rstarZ}{r_{_{*0}}}

\newcommand{\rs}{r_{_{\rm S}}}

\newcommand{\rc}{r_{_c}}

\newcommand{\OmegaK}{\Omega_{\rm K}}

\def\ellprime0{\ell'_0}

\newcommand{\begeq}{\begin{equation}}
\newcommand{\fineq}{\end{equation}}
\newcommand{\begfig}{\begin{figure}}
\newcommand{\finfig}{\end{figure}}
\newcommand{\begeqarray}{\begin{eqnarray}}
\newcommand{\fineqarray}{\end{eqnarray}}

\slugcomment{accepted for publication in \apj}

\shorttitle{Instability Analysis in ADAF disk} 

\shortauthors{Le, Kent, Wolff, Becker, \& Putney}

\begin{document}

\title{Standing Shock Instability in Advection-Dominated Accretion Flows}

\author{Truong Le$^{1}$, Kent S. Wood$^{2}$, Michael T. Wolff$^{2}$, Peter A. Becker$^{3}$, and Joy Putney$^{4}$}

\affil{$^{1}$Department of Physics, Astronomy \& Geology, Berry College, Mount Berry, GA 30149, USA; tle@berry.edu; \\$^{2}$High Energy Space Environment Branch, Space Science Division, Naval Research Laboratory, Washington, DC 20375, USA; \\$^{3}$Department of Physics \& Astronomy, George Mason University, Fairfax, VA 22030, USA; \\$^{4}$Department of Physics and Engineering, Washington and Lee University, Lexington, VA 24450, USA}

\begin{abstract}

Depending on the values of the energy and angular momentum per unit mass in the gas supplied at large radii, inviscid advection-dominated accretion flows can display velocity profiles with either preshock deceleration or preshock acceleration. Nakayama has shown that these two types of flow configurations are expected to have different stability properties. By employing the Chevalier \& Imamura linearization method and the Nakayama instability boundary conditions, we discover that there are regions of parameter space where disks/shocks with outflows can be stable or unstable. In regions of instability, we find that preshock deceleration is always unstable to the zeroth mode with zero frequency of oscillation, but is always stable to the fundamental mode and overtones. Furthermore, we also find that preshock acceleration is always unstable to the zeroth mode and that the fundamental mode and overtones become increasingly less stable as the shock location moves away from the horizon when the disk half-height expands above $\sim 12$ gravitational radii at the shock radius. In regions of stability, we demonstrate the zeroth mode to be stable for the velocity profiles that exhibit preshock acceleration and deceleration. Moreover, for models that are linearly unstable, our model suggests the possible existence of quasi-periodic oscillations (QPOs) with ratios 2:3 and 3:5. These ratios are believed to occur in stellar and supermassive black hole candidates, for example, in GRS 1915+105 and Sgr A*, respectively.  We expect that similar QPO ratios also exist in regions of stable shocks.

\end{abstract}



\section{INTRODUCTION}

X-ray quasi-periodic oscillations (QPOs) have been observed in binary accretion systems containing neutron stars and black holes (BHs), spanning a wide range of frequencies from a few tens of millihertz to a few hundred hertz~\citep[e.g.,][]{van85,str96,mrg97,woo00,tw02,ba13}. More recently, QPOs have also been observed both in intermediate-mass black holes~\citep[IMBHs; e.g.,][]{frk10,mid11} and supermassive black holes~\citep[SMBHs; e.g.][]{md10,miy11}. The phenomenological features of these QPOs suggest that such oscillations can be a probe of the accretion processes in the inner regions of the accretion disks around the central compact objects~\citep[for a review see][]{dgk07}. However, QPOs are sometimes observed from accreting white dwarf (WD) stars (e.g., AM Herculis objects), in which the strong magnetic field channels the accreting gas onto a magnetic pole without the need for an accretion disk \citep[e.g.,][]{lar89,wol99}. Nevertheless, the origin of QPOs and the nature of the instability from black hole binary, IMBH, and SMBH sources are complicated and still unclear~\citep{dgk07}.

Inviscid and viscous advection-dominated accretion flows (ADAFs) have been shown to display shocked solutions ~\citep[e.g.,][]{cha89,ky94,ly97,dcc01,lb05,dbl09}, and the possibility that shock instabilities may generate the QPOs observed in some sources containing BHs has been pointed out by a number of authors~\citep[e.g.,][]{cm95,lmc98,tw02,aok04,cdp09}. Furthermore, shock stability analysis for thin disks has been studied using both analytical~\citep[e.g.,][]{cm93,nak92,nak94,nh94,msc96,gl06,ny09} and numerical simulation approaches~\citep[e.g.,][]{cm93,nh94,msc96,gl06,ny08}. However, several of these studies  mainly focused on the assumption of a constant disk height~\citep[e.g.][]{cm93,nh94} or did not fully examine the multiple modes of oscillation~\citep[e.g.][]{cm93,nh94,msc96,gl06,ny08}, which are known to exist in the WD accretion shock case studied by, e.g., \citet{ci82} and \citet{iwd84}. 

Accretion flows generally suffer from several types of instabilities, such as the thermal instability,  which was first discovered in numerical calculations~\citep{lcs81} and later by \citet{ci82}, who performed a linear stability analysis for plane-parallel shock waves with cooling functions of the form $\Lambda \propto \rho^2 T^\alpha$, where $\rho$, $T$, and $\alpha$ are the gas density, temperature, and power-law index, respectively. \citet{ci82} found that such shocks were linearly unstable in a fundamental mode if $\alpha \lesssim 0.4$ and unstable to the first- and second-overtone modes if $\alpha \lesssim 0.8$. The linear analysis was subsequently confirmed by numerical simulations \citep{iwd84,wgw89}. Nevertheless, in ADAF disks the flow is generally radiatively inefficient; hence, global thermal instability does not exist in the flow~\citep[e.g.,][]{ny95,wu97}. ADAF disks are locally stable if the disk is optically thin but thermally unstable if the disk is optically thick~\citep[e.g.,][]{kac96}. One can always incorporate radiative cooling processes into the model; however, in this case the system is no longer an ADAF type~\citep[e.g.,][]{msc96}. Nonradiative (ADAF) disks are still the preferred model in a number of sources, such as Sgr A* and M87, and therefore we will not include any radiative cooling or viscous heating mechanisms in our model, since the stability of a shock flow, even with the simple inviscid hypothesis, has not yet been fully explored.

The studies of nonspherical accretion flows that include angular momentum have demonstrated the variety and complexity of structures forming around the central object~\citep[e.g.,][]{hsw84a,hsw84b,eck85,eck87,eck88,cha89,cm93,mlc94}. \citet{cha89} performed a detailed study of the structure of vertically hydrostatic, inviscid, adiabatic, steady accretion flows with Rankine-Hugoniot, isentropic, and isothermal standing shock waves. The qualitative properties of his solutions contain multiple critical points and two possible locations of the shock waves. \citet{cha89} further examined the linear and local stability of Rankine-Hugoniot, isentropic, and isothermal standing shock waves through a derived dispersion relation. They found that the Rankine-Hugoniot-type shocks are stable. The inner shock for the isentropic shocks becomes unstable with increasing angular momentum, but the outer shock is always stable. However, for the isothermal shock type, their analysis revealed no stability, and in this case, it was not possible to resolve the uniqueness of the shock location. 

Following the work of~\citet{cha89}, \citet{nak92,nak94} introduced a global spontaneous perturbation into a steady-state isothermal accretion flow with an isothermal standing shock wave. He showed that the growth of the perturbation is unstable if the steady-state velocity profile displays deceleration in the preshock region. In his work \emph{spontaneous instability} simply means that the system is not exposed to perturbations sustained by any external forcing. Furthermore, \citet{nak92,nak94} showed that the shock flow solution that contains the inner shock is unstable and that the solution containing the outer shock is stable because of \emph{preshock deceleration and acceleration}, respectively. Here, the term preshock deceleration or acceleration simply means that the flow is decelerated or accelerated toward the steady-state shock radius, respectively. They argued that the instability occurs owing to the energy input through the displacement of the shock wave, where the jump of the fluid density and displacement of the shock surface release the potential energy that amplifies the perturbations in the postshock flow. These results were later confirmed by~\citet{cm93} and \citet{nh94} through numerical simulations, assuming adiabatic flows with Rankine-Hugoniot standing shock waves and isothermal flows with isothermal standing shock waves, respectively.  

It is important to note here that these authors assumed a constant disk height in their work. Furthermore, \citet{nh94} showed that the fundamental and overtone modes with nonzero frequency are stable to both preshock acceleration and preshock deceleration, but the zeroth mode with zero frequency is stable to preshock acceleration and unstable to preshock deceleration. In other words, \citet{nh94} demonstrated that preshock deceleration and preshock acceleration are unstable and stable, respectively, using both linear and nonlinear treatments, consistent with the Nakayama instability. \citet{gf03} (assuming isothermal flows) and \citet{gl06} (assuming adiabatic flows) later demonstrated that under a \emph{constant disk height} assumption with isothermal standing shock waves the outer shock was generally non-axisymmetrically unstable when the shock Mach number was not close to unity, and argued that the advective-acoustic cycle between the corotation radius and the shock could be responsible for the instability.

Since \citet{nh94} and \citet{gl06} assumed a constant disk height in their work with either isothermal or adiabatic accretion flows and isothermal shock waves, it is therefore interesting to see whether shock flows with {\it varying} disk heights still obey the~\citet{nak92,nak94} instability criteria. In this paper, without invoking any type of cooling or heating for the system, we undertake a linear analysis of the global stability of ADAF adiabatic accretion flows with isothermal standing shock waves utilizing the~\citet{nak92,nak94} treatment of the boundary conditions and allowing for the disk half-thickness to be given by the standard hydrostatic prescription. In this sense, we are reexamining the original Chakrabarti (1989) adiabatic accretion flow with isothermal shock instability problem, since this instability problem is still unresolved. We will briefly discuss the fundamental parameters that allow us to determine the disk-shock structures, and we refer the readers to the previous series of papers by~\citet{lb04,lb05,lb07} for further details.

The paper is organized as follows. In Section~2 we discuss the~\citet{ci82} perturbation method and the~\citet{nak92,nak94} instability boundary conditions. In Section~3 we describe the time-dependent equations and the steady-state nonradiative shock wave structures in our model.  We then introduce a small perturbation to the shock velocity and determine a set of linear equations that allows us to solve for both the growth (or damping) rates in Section~4. In Sections~5 and 6 we discuss the results and the astrophysical application to QPOs, and we conclude with a discussion of the astrophysical significance of our results in Section~7.

\section{SHOCK INSTABILITY BOUNDARY CONDITIONS}

In our analysis we follow the method developed by~\citet{ci82}, in which we introduce a small perturbation to the shock velocity and determine a set of coupled differential equations that allow us to obtain both the growth or damping rates and the oscillatory periods of the perturbation between the inner sonic point and the shock location. Our inner boundary condition is different from that utilized by~\citet{ci82} because in their application, matter was accreting onto the solid surface of a WD, and the flow velocity vanishes at the surface. Since the radial velocity approaches zero at the surface, the perturbed radial velocity must vanish there as well. The inner boundary condition for the white WD is, therefore, easy to formulate. In a BH problem, near the event horizon the radial velocity approaches the speed of light~\citep[e.g.,][]{wei72}.

We need to look into the intrinsic properties of the shock and how the perturbation affects the nature of the flow from large radii to the event horizon to establish the inner boundary condition for a BH accretion disk stability problem. Because of the supersonic nature in the preshock (or upstream) region, any initial disturbances in the preshock region are transferred through the shock to the postshock (or downstream) region in a finite time, but any disturbances in the postshock region cannot affect the flow in the preshock region. Moreover, any perturbations downstream from the inner sonic point cannot affect the upstream region since the flow downstream of the inner sonic point is supersonic. Thus, as long as there are no perturbations sustained by external forcing, it is assumed that only the postshock region is perturbed, which is between the inner sonic radius and the shock location. 

Following this line of argument, a new global instability between a \emph{fixed inner sonic point and the location of a shock} for an axisymmetric inviscid disk was found, that is, \emph{preshock deceleration causes instability~\citep{nak92,nak94}.} There is no physical reason why the perturbed physical quantities (e.g., radial velocity) should vanish at a ``fixed'' inner sonic point radius. As demonstrated, e.g., by ~\citet{dcc01} and \citet{lb05}, in a steady-state solution, the position of the inner sonic radius moves toward the horizon as the location of the shock radius moves outward. This result implies that if the shock location is perturbed, the inner sonic radius must therefore be perturbed accordingly. However, because information about the perturbations occurring at the shock location takes a finite time to reach the inner sonic point, the inner sonic point position will only be perturbed at a later time. Based on this argument, we agree with \citet{nak92,nak94} that a ``fixed'' inner sonic point is an appropriate inner boundary condition. Since the inner sonic radius is fixed, this implies that the radial velocity perturbation must vanish at the inner sonic point radius in order to have a self-consistent formulation of the problem. In our analysis the growth or damping rate of the perturbation is obtained when the perturbed velocity goes to zero at the inner sonic radius. Furthermore, it is clear that if the perturbation radial velocity goes to zero at the inner sonic radius, then the gas sound speed perturbation will likewise go to zero, and as a result, the perturbed energy is radiated from the system at that point. This implies that the inner sonic radius has a stabilizing effect because the perturbation energies leak away through the sonic point~\citep{nak92,nak94}.

\begfig[t] \hskip-0.25in \epsscale{1.15} \plottwo{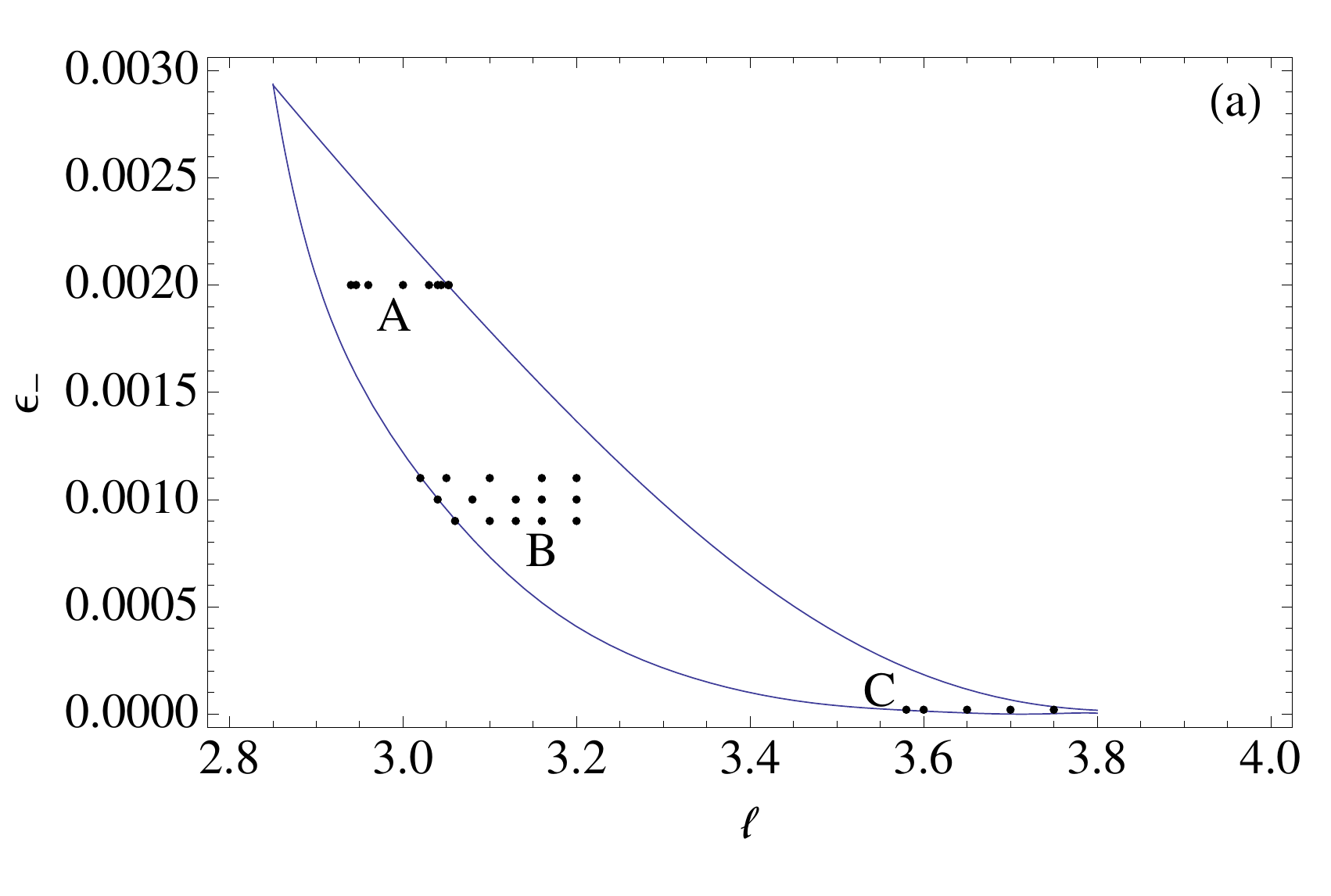}{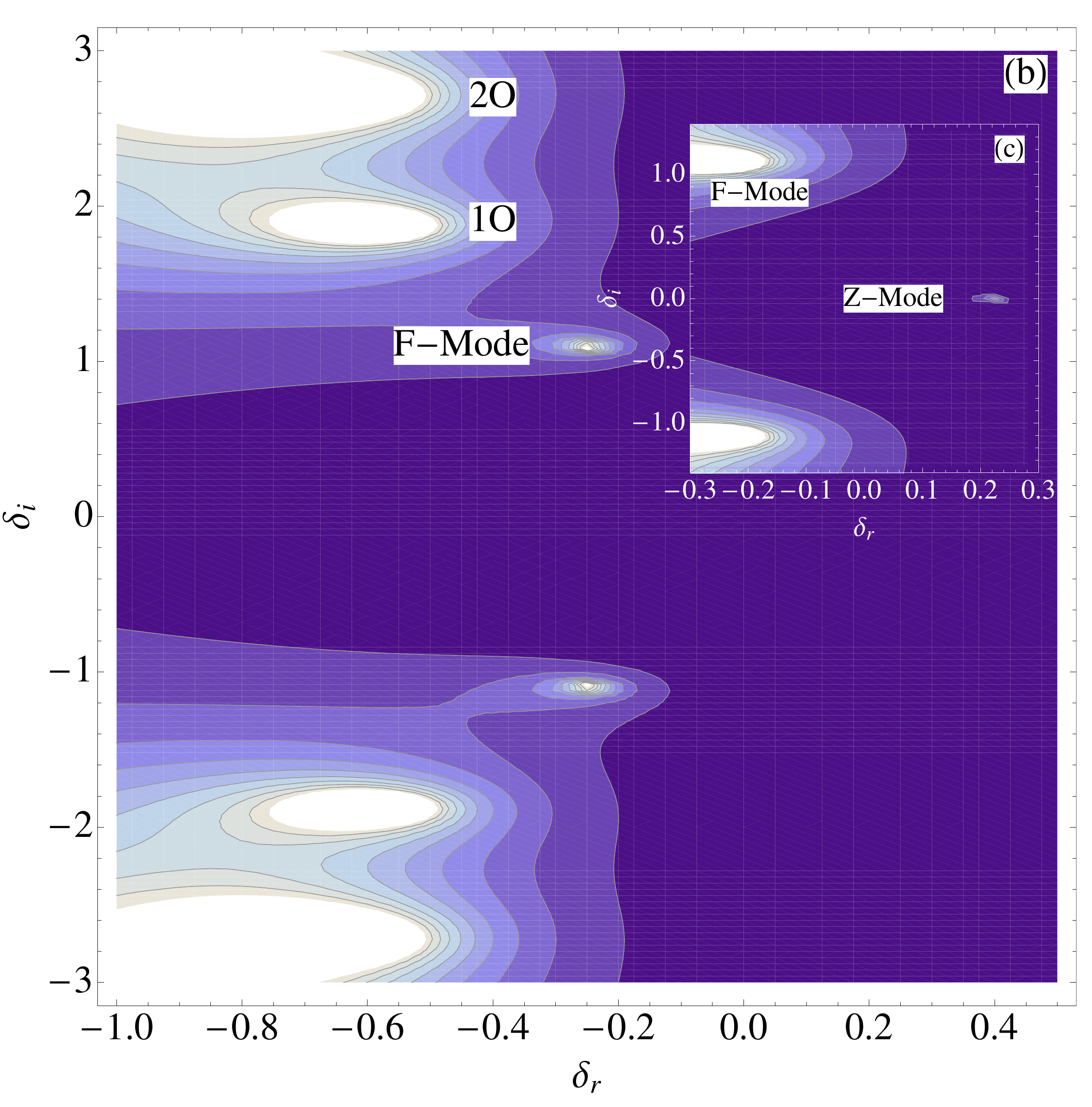}
\caption{\footnotesize (a) The wedge represents the ($\epsilon_{-}, \ell$)-parameter space where shocks can form for an isothermal shock model. The construction of this curve is discussed in the text. The $\epsilon_{_{-}}$ and $\ell$ parameter space for the 29 different models studied here is indicated with filled circles. We use regions A (9 models) and B (15 models) to study the disk/shock stability/instability and the direction of stability, respectively. Models in regions A and B and in region C (5 models) indicate the regions of unstable and stable disk/shock, respectively. (b) The eigenvalue contours of the inner shock for Model 2 are plotted to illustrate the identical solutions between the values $\delta_{_{i}}$ and -$\delta_{_{i}}$ with the inset (c) to show the eigenfrequencies of the Z-Mode and F-Mode at higher resolution. The white-colored contours indicate where $1/|\eta|$ blows up, and the blue-colored area in plots (b) and (c) indicates that $1/|\eta|$ is small ($< 1$). The small white-colored contour point ($\delta_{_{r}} = 0.16, \delta_{_{i}} = 0$) of inset (c) represents an unstable solution with zero frequency of oscillation.}
\label{fig1} \finfig

\begfig[t] \hskip-0.25in \epsscale{1.15} \plottwo{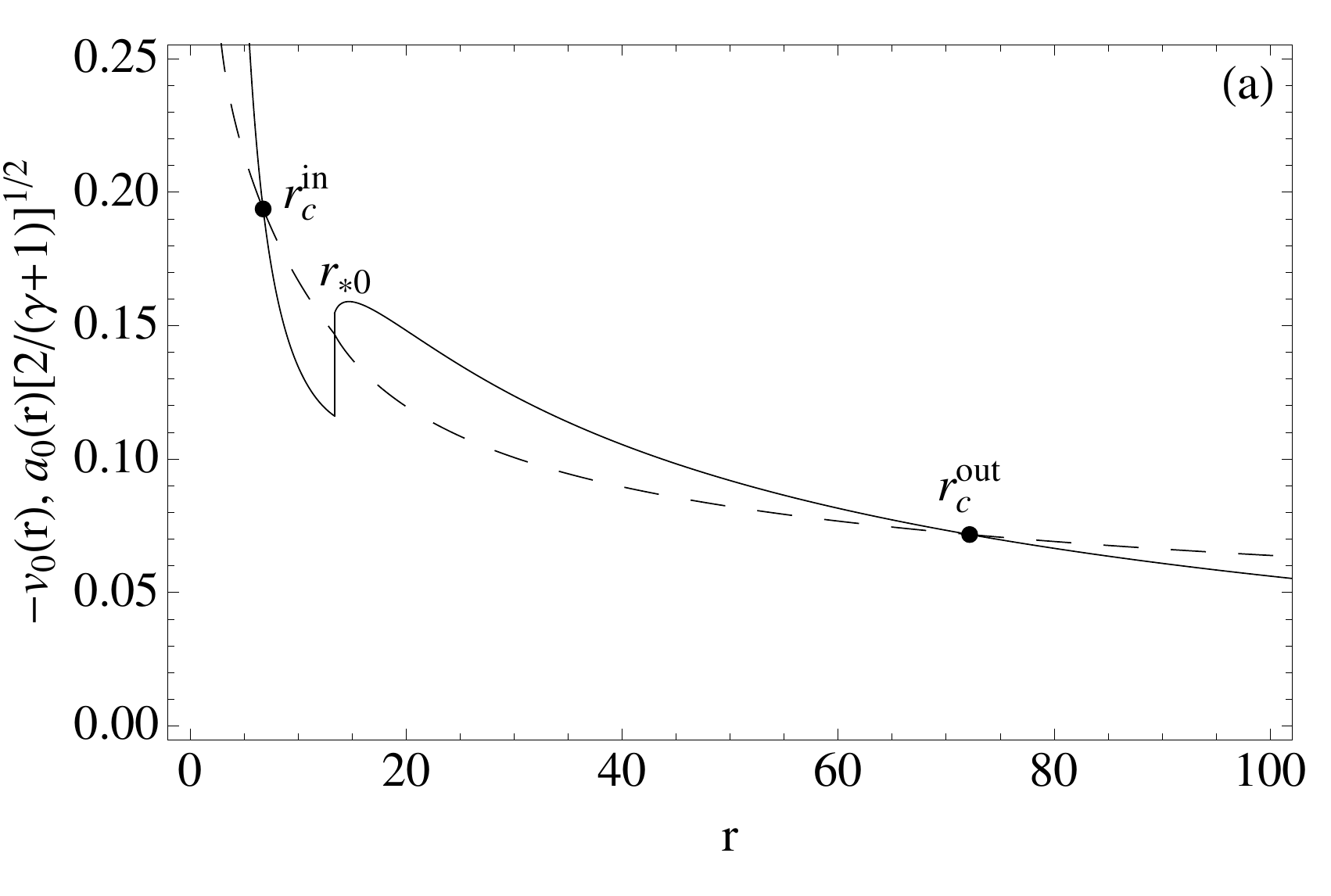}{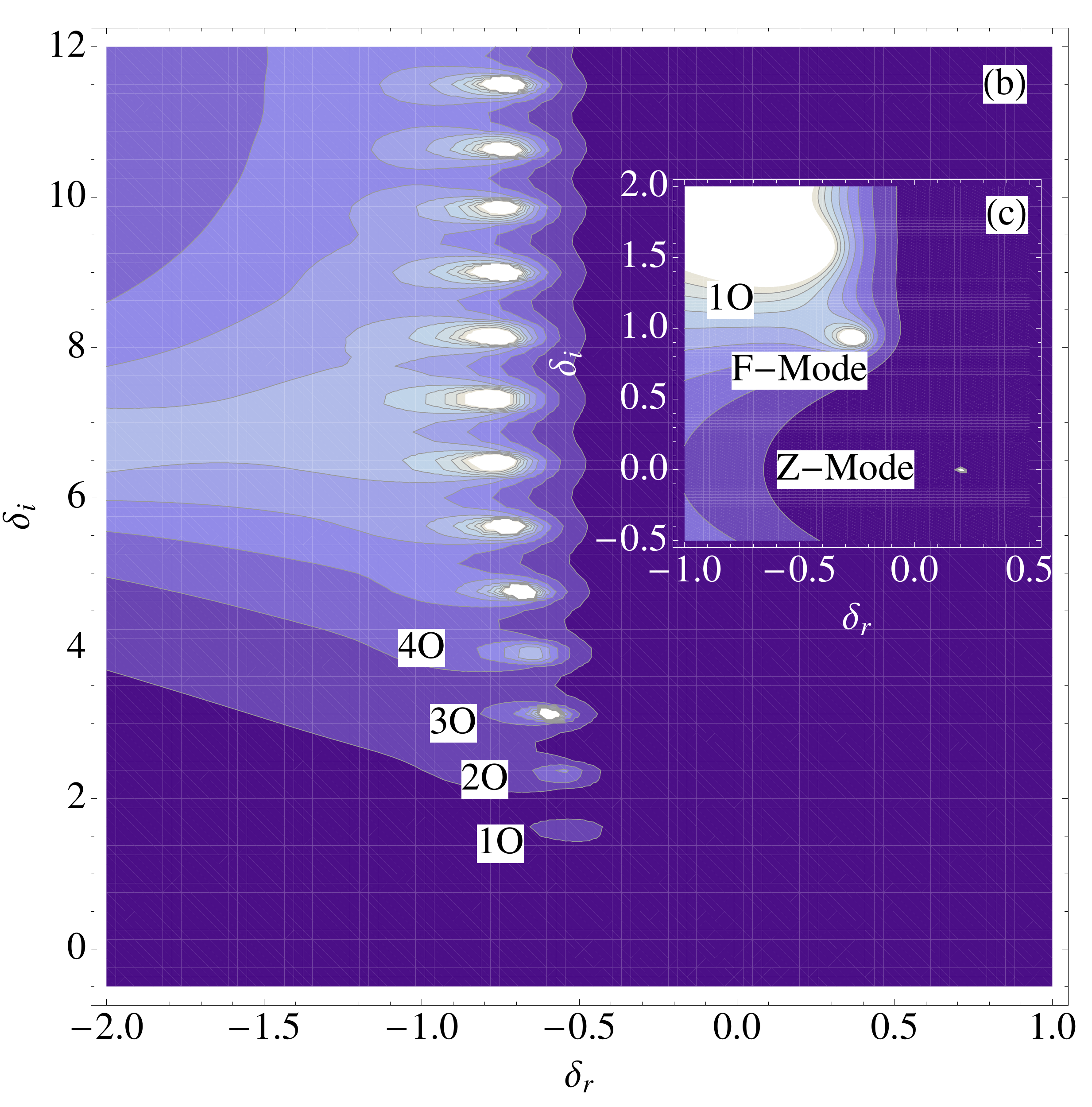}
\caption{\footnotesize Model 0 - only one shock flow solution is possible for this model: (a) inflow velocity $v$ (solid) and isothermal sound speed $a$ (dashed) are plotted as functions of the radius $r$ in units of $GM/c^2$. This is a shocked disk profile with preshock deceleration as displayed before the shock location, and (b) its eigenfrequencies with the inset (c) to show the eigenfrequencies of the Z-Mode, F-Mode, and 1O at higher resolution. Plot (b) is displayed at low resolution so that modes with small resolution can be seen; hence, the large white-colored contour looks unresolved above the fifth overtone.}
\label{fig2} \finfig

\begfig[t] \hskip-0.25in \epsscale{1.15} \plottwo{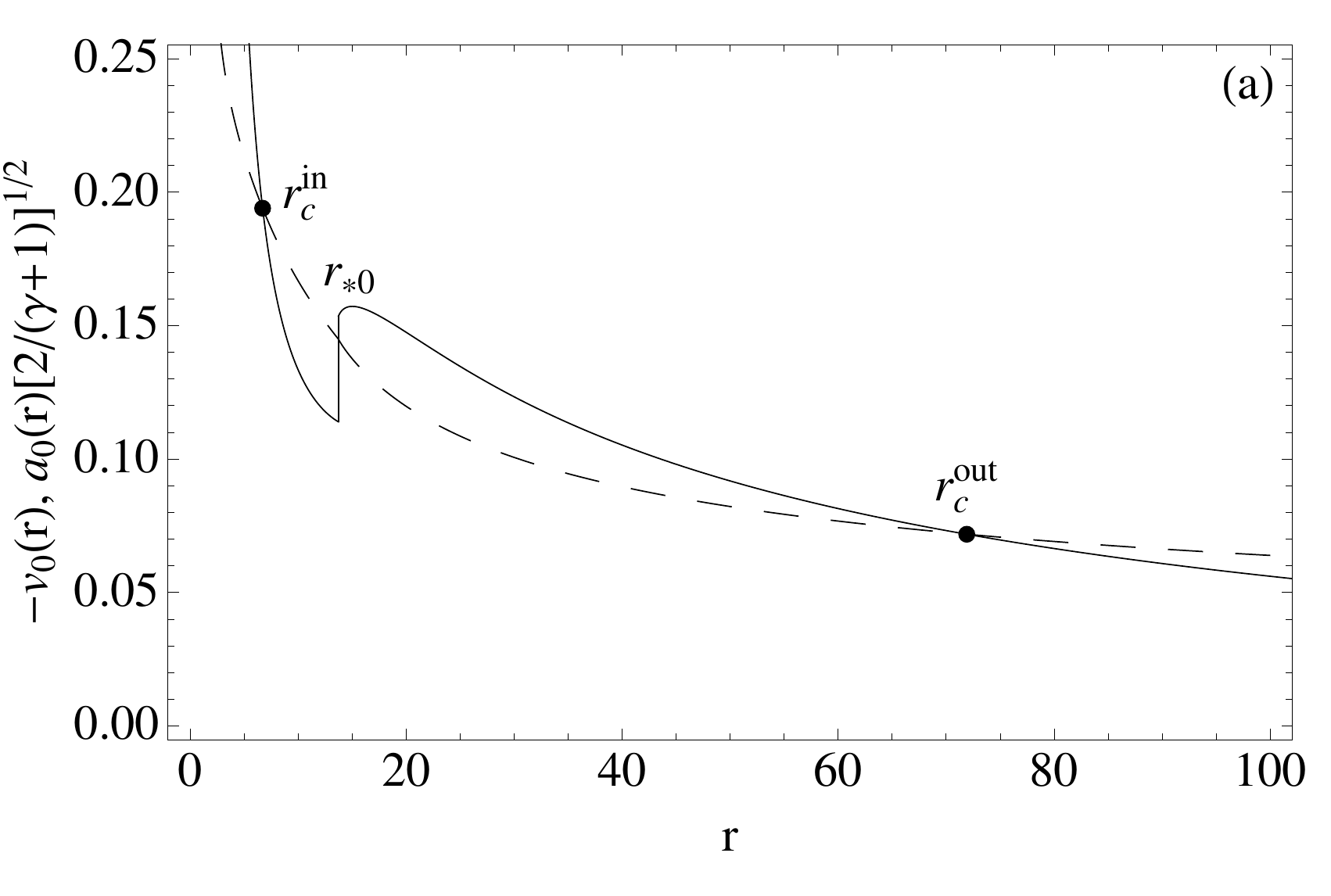}{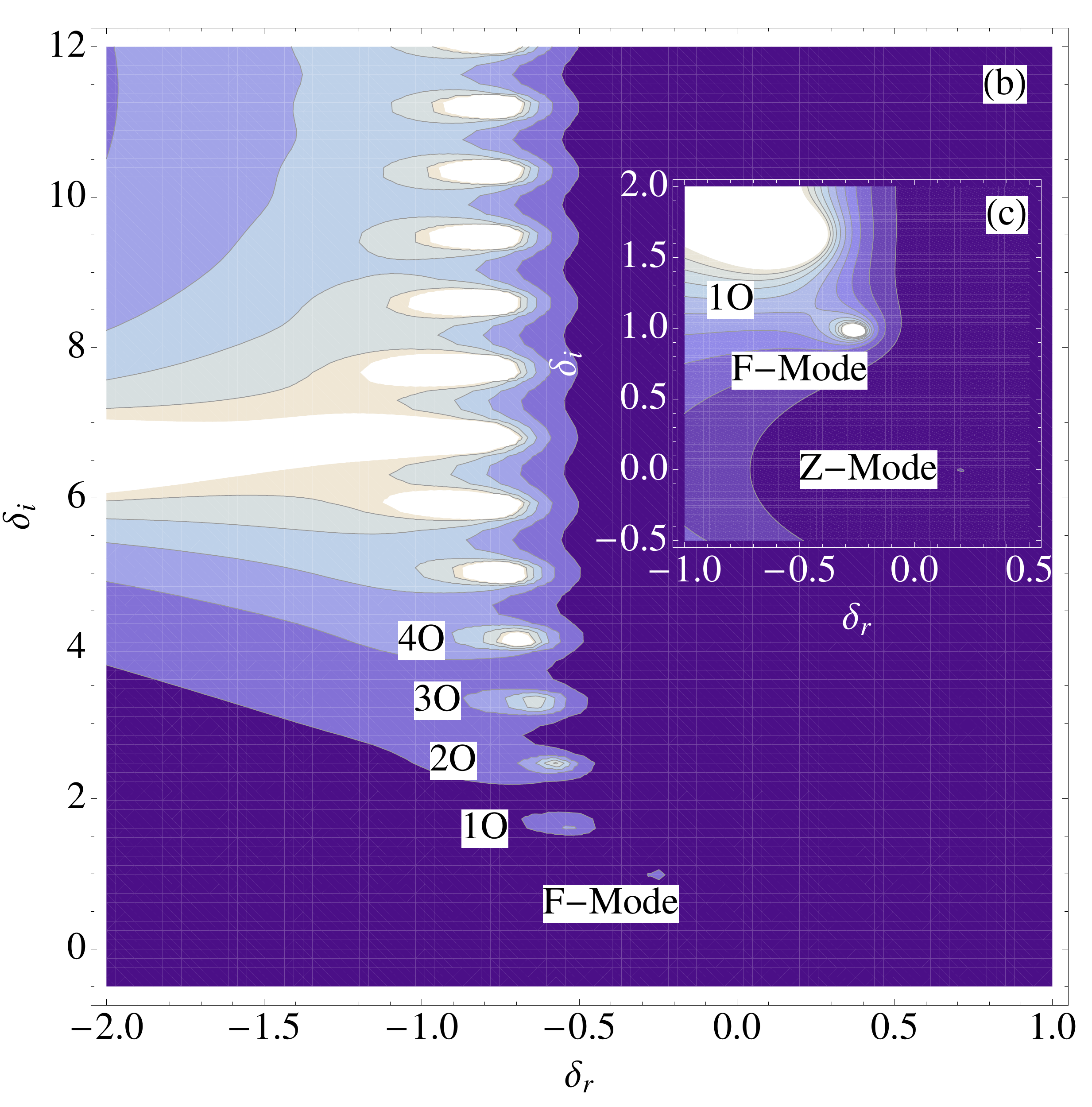}
\caption{\footnotesize Model 1 - inner shock flow: (a) inner shock disk profile with preshock deceleration, and (b) its eigenfrequencies with the inset (c) to show the eigenfrequencies of the Z-Mode, F-Mode, and 1O at higher resolution. This model has two possible shock flow solutions. The outer shock solution is depicted in Figure~\ref{fig17}.}
\label{fig3} \finfig

\begfig[t] \hskip-0.25in \epsscale{1.15} \plottwo{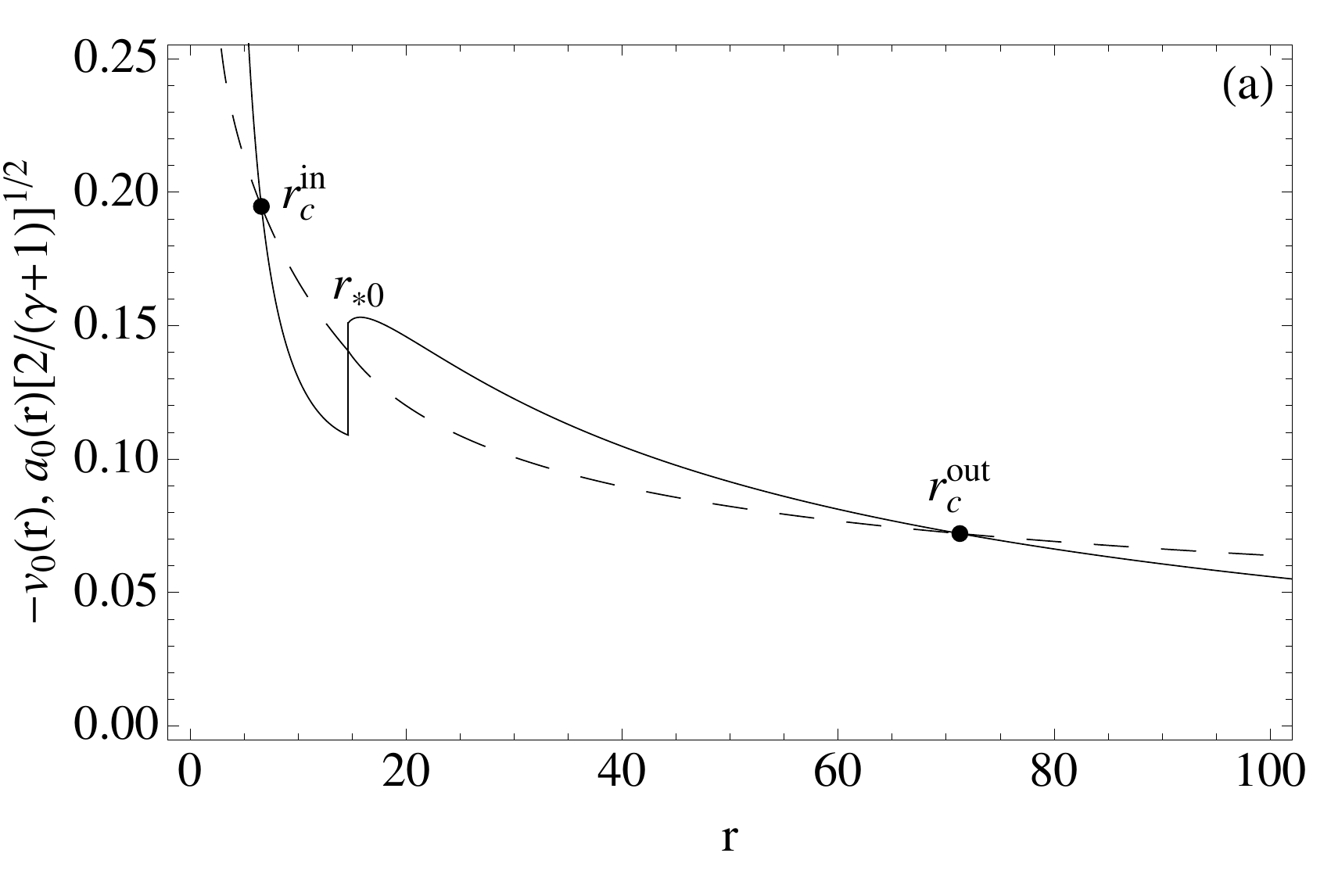}{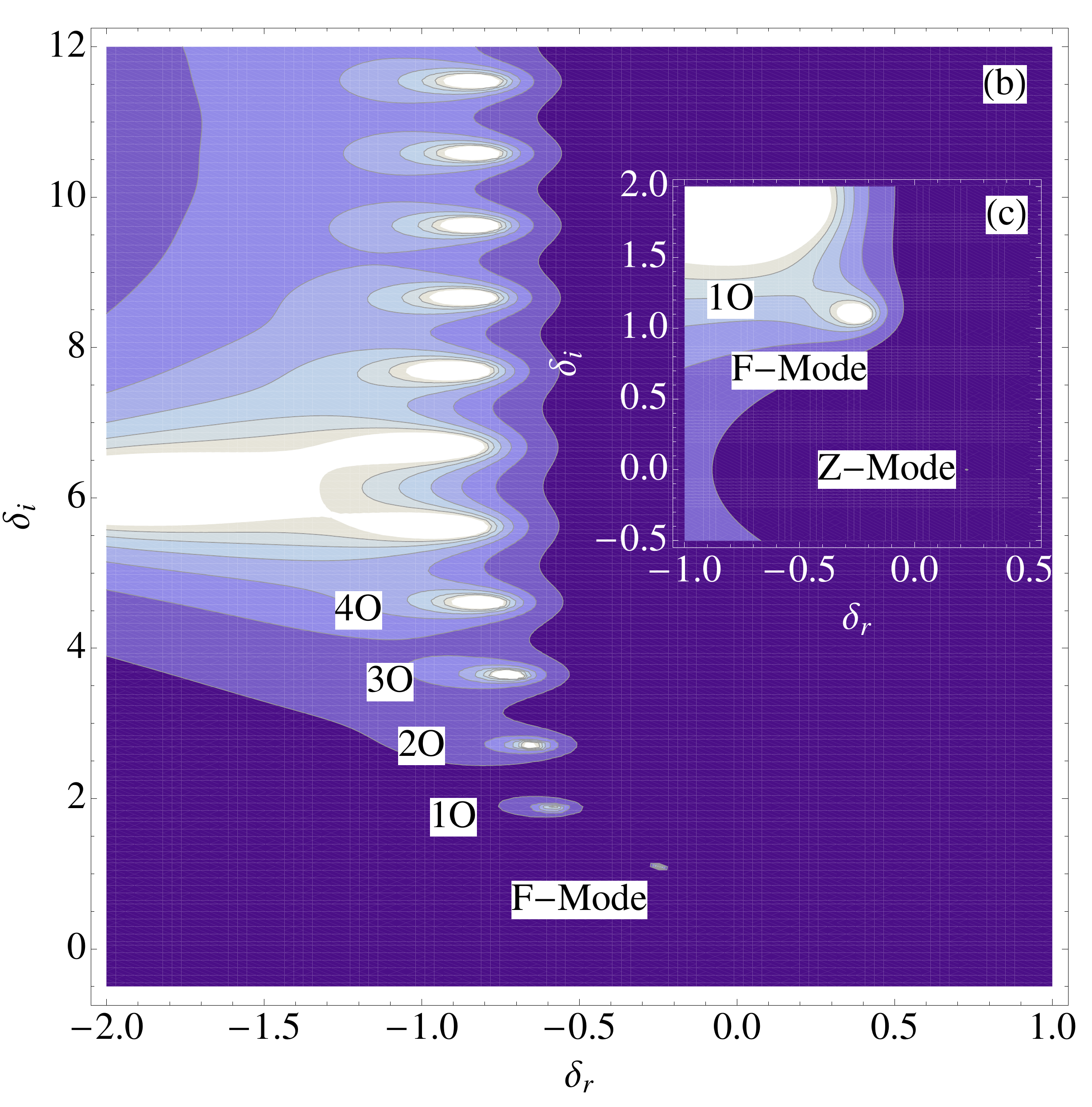}
\caption{\footnotesize Model 2 - inner shock flow: (a) inner shock disk profile with preshock deceleration, and (b) its eigenfrequencies with the inset (c) to show the eigenfrequencies of the Z-Mode, F-Mode, and 1O at higher resolution. This model has two possible shock flow solutions. The outer shock solution is depicted in Figure~\ref{fig16}.}
\label{fig4} \finfig

\begfig[t] \hskip-0.25in \epsscale{1.15} \plottwo{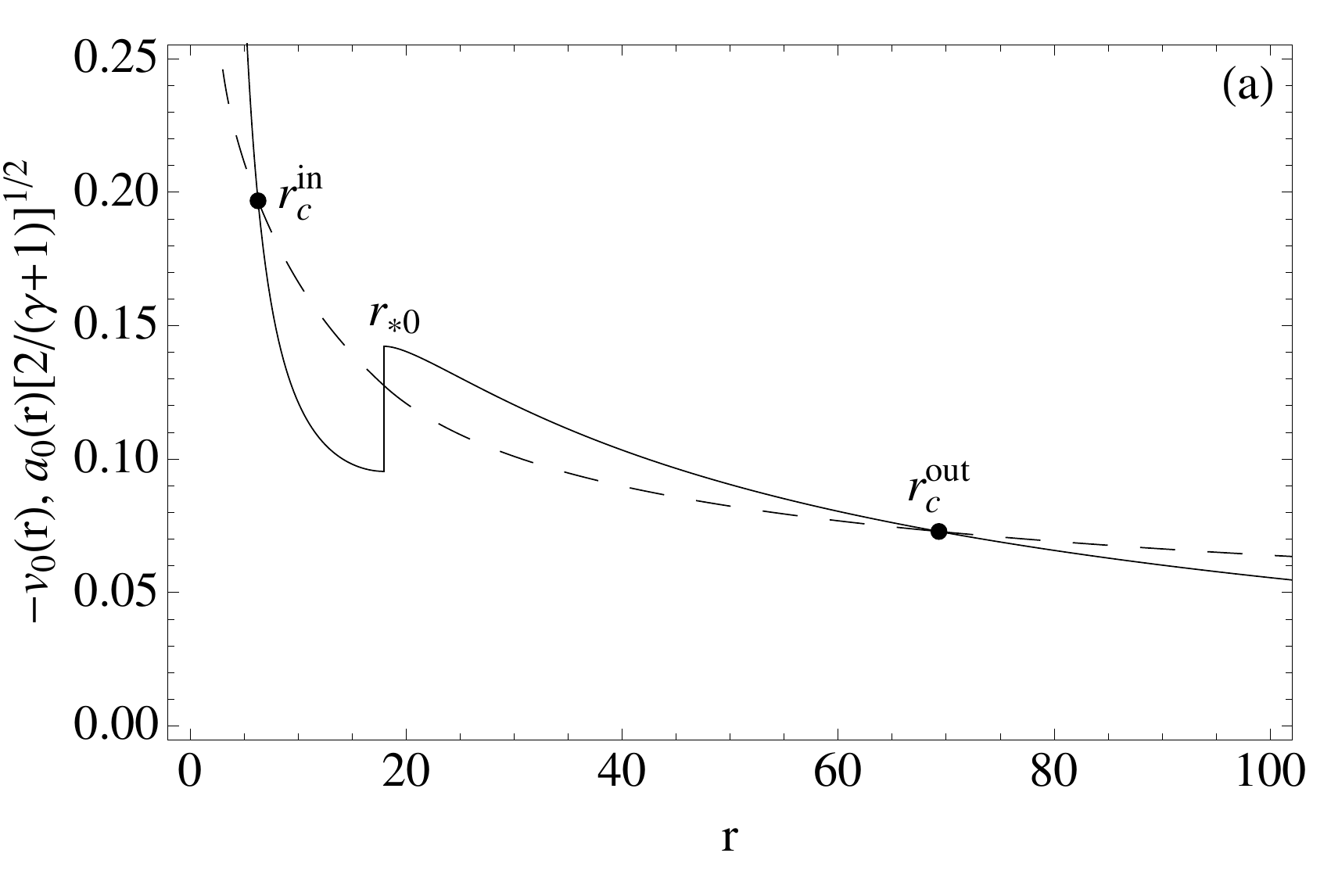}{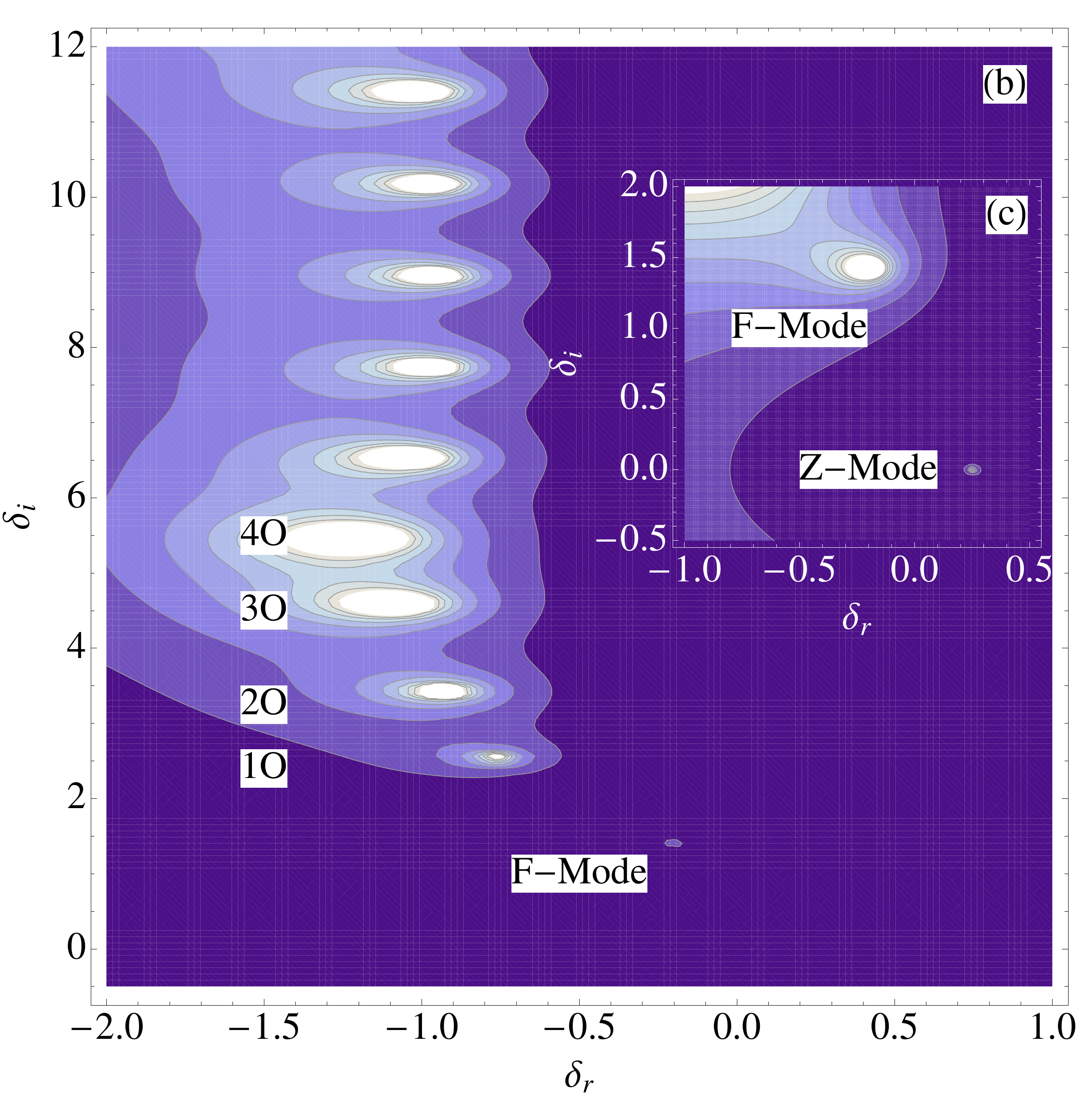}
\caption{\footnotesize Model 3 - inner shock flow: (a) inner shock disk profile with preshock deceleration, and (b) its eigenfrequencies with the inset (c) to show the eigenfrequencies of the Z-Mode, F-Mode, and 1O at higher resolution. This model has two possible shock flow solutions. The outer shock solution is depicted in Figure~\ref{fig15}.}
\label{fig5} \finfig

\begfig[t] \hskip-0.25in \epsscale{1.15} \plottwo{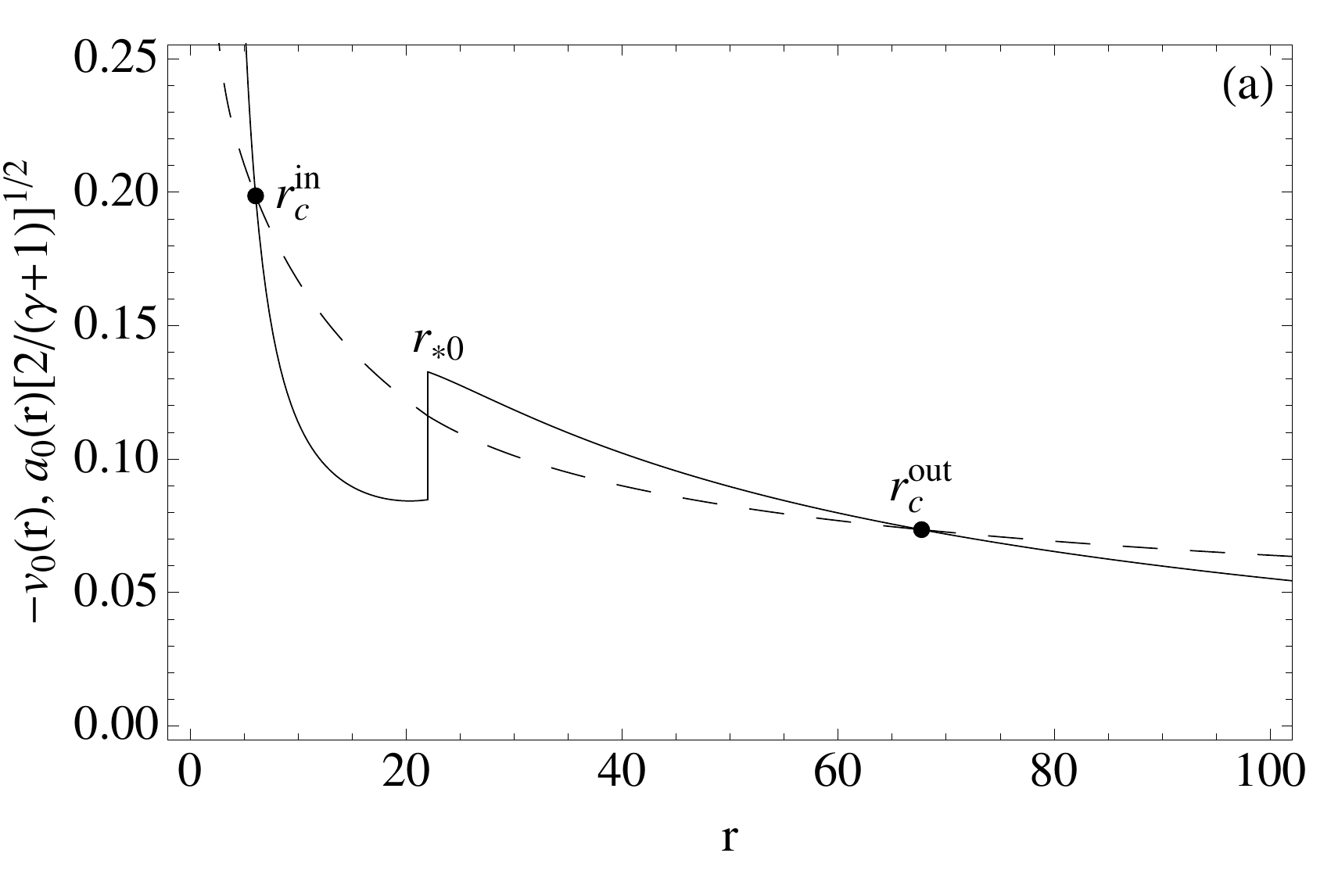}{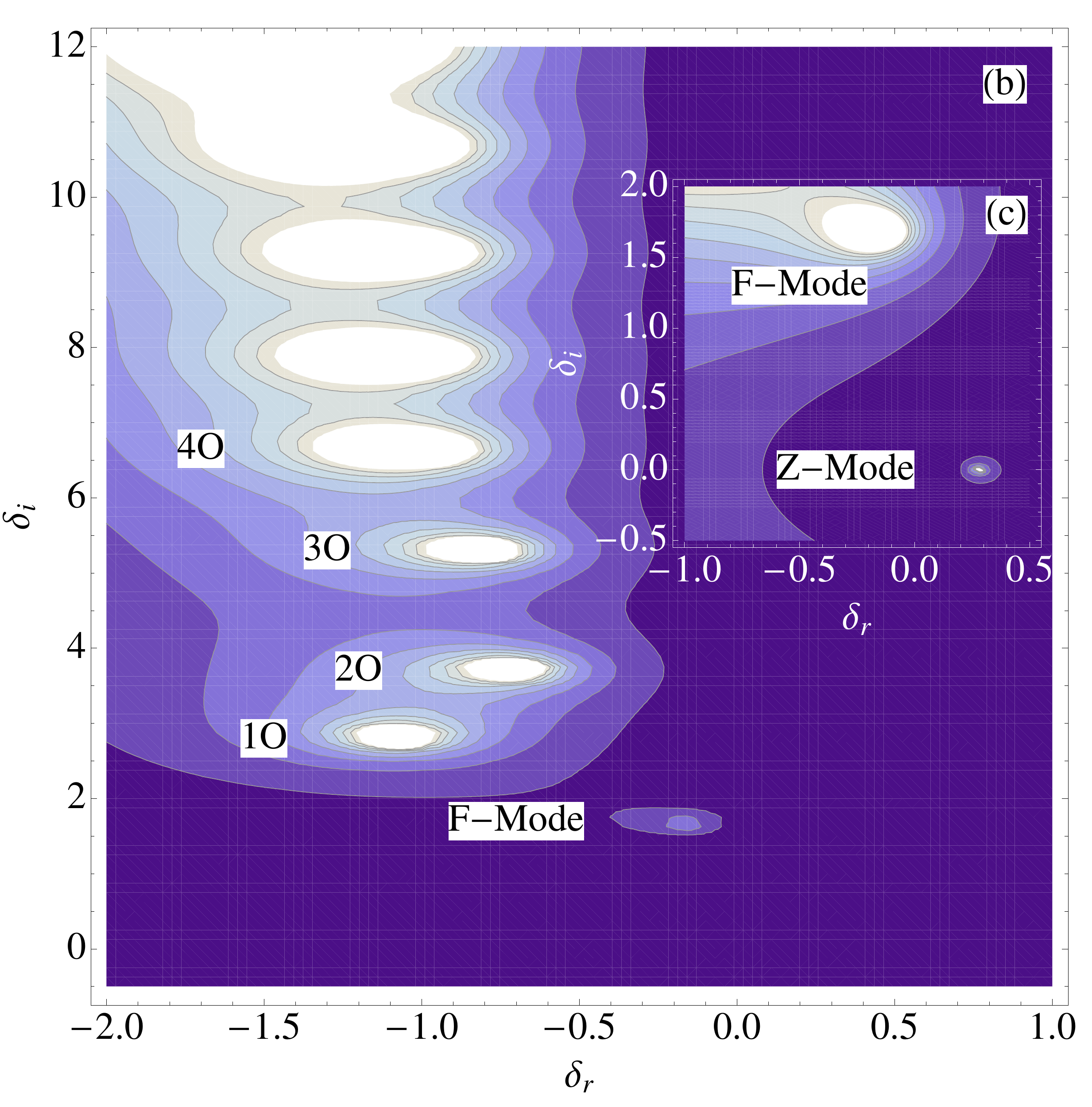}
\caption{\footnotesize Model 4 - inner shock flow: (a) inner shock disk profile with preshock acceleration, and (b) its eigenfrequencies with the inset (c) to show the eigenfrequencies of the Z-Mode and F-Mode at higher resolution. This model has two possible shock flow solutions. The outer shock solution is depicted in Figure~\ref{fig14}.}
\label{fig6} \finfig

\begfig[t] \hskip-0.25in \epsscale{1.15} \plottwo{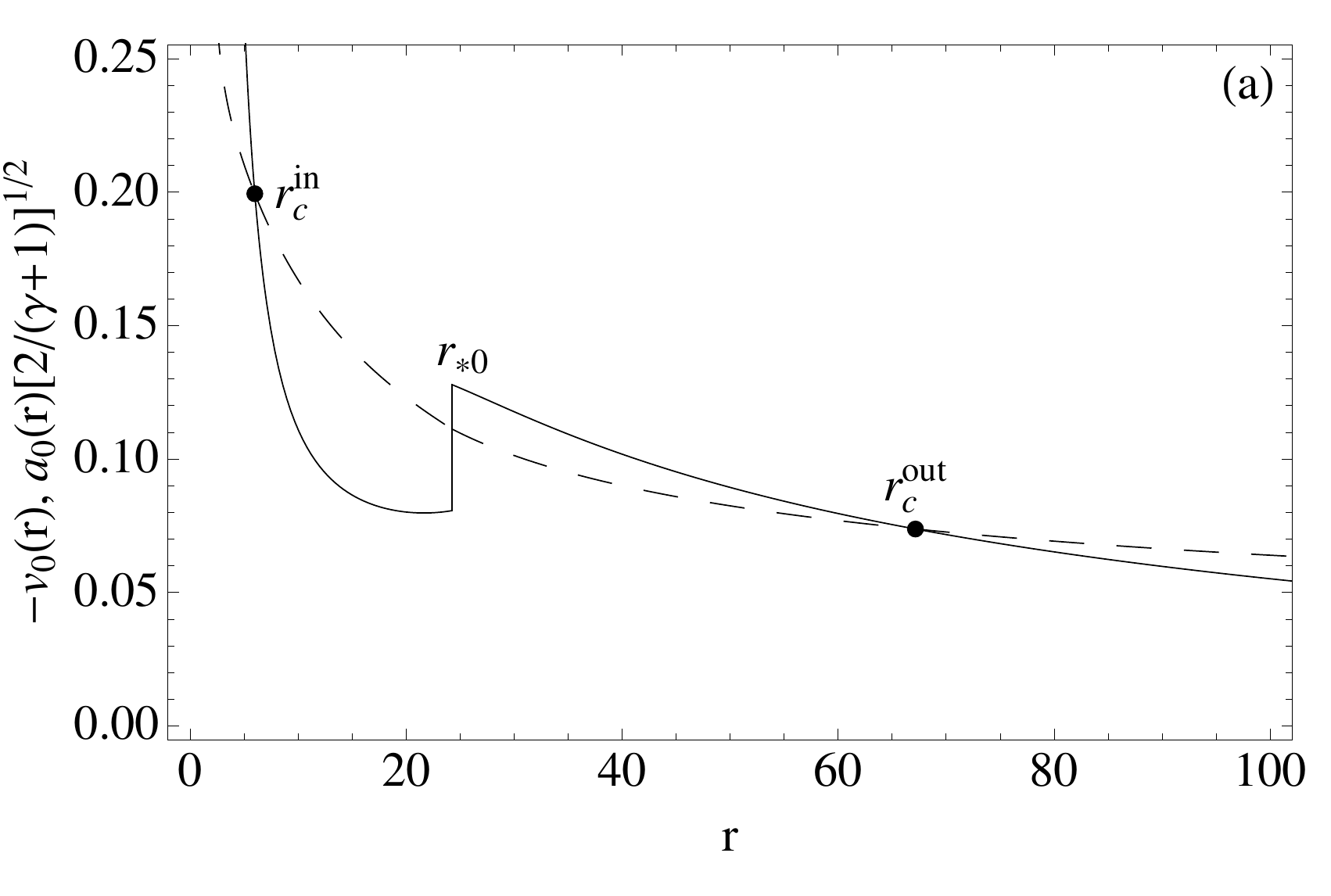}{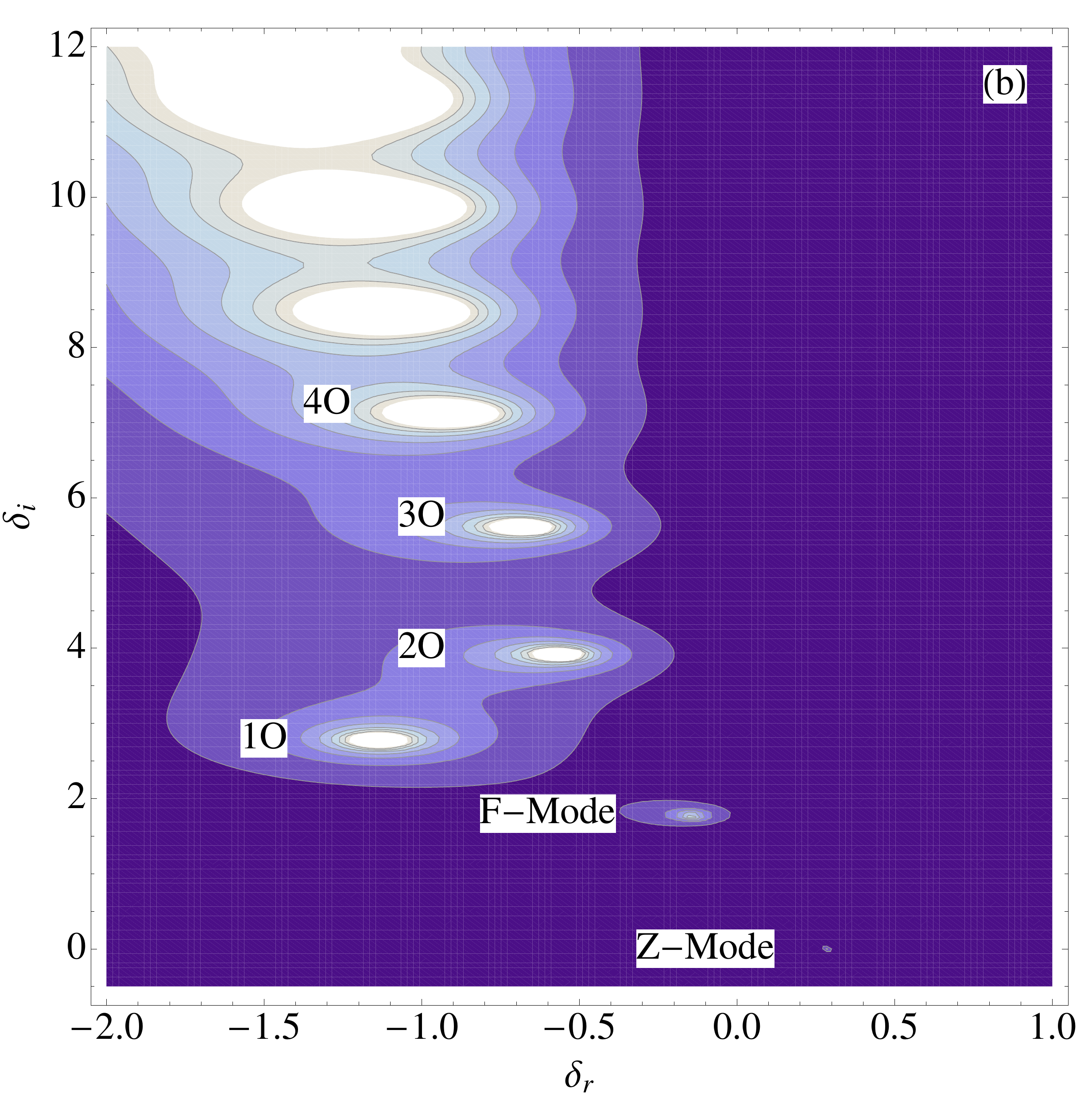}
\caption{\footnotesize Model 5 - inner shock flow: (a) inner shock disk profile with preshock acceleration, and (b) its eigenfrequencies with the inset (c) to show the eigenfrequencies of the Z-Mode and F-Mode at higher resolution. This model has two possible shock flow solutions. The outer shock solution is depicted in Figure~\ref{fig13}.}
\label{fig7} \finfig

\begfig[t] \hskip-0.25in \epsscale{1.15} \plottwo{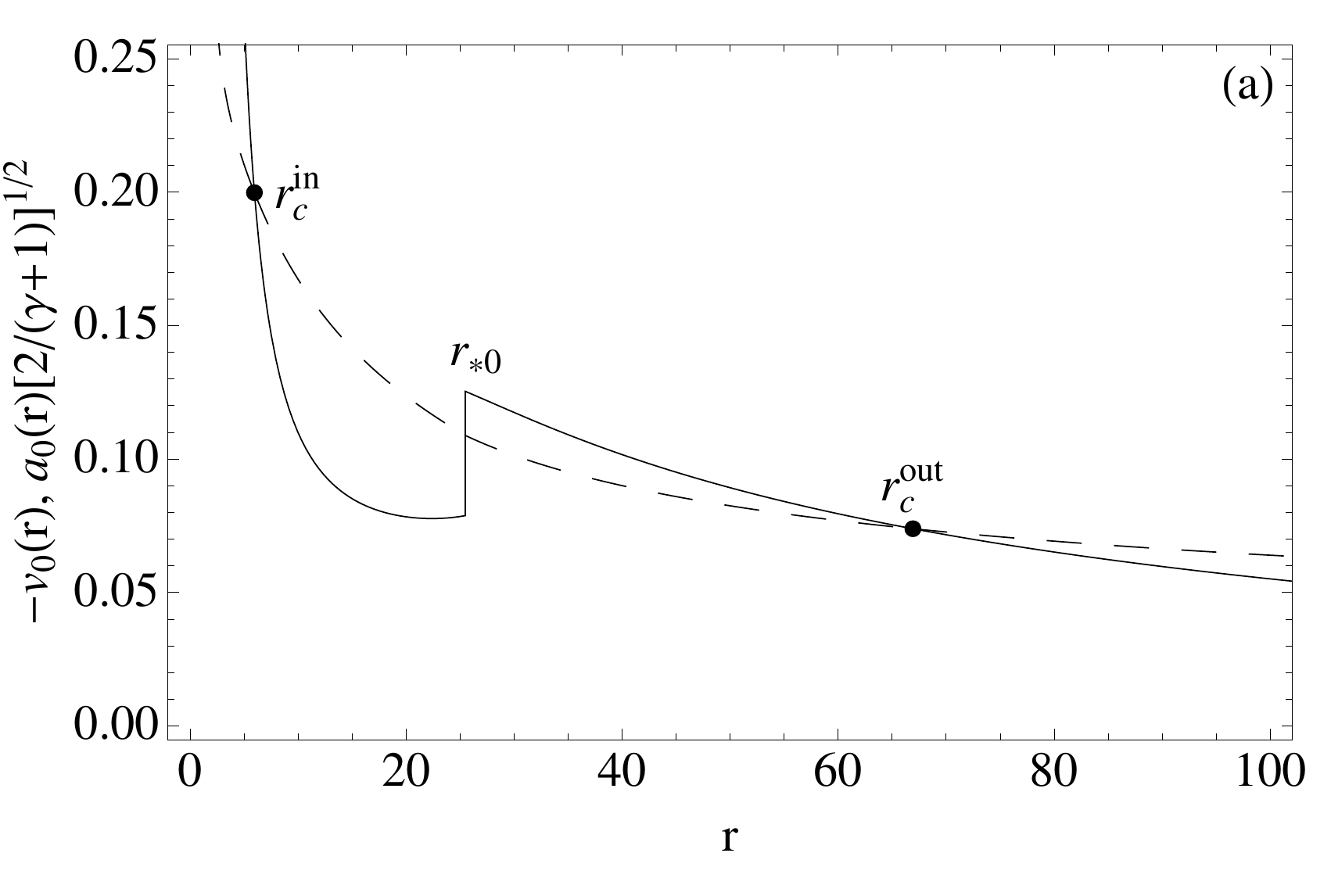}{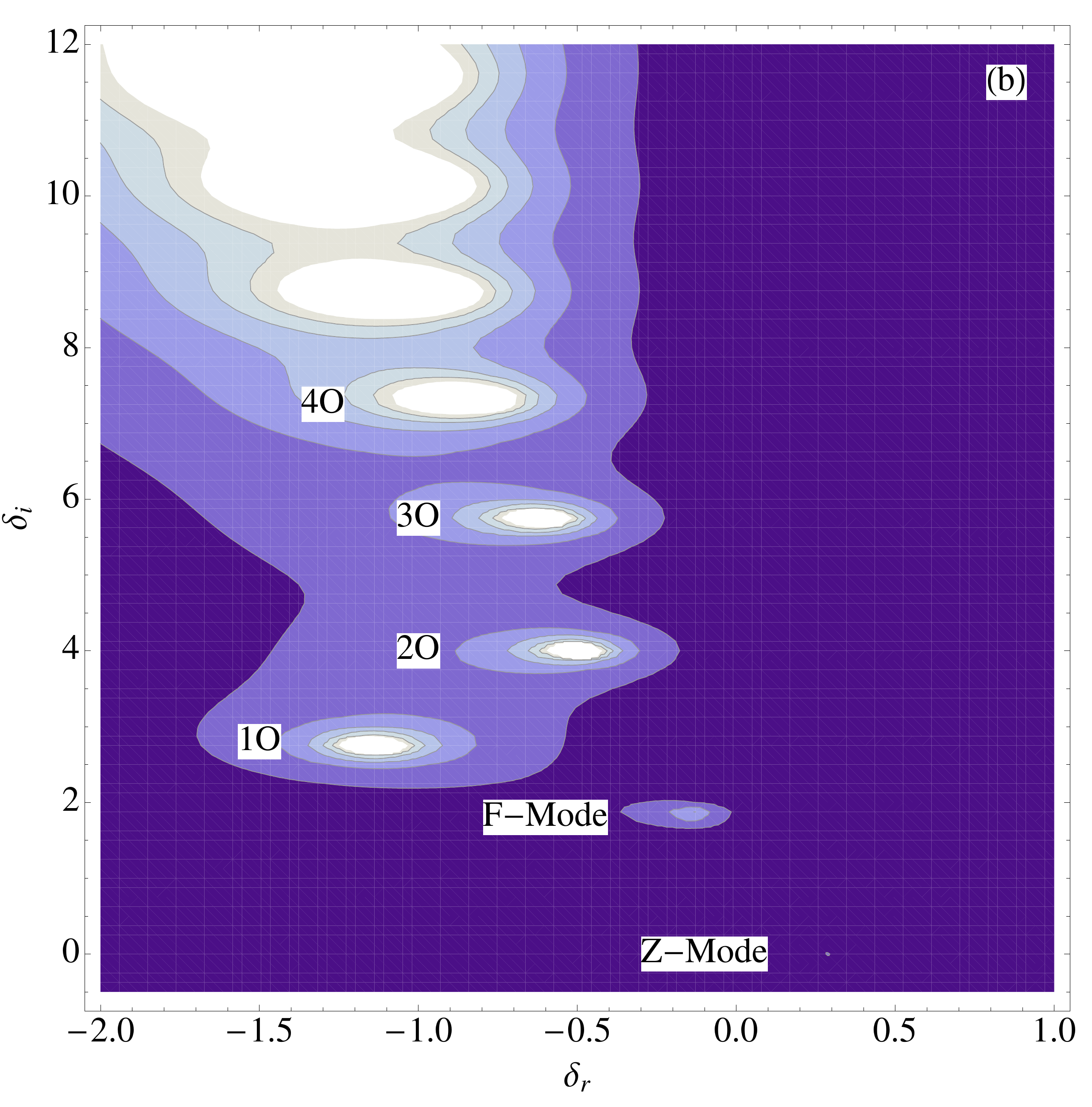}
\caption{\footnotesize Model 6 - inner shock flow: (a) inner shock disk profile with preshock acceleration, and (b) its eigenfrequencies with the inset (c) to show the eigenfrequencies of the Z-Mode and F-Mode at higher resolution. This model has two possible shock flow solutions. The outer shock solution is depicted in Figure~\ref{fig12}.}
\label{fig8} \finfig

\begfig[t] \hskip-0.25in \epsscale{1.15} \plottwo{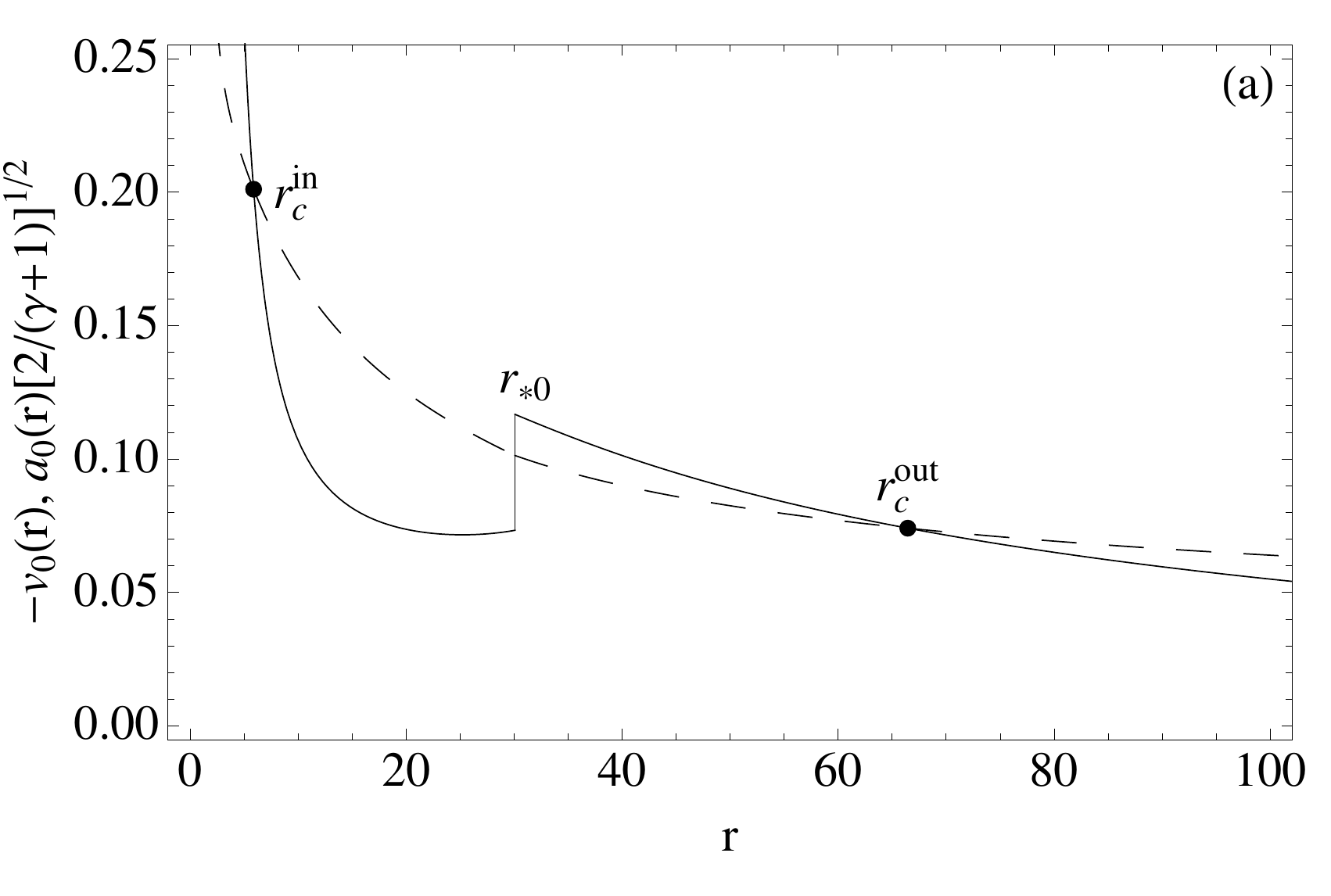}{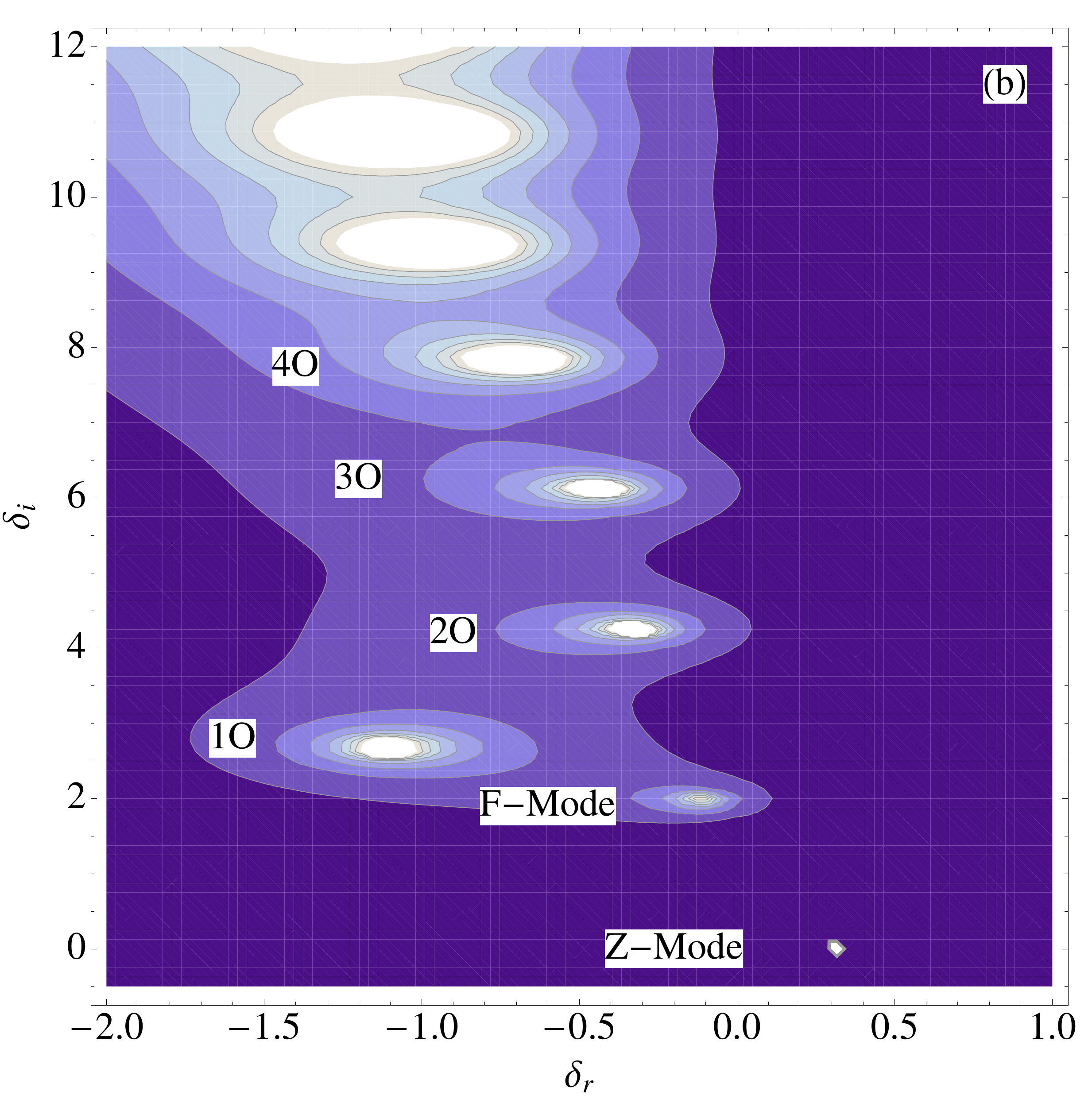}
\caption{\footnotesize Model 7 - inner shock flow: (a) inner shock disk profile with preshock acceleration, and (b) its eigenfrequencies. This model has two possible shock flow solutions. The outer shock solution is depicted in Figure~\ref{fig11}.}
\label{fig9} \finfig

\begfig[t] \hskip-0.25in \epsscale{1.15} \plottwo{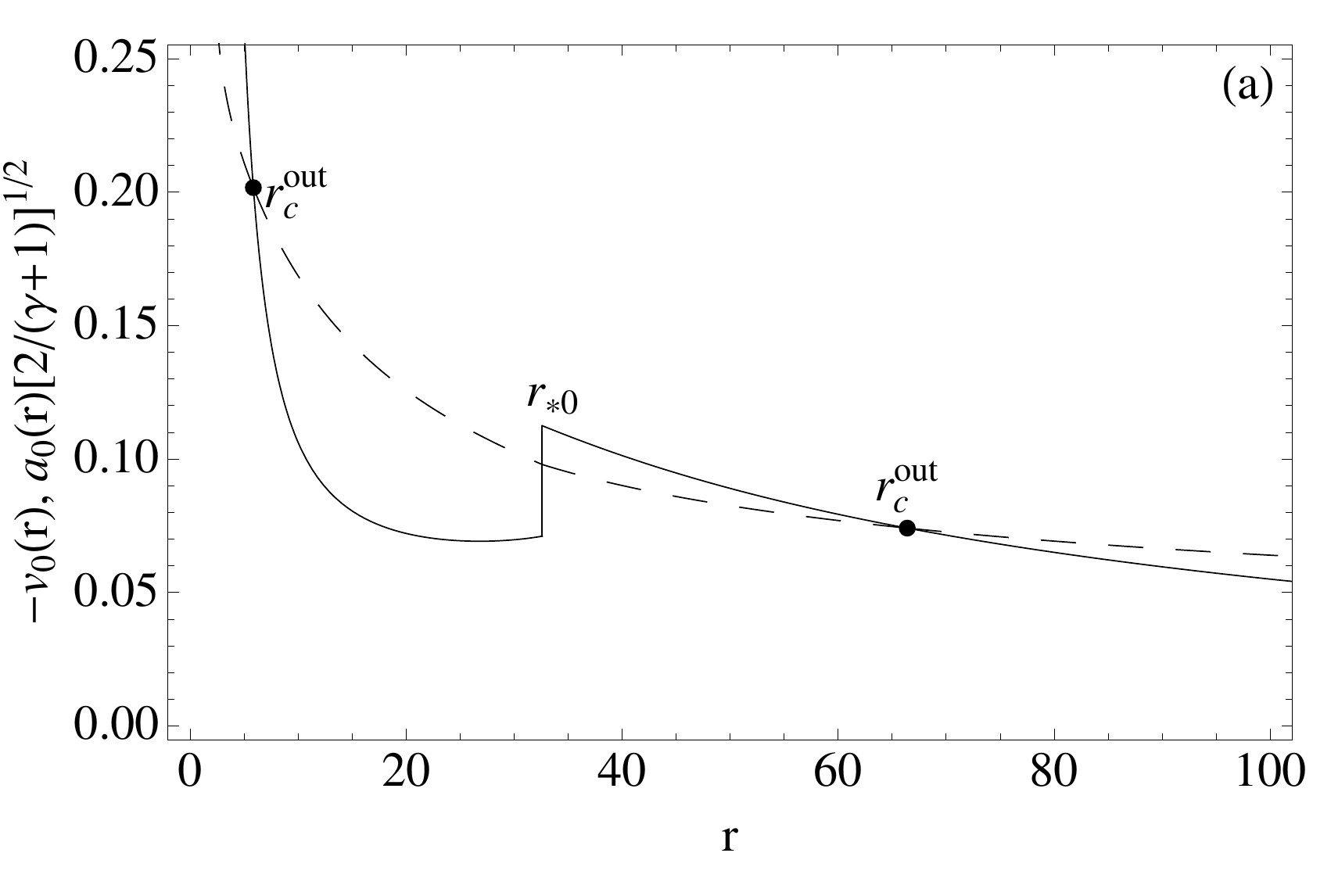}{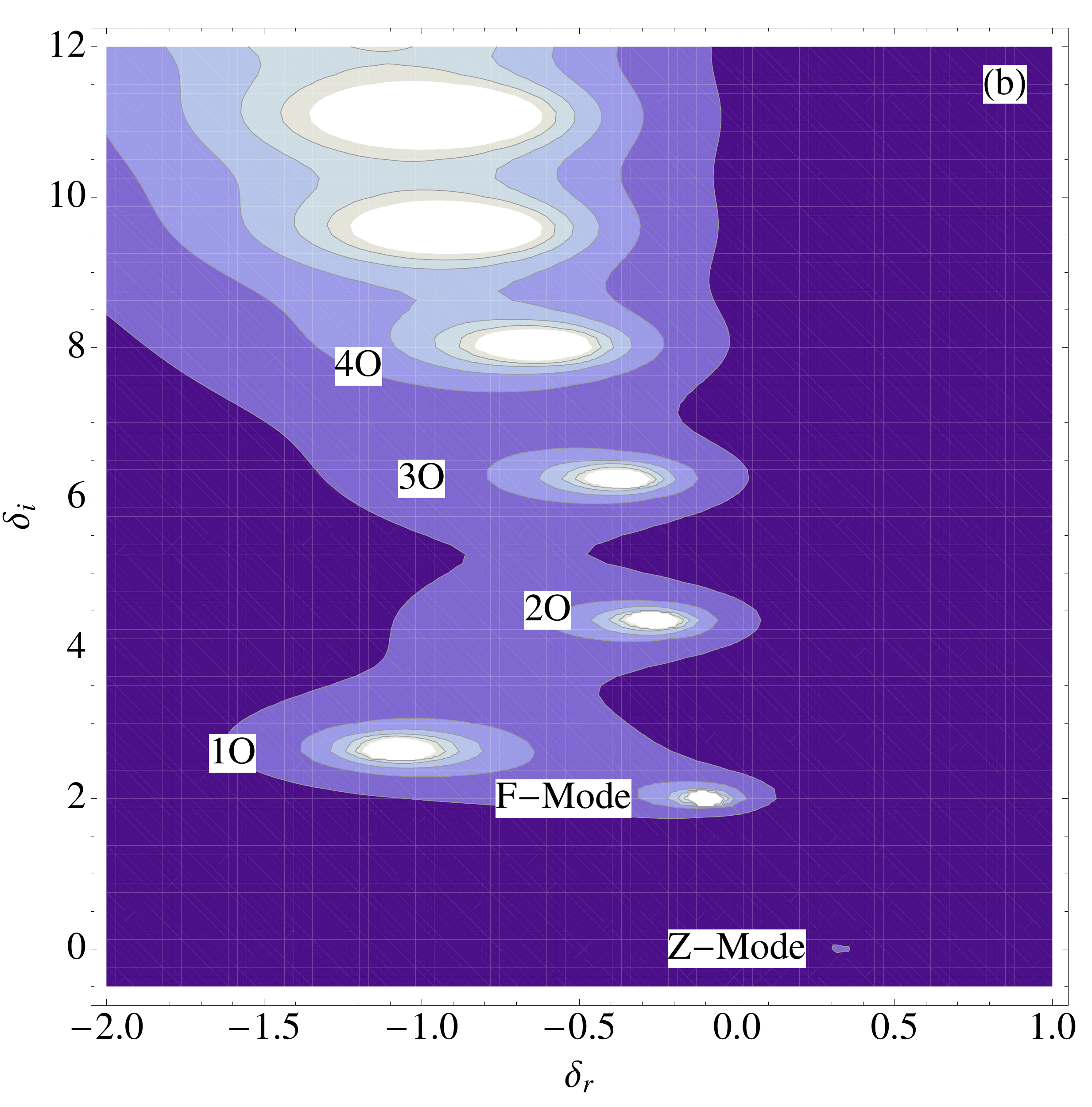}
\caption{\footnotesize Model 8: (a) disk profile with preshock acceleration, and (b) its eigenfrequencies. Only one shock flow solution is possible for this model.}
\label{fig10} \finfig

\begfig[t] \hskip-0.25in \epsscale{1.15} \plottwo{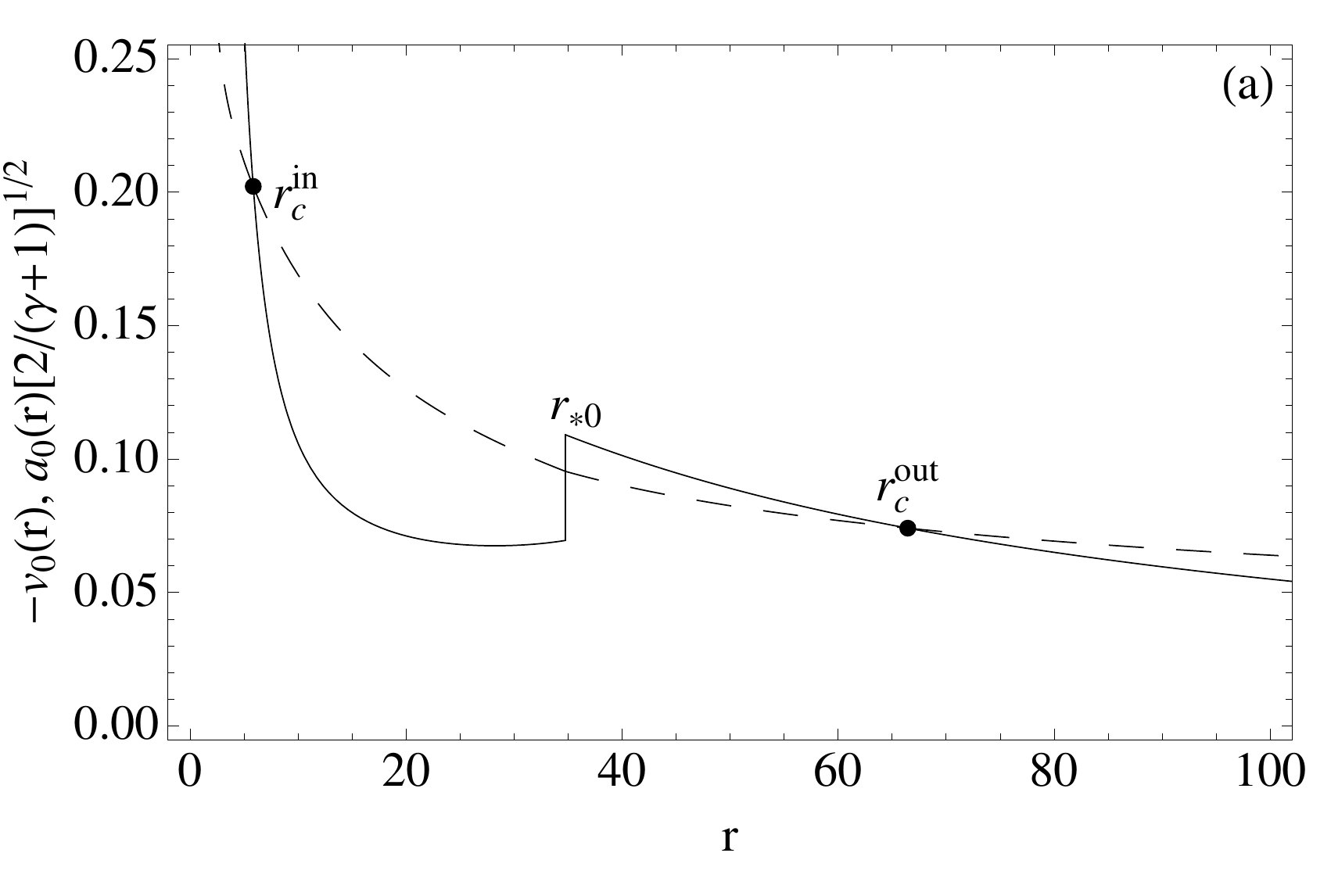}{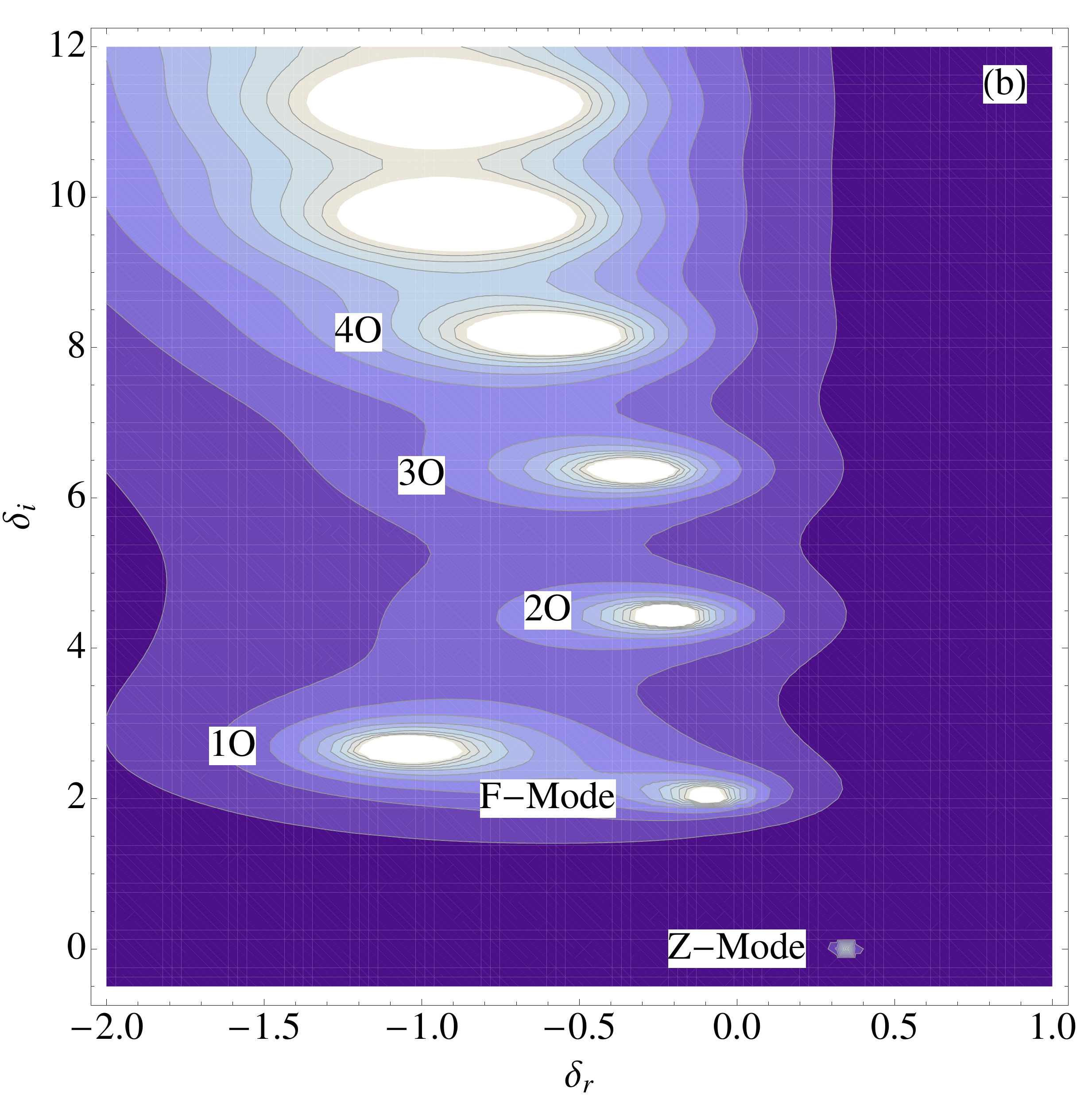}
\caption{\footnotesize Model 7 - outer shock flow: (a) outer shock disk profile with preshock acceleration, and (b) its eigenfrequencies.}
\label{fig11} \finfig

\begfig[t] \hskip-0.25in \epsscale{1.15} \plottwo{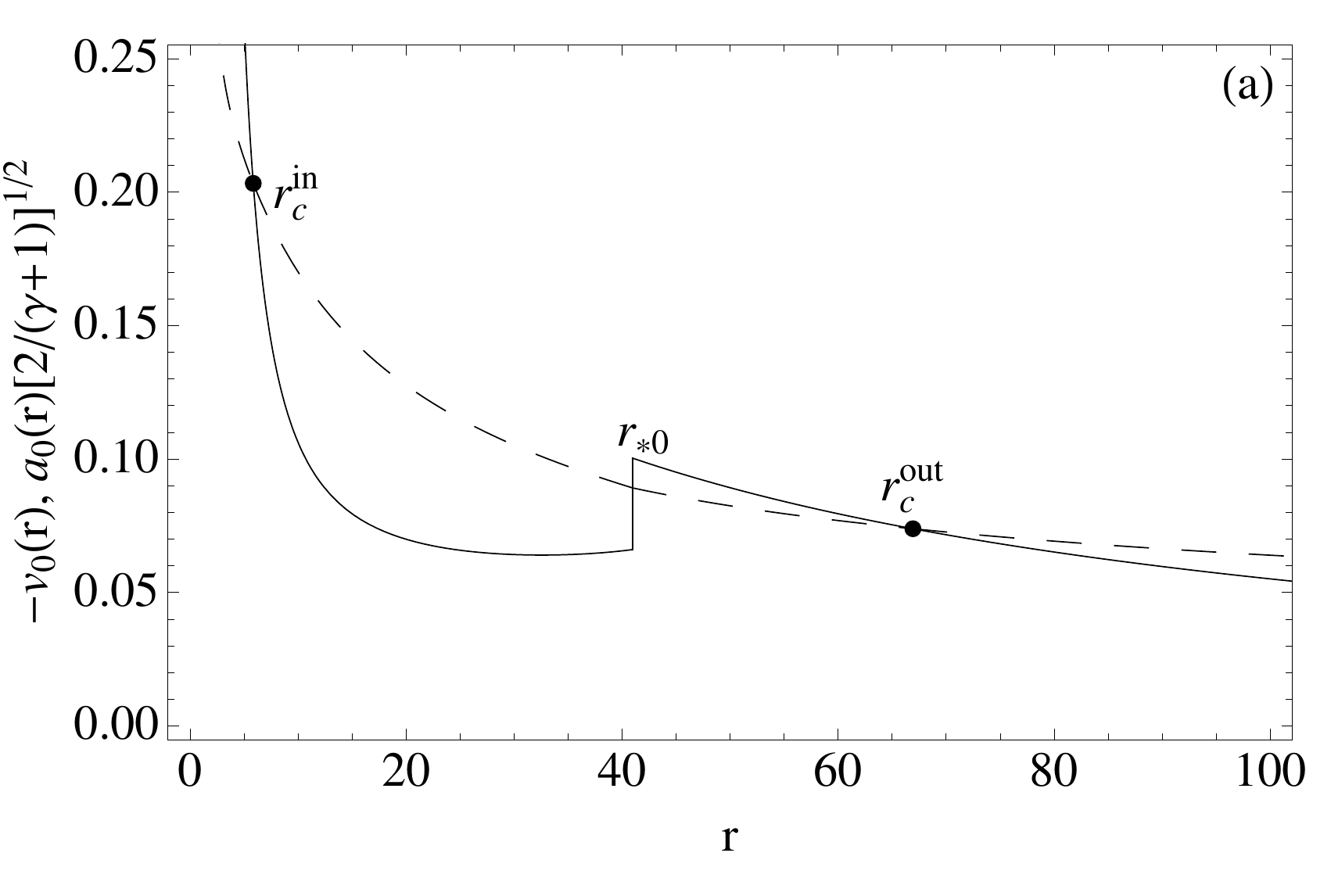}{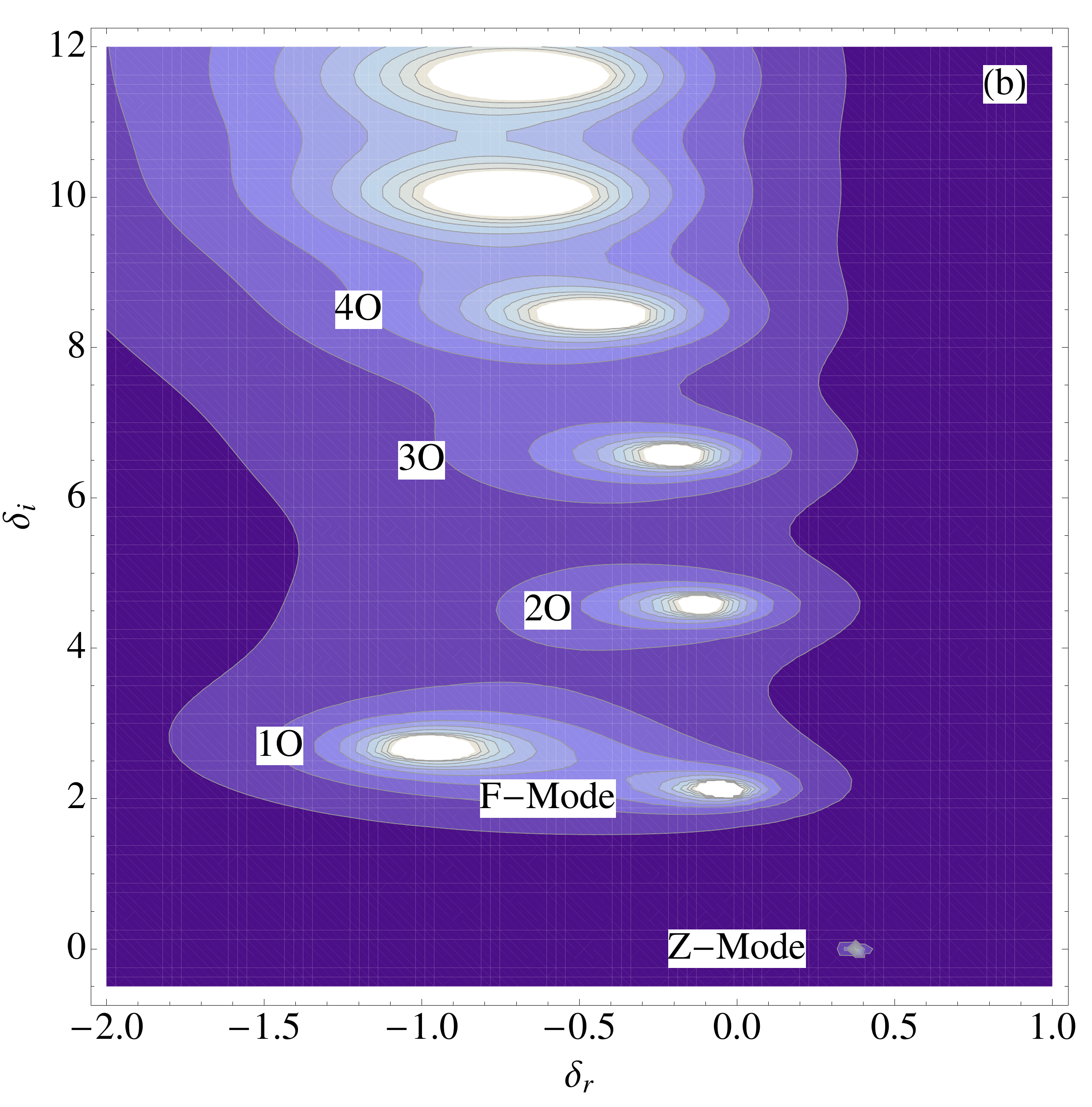}
\caption{\footnotesize Model 6 - outer shock flow: (a) outer shock disk profile with preshock acceleration, and (b) its eigenfrequencies.}
\label{fig12} \finfig

\begfig[t] \hskip-0.25in \epsscale{1.15} \plottwo{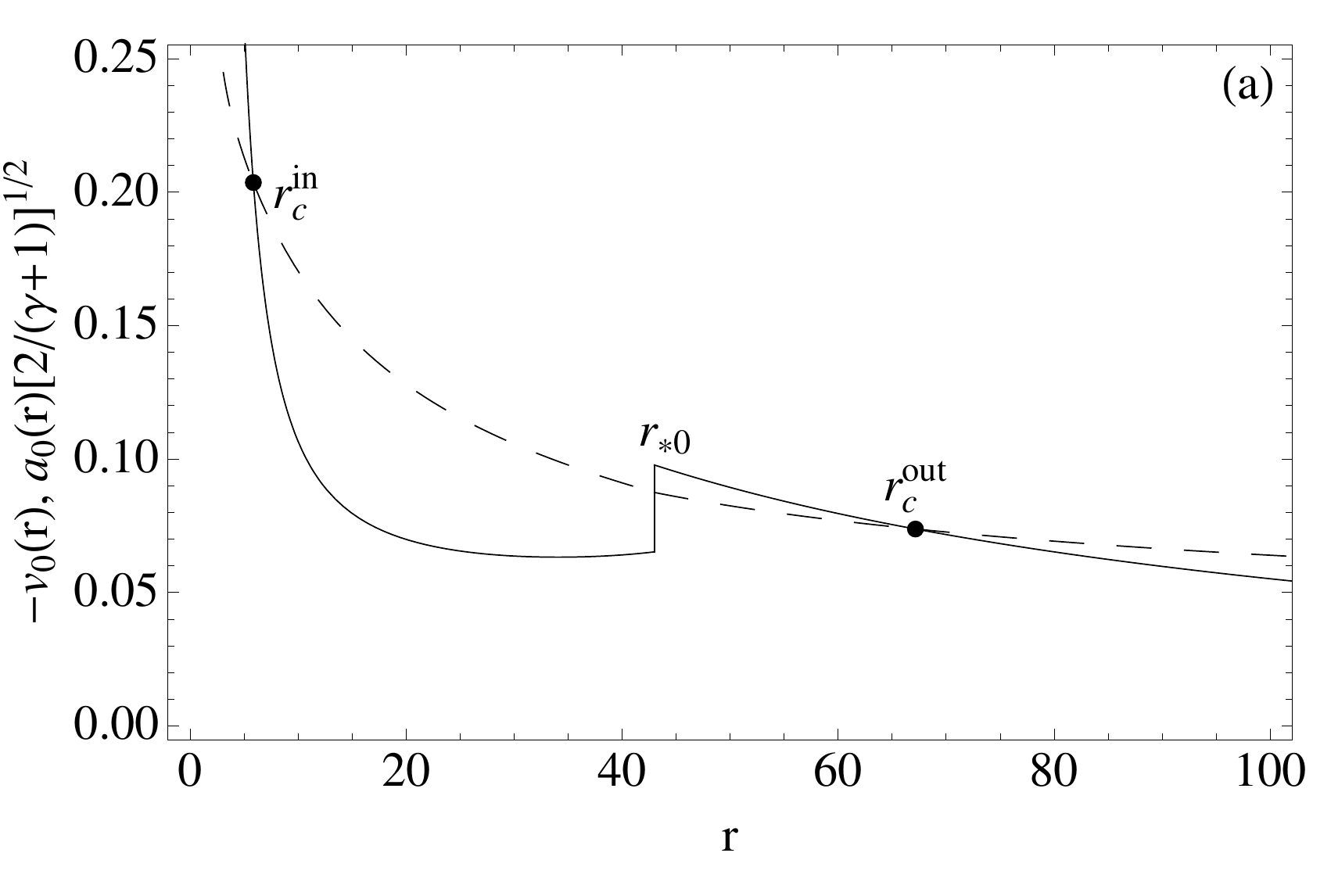}{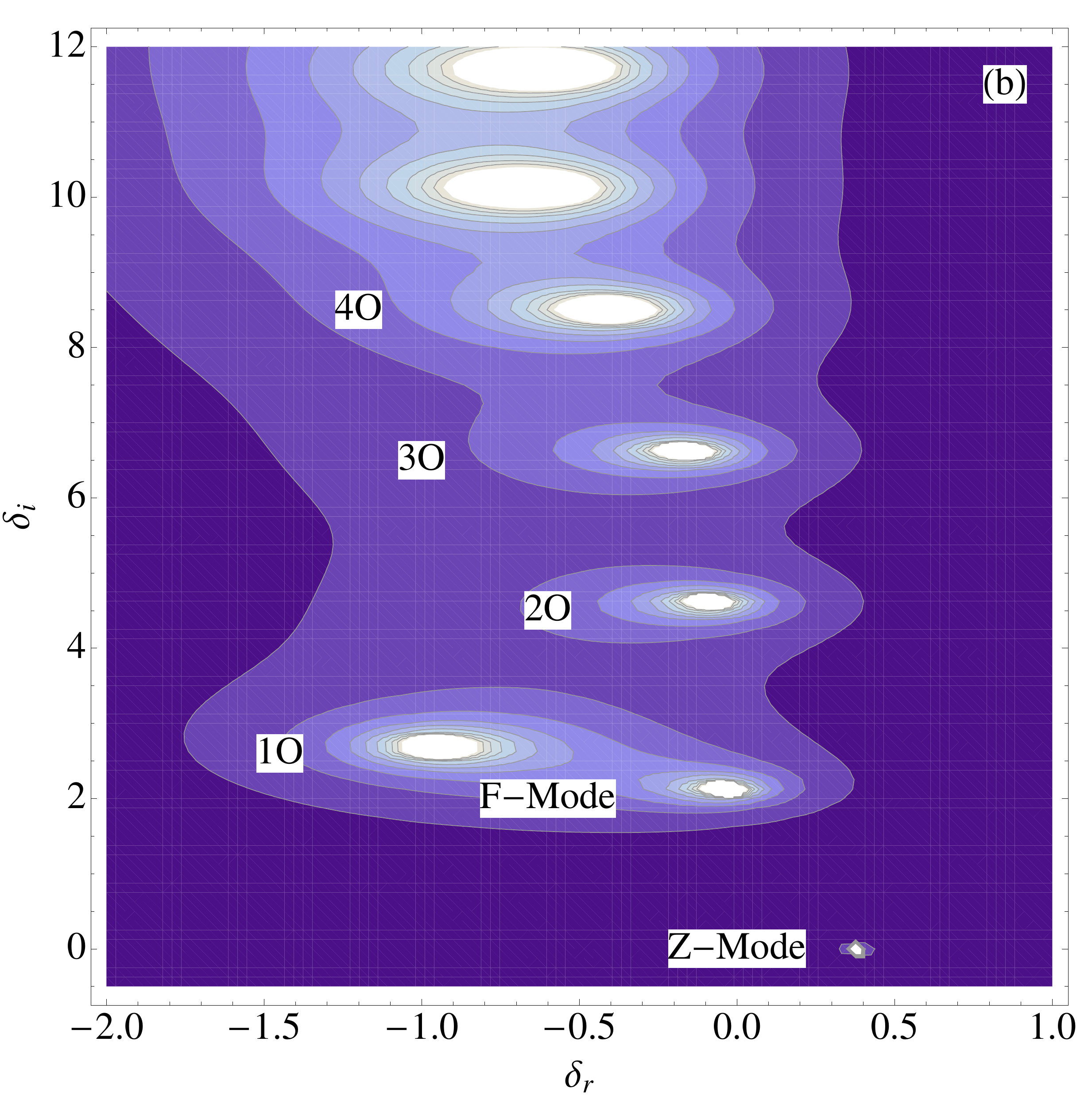}
\caption{\footnotesize Model 5 - outer shock flow: (a) outer shock disk profile with preshock acceleration, and (b) its eigenfrequencies.}
\label{fig13} \finfig

\begfig[t] \hskip-0.25in \epsscale{1.15} \plottwo{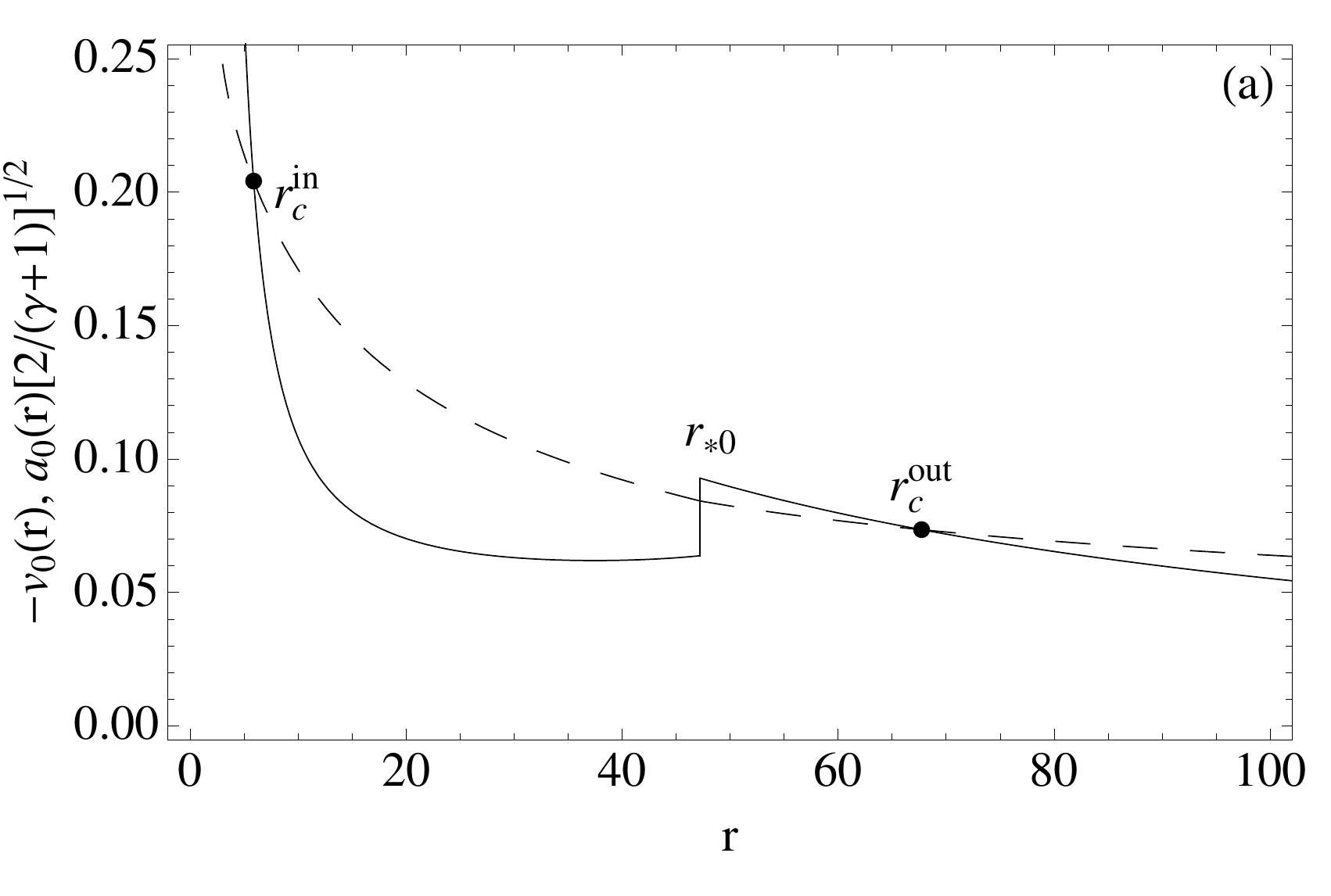}{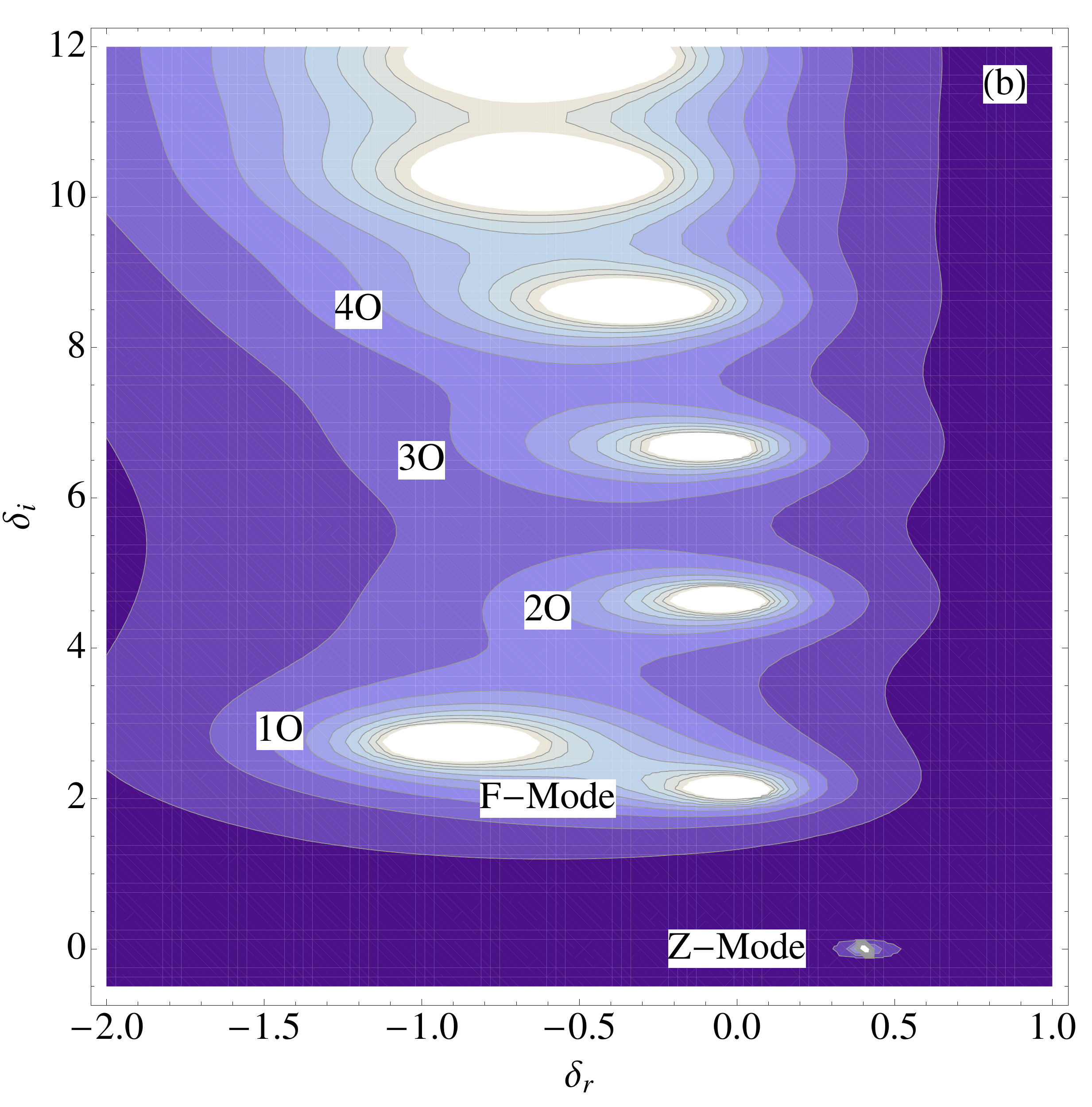}
\caption{\footnotesize Model 4 - outer shock flow: (a) outer shock disk profile with preshock acceleration, and (b) its eigenfrequencies.}
\label{fig14} \finfig

\begfig[t] \hskip-0.25in \epsscale{1.15} \plottwo{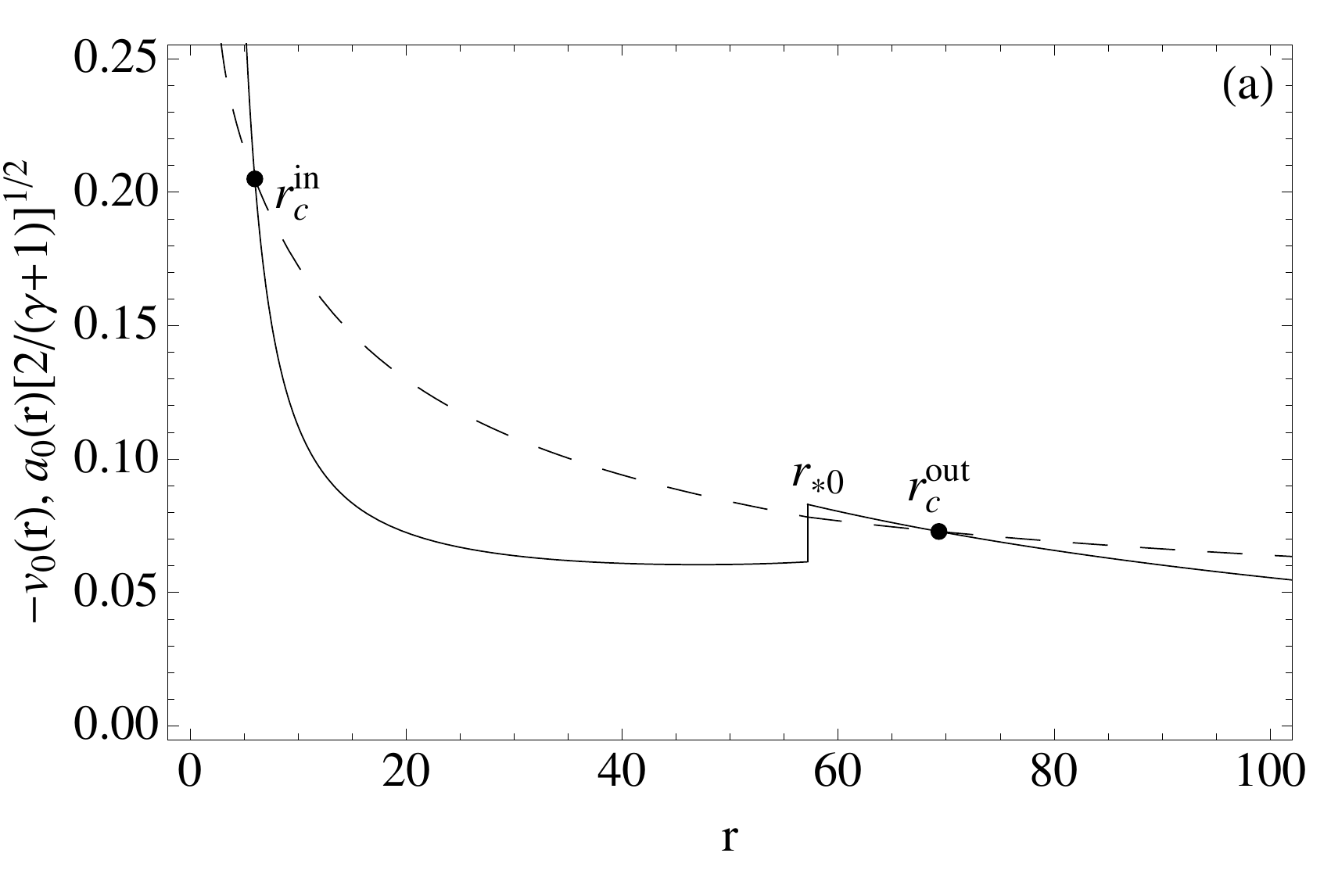}{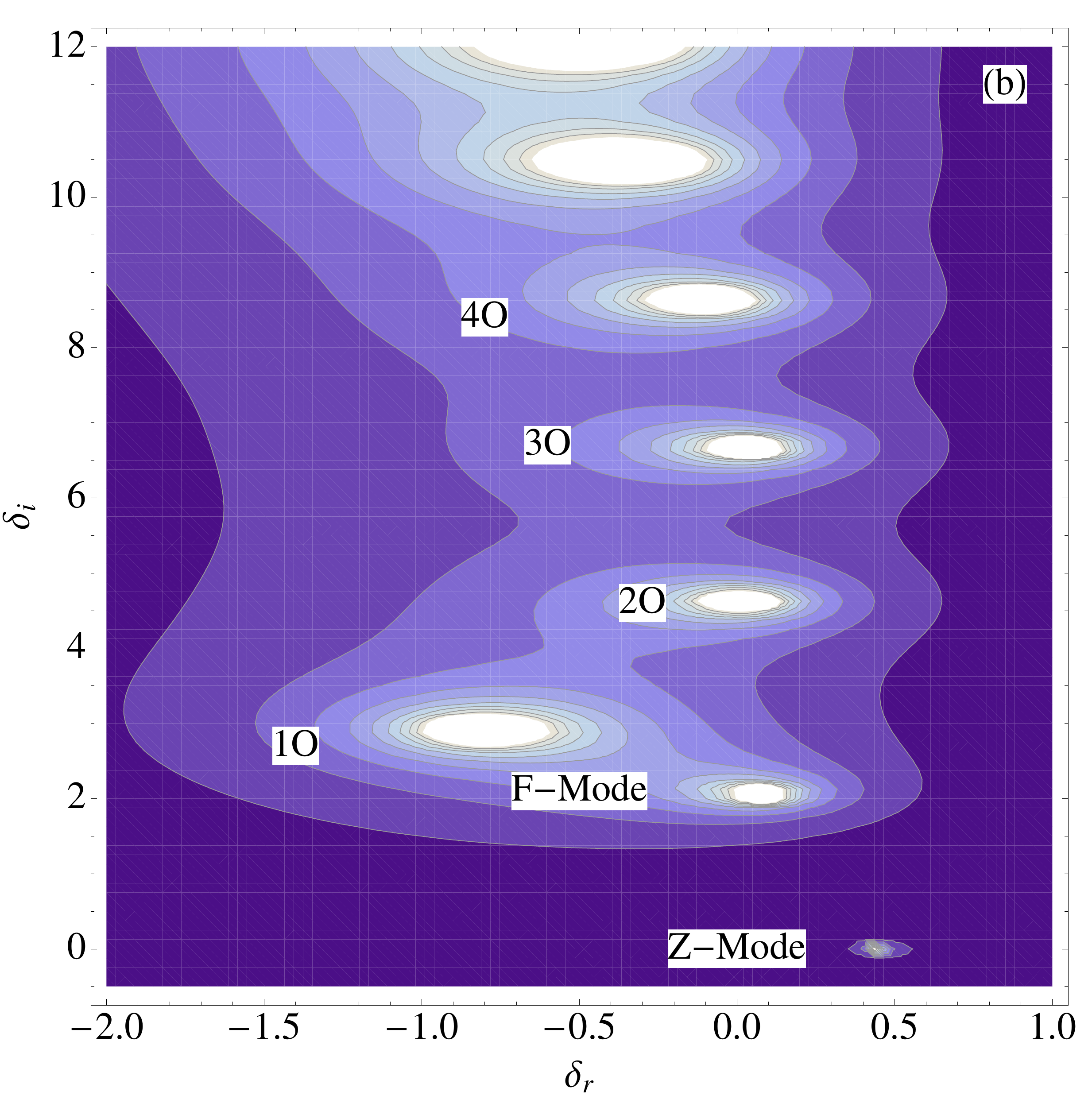}
\caption{\footnotesize Model 3 - outer shock flow: (a) outer shock disk profile with preshock acceleration, and (b) its eigenfrequencies.}
\label{fig15} \finfig

\begfig[t] \hskip-0.25in \epsscale{1.15} \plottwo{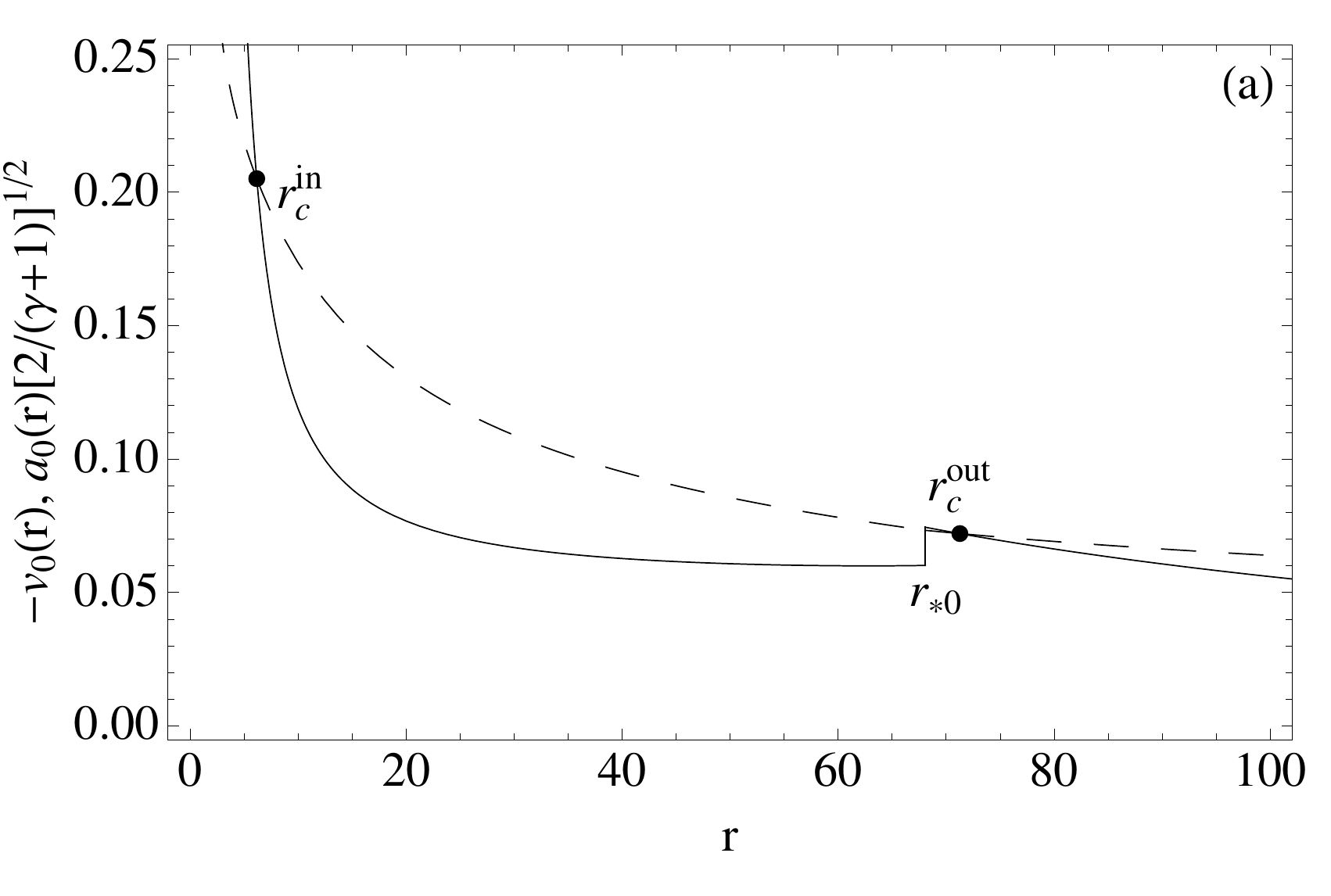}{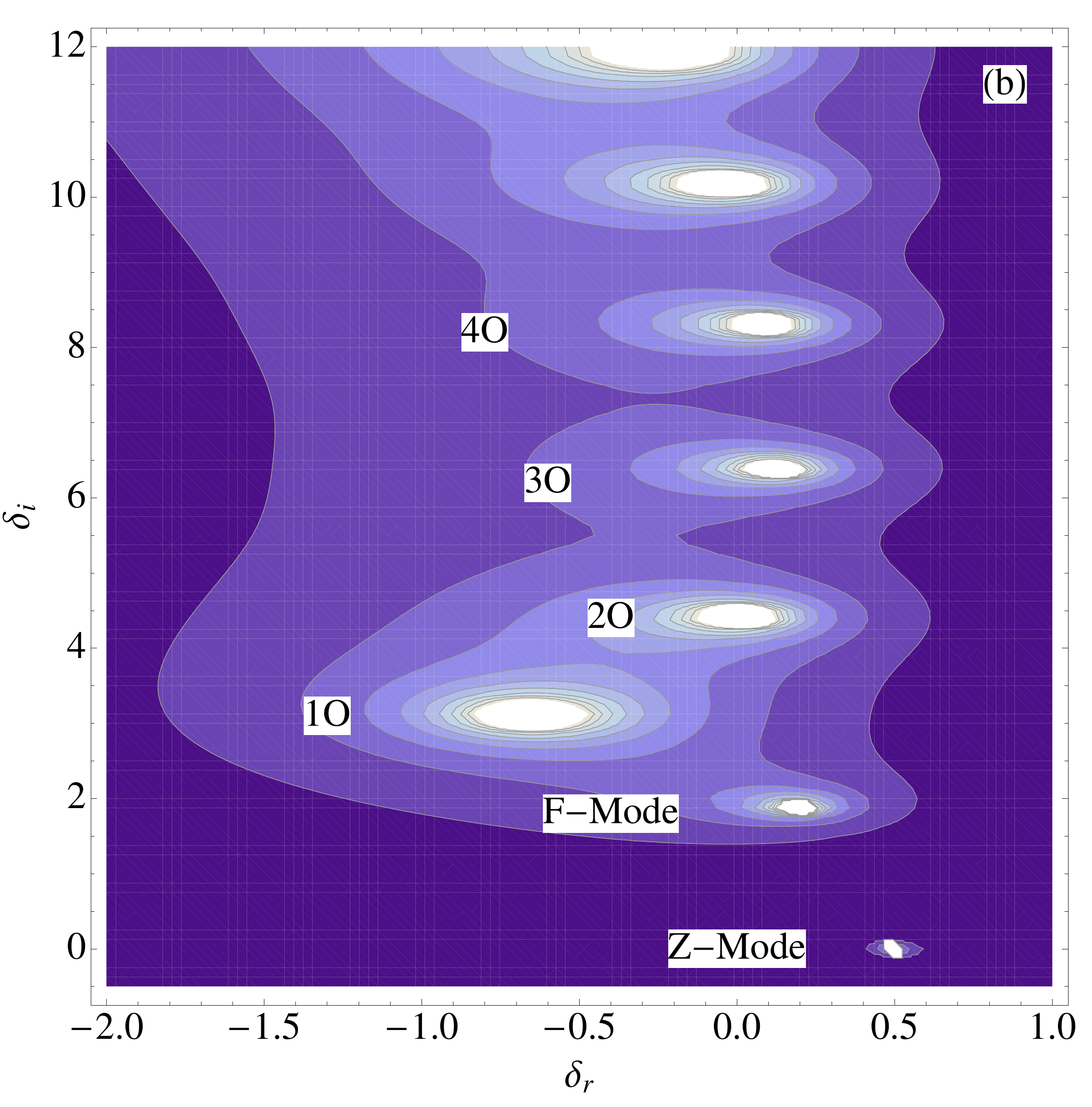}
\caption{\footnotesize Model 2 - outer shock flow: (a) outer shock disk profile with preshock acceleration, and (b) its eigenfrequencies.}
\label{fig16} \finfig

\begfig[t] \hskip-0.25in \epsscale{1.15} \plottwo{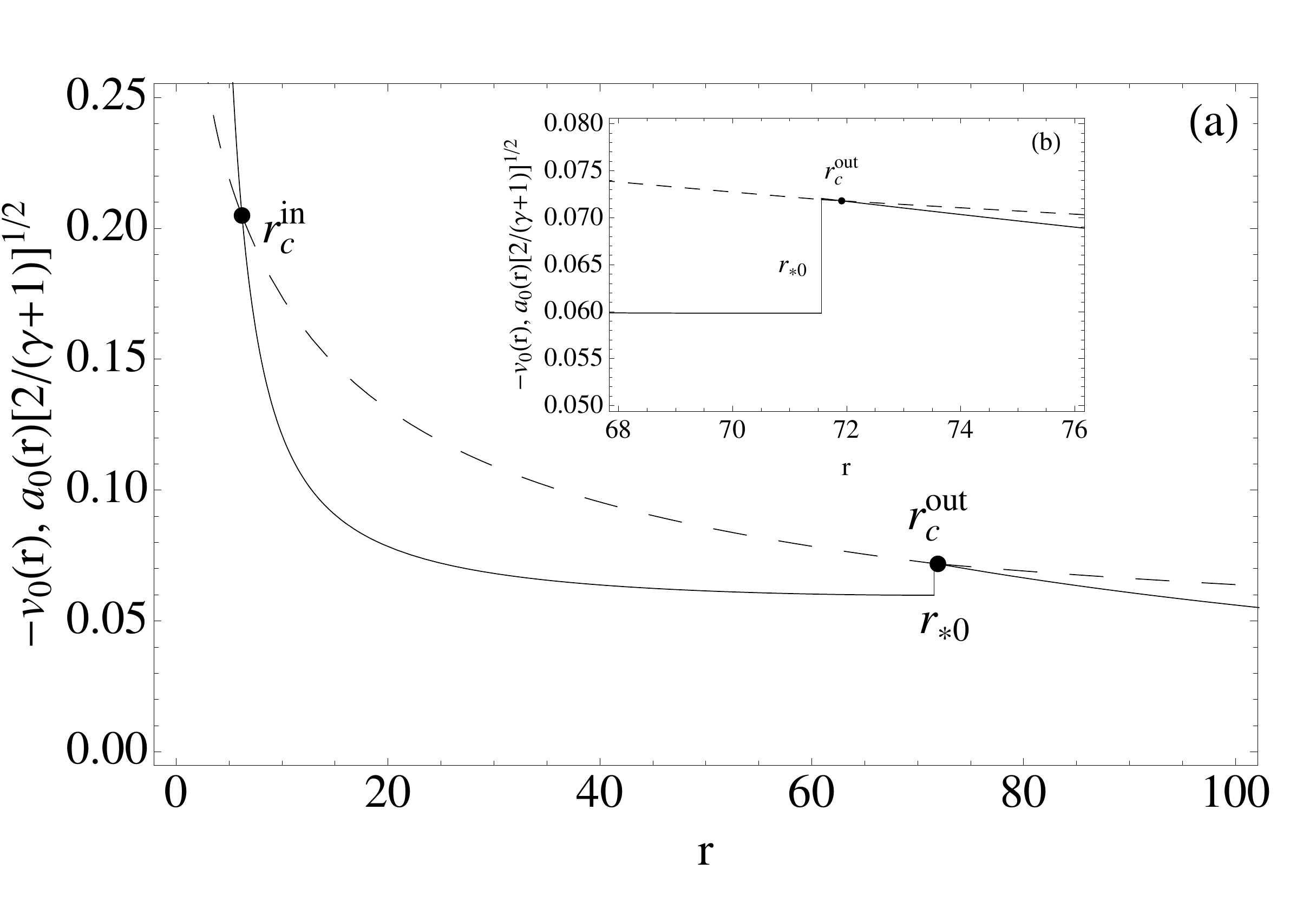}{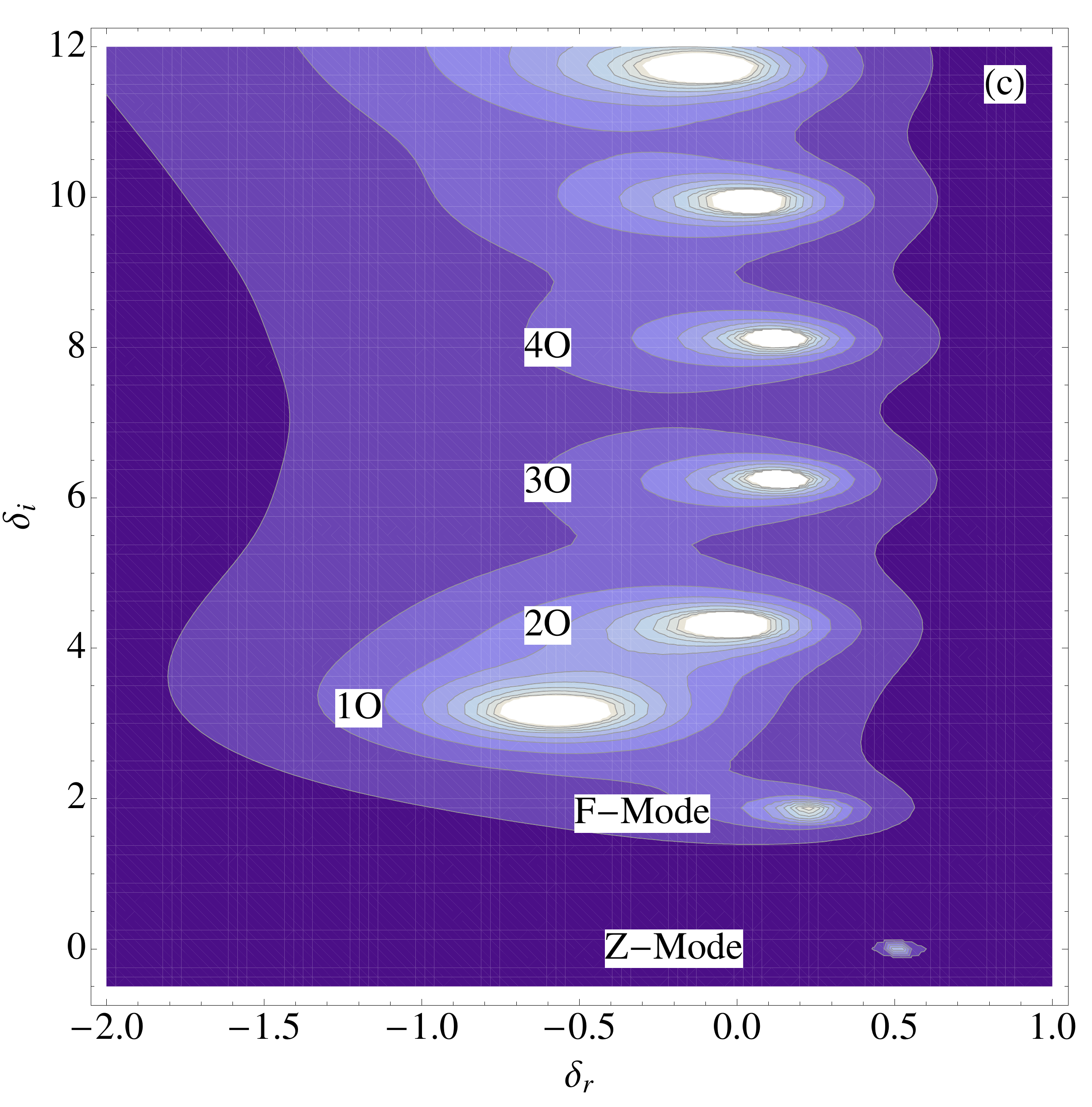}
\caption{\footnotesize Model 1 - outer shock flow: (a) outer shock disk profile with preshock acceleration, (b) a zoom-in at the outer sonic location to clarify that the shock location is not at the outer sonic point, and (c) its eigenfrequencies.}
\label{fig17} \finfig

\section{HYDRODYNAMIC EQUATIONS}

\subsection{Time-dependent Fluid Equations}

We assume a 1D axisymmetric inviscid accretion disk flow structure based on the conservation of mass, radial momentum, angular momentum, and internal energy, described by the equations
\begeq %
\frac{\partial}{\partial t} (H \rho) + \frac{1}{r} %
\frac{\partial}{\partial r} (r v \rho H) = 0, %
\label{eq1} %
\fineq

\begeq %
\frac{\partial v}{\partial t} + v \frac{\partial %
v}{\partial r} + \frac{1}{\rho} \frac{\partial P}{\partial r} - r( %
\Omega^2 - \Omega_{\rm K}^2) = 0, %
\label{eq2}
\fineq

\begeq  %
\ell \equiv r^2 \Omega =  {\rm constant} \, , %
\label{eq3} %
\fineq

\begeq %
\frac{\partial U}{\partial t} + v \frac{\partial %
U}{\partial r} - \frac{\gamma U}{\rho} \left(\frac{\partial %
\rho}{\partial t} + v \frac{\partial \rho}{\partial r} \right) = 0,  %
\label{eq4} %
\fineq
respectively, where $v$ denotes the radial velocity (defined to be negative for inflow), $\rho$ is the mass density, $\Omega$ is the angular velocity, $\ell$ is the accreted specific angular momentum, $H$ is the disk half-thickness, $\OmegaK$ is the Keplerian angular velocity, $U$ is the internal energy density, and $P=(\gamma -1) U$ is the gas pressure. Each of the various quantities represents a vertical average over the disk structure, and the ratio of specific heats, $\gamma$, is constant throughout the flow.

The effects of general relativity are incorporated by utilizing the pseudo-Newtonian gravitational potential per unit mass given by
\begeq  %
\Phi(r) = -\frac{G M}{r-\rs} \ , %
\label{eq5} %
\fineq
where $\rs = 2 GM/c^2$ is the Schwarzschild radius for a BH of mass $M$. The Keplerian angular velocity $\Omega_{\rm K}$ of matter in a circular orbit at radius $r$ in the pseudo-Newtonian potential is 
\begeq %
\Omega_{\rm K}^2 = \frac{G M}{r(r-\rs)^2} = \frac{1}{r}\frac{d\Phi}{dr} \ . %
\label{eq6} %
\fineq
The pseudo-Newtonian potential correctly reproduces the radii of the event horizon, the marginally bound orbit, and the marginally stable orbit~\citep{pw80}.  Hence, the use of such a potential allows one to approximate the spacetime geometry of a Schwarzschild BH.

The disk half-thickness $H$ in Equation (\ref{eq1}) is given by the standard hydrostatic prescription
\begeq %
H = \frac{a}{\Omega_{\rm K}} \ ,  %
\label{eq7} %
\fineq
where $a$ represents the adiabatic sound speed,
\begeq %
a \equiv \left(\frac{\gamma P}{\rho}\right)^{1/2} . \ %
\label{eq8} %
\fineq
Since the flow is purely adiabatic in the absence of viscosity, the pressure and density variations are coupled according to the standard relation
\begeq %
P = C_0 \rho^\gamma \ , %
\label{eq9} %
\fineq
where $C_0$ is a parameter related to the specific entropy  that remains constant  except at the location of an isothermal shock if one is present. 

\subsection{Inviscid Steady-state Equations}

The mass transport rate $\dot{M}$, the angular momentum transport rate $\dot{J}$, and the energy transport rate $\dot{E}$ have been demonstrated by~\citet{bl03} and \citet{bs05} to be conserved in steady ADAF disks in the absence of outflows. In the inviscid case treated here, the transport rates can be written as  
\begeq
\dot{M} = -4 \pi r H_{_0} \rho_{_0} v_{_0} \, ,
\label{eq10}
\fineq
\begeq
\dot{J} = \dot{M} r^2 \Omega \, ,
\label{eq11}
\fineq
and
\begeq
\dot{E} = \dot{M} \left(\frac{v^2_{\phi}}{2} + \frac{v^2_{_0}}{2} + \frac{P_{_0} + U_{_0}}{\rho_{_0}} + \Phi \right) \, ,
\label{eq12}
\fineq
where $\rho_{_0}$, $H_{_0}$, $v_{_0}$, $P_{_0}$, and $U_{_0}$ are defined as above but for a steady flow solution, and $v_{\phi} = r \Omega$ is the azimuthal velocity. Under the inviscid flow assumption, the disk/shock model depends on three fundamental parameters, namely, the conserved energy transport rate per unit mass $\epsilon \equiv \dot{E}/\dot{M}$, the conserved angular momentum transport per unit mass $\ell \equiv \dot{J}/\dot{M}$ (see Equation (\ref{eq3})), and the gas specific heat ratio $\gamma$ as discussed above. We assume that $\gamma=1.5$ in the present paper, to reflect the contributions to the pressure from the gas and the equipartition magnetic field \citep[e.g.,][]{nkh97}. The value of $\epsilon$ is constant in ADAF-type flows since there are no radiative losses, although it will jump at the location of an isothermal shock if one is present. On the other hand, the value of $\ell$ remains constant  throughout the disk since the flow is inviscid. By combining our previous relations, we can rewrite Equation (\ref{eq12}) as
\begeq %
\epsilon \equiv {1 \over 2} \, {\ell^2 \over r^2} + {1
\over 2} \, \vZ^2 + {1 \over \gamma-1} \, \aZ^2 - {GM \over r-\rs} \, ,%
\label{eq13} %
\fineq
where the steady-state solution is denoted by the subscript ``0.'' Note that $\epsilon$ is defined here to be positive for an inward flow of energy into the BH. 

Dissipation and radiative losses are unimportant in inviscid  advection-dominated disks, and therefore the flow is in general adiabatic and isentropic, except at the shock location. In this situation, the ``entropy parameter,''
\begeq %
K \equiv -\vZ \, \aZ^{(\gamma+1)(\gamma-1)} \, r^{3/2} \, (r-\rs) \ , %
\label{eq14} %
\fineq
is conserved throughout the flow, except at the location of a shock~\citep{bl03}. When a shock is present, we shall use the subscripts ``$-$'' and ``$+$'' to refer to quantities measured just upstream (preshock)  and just downstream (postshock) from the shock, respectively. In the isothermal shock model, $K$ and $\epsilon$ have smaller values in the postshock region ($K_{+}$, $\epsilon_{+}$) compared with the preshock region ($K_{-}$, $\epsilon_{-}$)~\citep[e.g.,][]{lb05}. Equation~(\ref{eq14}) is also used to discriminate different shock models, as will be discussed in Section~3.5 and also in~\citet{lb05}.  

\subsection{Steady-state Wind Equation}

By combining Equations~(\ref{eq13}) and (\ref{eq14}) and differentiating with respect to radius, we can obtain a ``wind equation'' of the form
\begeq %
\frac{1}{\vZ} \, \frac{d\vZ}{dr} = \frac{N_{_0}}{D_{_0}} \, %
\label{eq15} %
\fineq
where the numerator and denominator functions $N_{_0}$ and $D_{_0}$ are given by
\begeq %
N_{_0} \equiv - \, {\ell^2 \over r^3} + {GM \over
(r-\rs)^2} + {\aZ^2 \, (3 \rs - 5 r) \over (\gamma+1) \, r \,
(r-\rs)} \ , \ \ \ \ \ D_{_0} \equiv {2 \, \aZ^2 \over \gamma+1} - \vZ^2 \ . %
\label{eq16} %
\fineq
Critical points occur where the numerator $N_{_0}$ and the denominator $D_{_0}$ vanish simultaneously. The associated critical conditions can be combined with the energy Equation~(\ref{eq13}) to express the critical radial speed $v_c$, the critical sound speed $a_c$, and the critical radius $r_c$ as functions of $\epsilon$ and $\ell$. The value of $K$ is then obtained by substituting $r_c$, $v_c$, and $a_c$ into Equation~(\ref{eq14})~\citep[see Section~3.3 of ][]{lb05}. In general, one obtains four solutions for the critical radius, denoted by $r_{c4}$, $r_{c3}$, $r_{c2}$, and $r_{c1}$ in order of increasing radius. 

As discussed in Section~3.3 of \citet{lb05}, the critical radius $r_{c4}$ always lies inside the event horizon and is therefore not physically meaningful, but the other three are located outside the horizon. There are three possible types of critical points, namely,  the O-type, X-type, and $\alpha$-type. The type of each critical point is determined by computing the two possible values for the derivative $d\vZ/dr$ at the corresponding location using L'H\^opital's rule and then checking to see whether they are real or complex~\citep{ac90}. We find that $r_{c2}$ is an O-type where the values for the derivatives are complex, and therefore this does not yield a physically acceptable solution. The remaining roots $r_{c1}$ and $r_{c3}$ each possess real derivatives and are therefore physically acceptable sonic points, although the types of accretion flows that are associated with them are different. Specifically, $r_{c3}$ is an X-type critical point where a smooth, global, shock-free solution always exists. On the other hand, $r_{c1}$ (hereafter $r^{\rm out}_{c} = r_{c1}$ is the outer critical point) is an $\alpha$-type critical point.  Consequently, any accretion flow originating at a large distance that passes through this $\alpha$-type critical point must undergo a shock transition~\citep{ac90}. Hereafter, we shall denote the shock location for a steady-state solution as $r_{_{*0}}$. After crossing the shock, the subsonic gas must pass through another ($\alpha$-type) critical point, which we shall denote as $r^{\rm in}_{c}$ (the inner critical point), before it enters the BH since the flow has to be supersonic at the event horizon~\citep{wei72}. A specific example of a steady-state shock flow solution is  discussed in Section~3.6.

\subsection{Isothermal Shock Jump Conditions}

The radius of the steady isothermal shock, denoted by $\rstarZ$, is determined self-consistently along with the structure of the disk by satisfying the velocity and energy jump conditions~\citep[e.g.,][]{cha89,lb05}
\begeq %
{\cal R}_{_{*0}}^{-1} \equiv {v_+ \over v_-} = {1 \over \gamma \,
{\cal M}_-^2} \ , \ \ \ \ \ \Delta\epsilon \equiv
\epsilon_+-\epsilon_- = {v_+^2 - v_-^2 \over 2} \ , %
\label{eq17} %
\fineq
where ${\cal M}_- \equiv u_-/a_-$ is the steady-state upstream Mach number at the unperturbed shock location, ${\cal R}_{_{*0}}$ is the steady-state shock compression ratio, and $u_-  = -v_->0$. For an isothermal shock, the energy transport rate $\epsilon$ drops from the upstream value $\epsilon_-$ to the downstream value $\epsilon_+$, and consequently $\Delta \epsilon< 0$. The downstream flow must therefore pass through a new inner critical point located at $r^{\rm in}_{c}< \rstarZ$ as discussed above. This inner critical point is computed using the downstream value $\epsilon_+$~\citep[see][]{lb05}.

Depending on the values of ($\epsilon_-$, $\ell$, and $\gamma$), as will be discussed in Section~3.6, one shock flow solution or two different  shock flow solutions that go through the same outer sonic point ($r^{\rm out}_{c}$) can be obtained~\citep[see discussion in Sections~3 and 4 of][for more details]{lb05}. Furthermore, the numerical steady-state solution for the inflow speed $\vZ(r)$, gas density $\RhoZ(r)$, and pressure $\PZ(r)$ must satisfy the following isothermal shock jump conditions at $\rstarZ$~\citep[see also][]{lb05}:
\begeqarray %
{\RhoZ(\rstarZ) \over \rho_{_-}} & = & \gamma \,{\cal M}_{_-}^2 \ , 
\label{eq18} \\ %
\nonumber \\ %
{\vZ(\rstarZ) \over \uMinus} & = & - \gamma^{-1} \, {\cal M}_{_-}^{-2} \ , 
\label{eq19}  \\%
\nonumber \\%
\frac{\PZ(\rstarZ)}{P_{_-}} & = & \gamma {\cal M}_{_-}^2  \ , 
\label{eq20} %
 \fineqarray
where ${\cal M}_{_-}$ is the incident Mach number of the shock as stated above. We also define the following dimensionless variables $\w0$, $\y0$, and $\z0$ for the steady-state velocity, density, and pressure solutions, respectively, as
\begeqarray %
\w0 & \equiv & \frac{\vZ}{\uMinus} \ , \label{eq21} \\ %
\nonumber \\ %
\y0 & \equiv & \frac{\RhoZ}{\rhoMinus} \ , \label{eq22} \\ %
\nonumber \\ %
\z0 & \equiv & \frac{\PZ}{\uMinus^2 \rhoMinus} \ . %
\label{eq23} %
\fineqarray
These equations will be used when we linearize the time-dependent mass, linear momentum, and energy conservation equations, which will be discussed in Section~4.2.

\begin{deluxetable}{lcccccccccccccccr}
\tabletypesize{\scriptsize} \tablecaption{Disk Structure
Parameters \label{tbl1}} \tablewidth{0pt} \tablehead{
\colhead{Model}  %
&\colhead{Shock Branch} %
&\colhead{$\ell$} %
&\colhead{$\epsilon_{_{-}}$} %
&\colhead{$\epsilon_{_{+}}$} %
&\colhead{$r_{_{out}}$} %
&\colhead{$\rstarZ$} %
&\colhead{$r_{_{in}}$} %
&\colhead{$u_{_{-}}$} %
&\colhead{$\xstarZ$} %
&\colhead{$H_{_{*0}}$} %
&\colhead{${\cal R}_{_{*0}}$} %
&\colhead{${\cal{M}}_{_{*0}}$}} %
\startdata %
0
&Inner shock %
&2.94 %
&0.002%
&-0.003216 %
&72.180%
&13.359    %
&6.744     %
&0.15463    %
&6.615    %
&6.814   %
&1.332    %
&0.942 \\ %
1
&Inner shock %
&2.946%
&0.002%
&-0.0033 %
&71.913%
&13.722    %
&6.695     %
&0.1535    %
&7.027    %
&7.034    %
&1.348    %
&0.948 \\ %
2
&Inner shock %
&2.96%
&0.002%
&-0.00345 %
&71.277%
&14.627    %
&6.582     %
&0.1509    %
&8.044    %
&7.584    %
&1.385    %
&0.961 \\ %
3
&Inner shock %
&3.00%
&0.002%
&-0.00356 %
&69.334%
&17.925    %
&6.271     %
&0.14220    %
&11.654    %
&9.616    %
&1.491    %
&0.997 \\ %
4
&Inner shock %
&3.03%
&0.002%
&-0.00321 %
&67.729%
&21.993    %
&6.048     %
&0.13266    %
&15.945    %
&12.176    %
&1.565    %
&1.022 \\ %
5
&Inner shock %
&3.04%
&0.002%
&-0.00292 %
&67.159%
&24.232    %
&5.971     %
&0.12786    %
&18.260    %
&13.611    %
&1.585    %
&1.028 \\ %
6
&Inner shock %
&3.044%
&0.002%
&-0.00276%
&66.926%
&25.457    %
&5.939     %
&0.12537    %
&19.518    %
&14.404    %
&1.592    %
&1.030 \\ %
7
&Inner shock %
&3.052%
&0.002%
&-0.00213 %
&66.449%
&30.069    %
&5.860     %
&0.11673    %
&24.209    %
&17.435    %
&1.593    %
&1.031 \\ \hline %
8
& Outer shock%
&3.0528%
&0.002%
&-0.00180 %
&66.401%
&32.588    %
&5.838    %
&0.11248    %
&26.750     %
&19.120    %
&1.583    %
&1.027 \\ %
7
& Outer shock%
&3.052%
&0.002%
&-0.00153 %
&66.449%
&34.740    %
&5.827    %
&0.10907    %
&28.913     %
&20.575    %
&1.570    %
&1.023 \\ %
6
& Outer shock%
&3.044%
&0.002%
&-0.00849 %
&66.926%
&40.959    %
&5.828    %
&0.10031    %
&35.131     %
&24.856    %
&1.519    %
&1.006 \\ %
5
& Outer shock%
&3.04%
&0.002%
&-0.00065 %
&67.159%
&42.983    %
&5.837    %
&0.09775    %
&37.147     %
&26.273    %
&1.499    %
&0.9997 \\ %
4
& Outer shock%
&3.03%
&0.002%
&-0.00028 %
&67.729%
&47.214    %
&5.864    %
&0.09282    %
&41.350     %
&29.268    %
&1.456    %
&0.985 \\ %
3
& Outer shock%
&3.00%
&0.002%
&0.00044 %
&69.334%
&57.167    %
&5.972    %
&0.08300    %
&51.195     %
&36.488    %
&1.350    %
&0.949 \\ %
2
& Outer shock%
&2.96%
&0.002%
&0.00104 %
&71.277%
&68.069    %
&6.152    %
&0.0744    %
&61.917     %
&44.652    %
&1.238    %
&0.909 \\ %
1
& Outer shock%
&2.946%
&0.002%
&0.00119 %
&71.913%
&71.558    %
&6.223    %
&0.0721    %
&65.335     %
&47.317    %
&1.204    %
&0.896 \\ \hline%
\enddata


\hskip-2.8 truein Note. -- All quantities are expressed in
gravitational units ($GM=c=1$).

\end{deluxetable}

\subsection{Parameter Space Configuration}

In this section and Section~3.6 we briefly discuss the construction of the steady-state solutions and refer the reader to~\citet{lb05} for further details. The process of finding a steady-state solution for a specific disk/shock system begins with the selection of values for the fundamental parameters ($\epsilon_{_{-}}, \ell, \gamma$). In this paper we set $\gamma = 1.5$; hence, only $\epsilon_{_{-}}$ and $\ell$ remain to be determined. The wedge in Figure~\ref{fig1}(a) represents the ($\epsilon_{_{-}}$, $\ell$)-parameter space where shocks can form for an isothermal shock type as discussed in \citet{lb05} for $\gamma=1.5$. For an isothermal shock model, the postshock entropy value is smaller than the preshock entropy value as discussed earlier. To be consistent with \citet{nak92,nak94} and \citet{nh94}, in this paper we also adopt the isothermal shock model, Figure~\ref{fig1}(a), to explore a different parameter space of ($\epsilon_{_{-}}, \ell$) within the wedge to understand the disk/shock stability and instability of our model. Briefly, Figure~\ref{fig1}(a) is constructed first by selecting a pair of values of the conserved accreted specific angular momentum $\ell$ and the preshock accreted specific energy rate $\epsilon_{-}$. After utilizing the sonic conditions (see Equation \ref{eq16}) and the accreted specific energy equation (see Equation~\ref{eq13}), we obtain the outer sonic radius $r_{c}^{\rm out}$. The entropy value at the outer critical point is then determined by using Equation (\ref{eq14}). Since entropy is conserved throughout the disk, except at the location of an isothermal shock, the entropy value at the preshock location is the same as the entropy value at the outer critical point. The postshock entropy value is then determined by using the entropy jump condition using Equation~(48) in~\citet{lb05}. Using the same specific angular momentum $\ell$, the velocity and energy jump conditions (see Equations~\ref{eq17}), the sonic conditions, the entropy equation, and  the postshock entropy value, the steady shock radius $r_{*0}$ and the inner sonic point $r_{c}^{\rm in}$ can be obtained self-consistently. We refer the reader to Section~4.2 in ~\citet{lb05} for further details.   

In~Figure~\ref{fig1}(a), region A, we select nine different models with the same accreted specific energy transport rate $\epsilon_{_-}$ but different accreted specific angular momentum $\ell$ to demonstrate the nature of the flows and the associated stability and instability, and the possibility of the existence of QPOs in different disk/shock scenarios. We also explore 20 additional models (15 models in region B and 5 models in  region C) to understand the behavior of the zero-frequency mode, as will be discussed in Section~4.3.

\subsection{Steady-state Solutions}

As discussed in Section~3.4, depending on the values of $\epsilon_{_{-}}$ and $\ell$, it is possible to obtain one single-shock or two different single-shock solutions. For a case where two possible single-shock solutions can be found, we refer to the solution that has a shock radius that is closer to the horizon as the inner shock solution and the other as the outer shock solution. From the selected ($\epsilon_{_{-}}$, $\ell$) values, model 0 and model 8 each have only one possible single-shock solution, while the other models can have two possible single-shock solutions. Figures~\ref{fig2}(a) and \ref{fig10}(a) show the structures of models 0 and 8, respectively. In our numerical examples, we utilize natural gravitational units ($GM = c = 1$ and $\rs = 2$), except as noted. For illustrative purposes, we use model 2 values with $\epsilon_{_{-}} = 0.002$ and $\ell = 2.96$ to discuss the nature of the flows, and with these values two single-shock solutions are possible with $r^{\rm out}_{_{c}} = 71.277$ as their outer sonic radius. The values for the various parameters
$\ell$, $\epsilon_{_{-}}$, $\epsilon_{_{+}}$, $r^{\rm out}_{_{c}}$, $r_{_{*0}}$, $r^{\rm in}_{_{c}}$, $u_{_{-}}$, $x_{_{*0}}$, $H_{_{*0}}$, ${\cal R}_{_{*}}$, and $\cal{M}_{_{-}}$ are reported in Table~\ref{tbl1}, where  $x_{_{*0}}$ is the steady-state thickness of the effective perturbation region between the shock radius and the inner critical point, which will be discussed in~Section~4. Using the values from model 2, we obtain two possible transonic profiles (see Figures~\ref{fig4}(a) and \ref{fig16}(a)), and each one contains a specific shock radius and an associated inner sonic point location. In Figures~\ref{fig4}(a) and \ref{fig16}(a), we plot the steady-state velocity $v_0$ (solid line) and the sound speed $a_0$ (dashed line) profiles that have the inner shock at $\rstarZ = 14.627$ and the outer shock at $\rstarZ = 68.069$, respectively, where $\rstarZ$ denotes a steady-state shock location of each profile. At the shock location the accreted specific energy drops from the upstream value $\epsilon_{_-} = 0.002$ to the downstream values $\epsilon_{_+} = -0.00345$ or $\epsilon_{_+} = 0.00104$ with the inner critical sonic points $r^{\rm in}_{_c} = 6.582$ or $r^{\rm in}_{_c} = 6.152$ associated with the inner shock and outer shock, respectively. It is important to mention here that even though we find two possible single-shock flow solutions, only one actual shock flow can occur in a given accretion flow. 

Note that for model 2 both the inner and outer shock profiles possess preshock deceleration and preshock acceleration, respectively (see Figures~\ref{fig4}(a) and \ref{fig16}(a)). However, not all inner shock solutions display preshock deceleration, and this is one basic difference between our disk dynamical profiles and the ones studied by \citet{nak92,nak94}, \citet{nh94}, \citet{cm93}, and \citet{gl06}. Furthermore, this result was never discussed by \citet{cha89}, \citet{dcc01}, or  \citet{lb05} because different disk structures for different ($\ell,\epsilon_-$) parameter values were never examined in detail. 

As demonstrated in Figures~\ref{fig2}(a) and all through \ref{fig9}(a), a preshock deceleration is highly dependent on our selected parameters $\epsilon_{_-}$ and $\ell$, and that they display a transition from a preshock deceleration to a preshock acceleration with increasing specific angular momentum. \citet{cm93} and \citet{nh94} have shown that under axisymmetric and constant disk-height assumptions, the pre-shock deceleration and acceleration flows are unstable and stable, respectively, using either adiabatic flows with Rankine-Hugoniot shock waves or isothermal flows with isothermal shock waves. We will show later in Section~5 that when the disk height assumption is relaxed, the preshock acceleration is also unstable for adiabatic accretion flows with isothermal standing shock waves. However, the condition for stability/instability is also subject to different modes of oscillation as in the WD instability problems~\citep[e.g.,][]{ci82} and in different regions of the isothermal shock ($\epsilon_{-}, \ell$)-parameter space, which will be discussed in Section~5.

\section{PERTURBATION EQUATIONS}

In this analysis, we assume $H$ to be time dependent and modified according to Equation (\ref{eq7}); hence, using Equations  (\ref{eq3}) and (\ref{eq6})-(\ref{eq9}), the time-dependent conservation of mass, momentum, and internal energy equations (Equations (\ref{eq1}), (\ref{eq2}), and (\ref{eq4}), respectively) can be expressed as
\begeq %
\frac{\partial \rho}{\partial t} + v \frac{\partial \rho}{\partial r} + \frac{2}{\gamma + 1} \rho \frac{\partial v}{\partial r} + \frac{2}{\gamma + 1} \rho v \left(\frac{3}{2 \, r} + \frac{1}{r-\rs} \right) = 0 \ , %
\label{eq24} %
\fineq
\begeq %
\rho \left(\frac{\partial v}{\partial t} + v \frac{\partial 
v}{\partial r}\right) + \frac{\partial P}{\partial r} - \rho \left[
\frac{\ell^2}{r^3} + \frac{G M}{(r-\rs)^2} \right]= 0 \ ,%
\label{eq25} %
\fineq
\begeq %
\frac{\partial P}{\partial t} + v \frac{\partial
P}{\partial r} - \frac{\gamma P}{\rho} \left(\frac{\partial
\rho}{\partial t} + v \frac{\partial \rho}{\partial r} \right) = 0 \ , %
\label{eq26} %
\fineq
where the disk half-height thickness $H$ in Equation (\ref{eq1}) is eliminated by using Equations (\ref{eq6})-(\ref{eq9}). 

In developing the equations governing the perturbation from Equations (\ref{eq24})-(\ref{eq26}), we first eliminate the radial coordinate $r$ in favor of a dimensionless radial coordinate $\xi$, which measures the relative distance between the fixed inner sonic radius and the shock location. The new radial coordinate $\xi$ is defined by
\begeq %
\xi(r,t) \equiv \frac{x}{\xstar} = \frac{r-\rc}{\rstar - \rc} \ , %
\label{eq27} %
\fineq
where $\rc$ is the inner sonic radius and $\xstar$ is the new shock position relative to the inner sonic radius. With this definition, the inner sonic point is located at $\xi = 0$ and the shock is located at $\xi = 1$ where $r = \rc$ and $r = \rstar$, respectively. 

Following~\citet{ci82}, we perturb the shock position by writing
\begeq %
\frac{d\xstar}{dt} = \vstar1 e^{\sigma t} \ , %
\label{eq28} %
\fineq
where $\sigma = \sigma_r + i \sigma_i$ is the angular frequency and $\vstar1$ is real. Upon integration, the shock position is now given as
\begeq %
\xstar = \xstarZ +  \xstarO e^{\sigma t} \ , %
\label{eq29} %
\fineq
where $\xstarO \equiv \vstar1/\sigma$ is the perturbed shock value, $\xstarZ \equiv \rstarZ - \rc$ is the steady-state thickness of the effective perturbation region, and $\rstarZ$ is the steady-state shock location. The spatial coordinate in Equation (\ref{eq27}) can be approximated by
\begeqarray %
\xi & \equiv & \frac{x}{\xstar} = \frac{x}{\xstarZ} \left(1
-\frac{\xstarO}{\xstarZ}e^{\sigma t}\right) \ , %
\label{eq30} %
\fineqarray
where here and below only the first-order terms of $\xstarO$ are retained. Hence, Equation (\ref{eq30}) gives
\begeq %
\frac{\partial \xi}{\partial r} =
\frac{1}{\xstarZ}\left(1-\frac{\xstarO}{\xstarZ} e^{\sigma t} \right) %
\label{eq31} %
\fineq
and
\begeq %
\frac{\partial \xi}{\partial t} = -
\frac{x\xstarO}{\xstarZ^2} \sigma e^{\sigma t} \ . %
\label{eq32} %
\fineq
It is important to note here that we are performing a coordinate transformation from variables ($r,t$) to ($\xi,t$), where $\xi \equiv \xi(r,t)$, as indicated in Equations (\ref{eq27}) and (\ref{eq30}). This means that in our $\xi$-coordinate any scalar variable $Q$ that transforms from $Q(r,t)$ to $\tilde{Q}(\xi,t)$ must transform according to the following rules:
\begeq %
\left. \frac{\partial Q}{\partial t} \right |_r = \left. \frac{\partial \tilde{Q}}{\partial t} \right |_{\xi} +
\left. \frac{\partial \xi}{\partial t} \right |_r \left. \frac{\partial \tilde{Q}}{\partial \xi} \right |_t \; \;
{\rm and} \; \; \left. \frac{\partial Q}{\partial r} \right |_t = \left. \frac{\partial
\xi}{\partial r} \right |_t \left. \frac{\partial \tilde{Q}}{\partial \xi} \right |_t %
\label{eq33} %
\fineq %
with the perturbed quantities for $\rho$, $v$, and $P$ up to first-order terms of $\Rho1$, $\v1$, and $\P1$, respectively, as
\begeq %
\rho(\xi, t) = \RhoZ(\xi) + \Rho1(\xi) e^{\sigma t}  \ , %
\label{eq34} %
\fineq
\begeq %
v(\xi, t) = \vZ(\xi) + \v1(\xi) e^{\sigma t} \ , %
\label{eq35} %
\fineq
\begeq %
P(\xi, t) = \PZ(\xi) + \P1(\xi) e^{\sigma t} \ .  %
\label{eq36} %
\fineq
The quantities with subscripts ``1'' and ``0'' represent the small perturbed factors and the steady-state solutions, respectively, and $\Rho1$, $\v1$, and $\P1$ are complex functions of $\xi$. Furthermore, all quantities are assumed to be vertically averaged over the disk height.

\subsection{Boundary Conditions for the Physical Perturbed Quantities at the Shock Radius}

To determine the boundary conditions for $\Rho1$, $\v1$, and $\P1$ at the shock location $\xi=1$, we assume that the velocity of the shock is $v_{_*} = \vstar1 e^{\sigma t}$ (see Equation (\ref{eq28})) with respect to the stationary observer, and the moving shock can be transformed to a rest frame by using a Galilean transformation such that the upstream velocity changes to
\begeq %
\vIn = - \uMinus - v_{_*} \ , %
\label{eq37}
\fineq
where $\vIn$ is the incoming gas velocity and $\uMinus> 0$. Hence, the boundary conditions of the moving shock wave can be obtained by taking the derivatives of the steady-state boundary values at the shock location (Equations (\ref{eq38})-(\ref{eq40})) with respect to $\uMinus$ to obtain
\begeqarray %
\Rho1 & = & \vstar1 \frac{d\RhoZ}{d\uMinus} = 2 \; \gamma \; \vstar1
\left(\frac{\rhoMinus}{\uMinus}\right) {\cal M}_{_-}^2 \ , %
\label{eq38} \\
\nonumber \\
\v1 & = &  \vstar1 + \vstar1 \frac{d\vZ}{d\uMinus} = \vstar1
\left(1+\frac{1}{\gamma {\cal M}_{_-}^2}\right) \ , %
\label{eq39} \\
\nonumber \\
 \P1 & = & \vstar1 \frac{d\PZ}{d\uMinus} = 2 \; \vstar1 \; \uMinus \;
 \rhoMinus \ , %
 \label{eq40} %
 \fineqarray
where the first term in $\v1$ comes from the inverse Galilean transformation (see the appendix for an alternative derivation). Note that the results given in Equations (\ref{eq38})-(\ref{eq40}) also account for the perturbation to the steady-state Mach number (see the appendix). 

\subsection{Mass Continuity, Linear Momentum, and Internal Energy Linearization Equations}

By utilizing Equations (\ref{eq31})-(\ref{eq36}), we linearize Equations (\ref{eq24})-(\ref{eq26}) and retain only quantities that are linear in first order to obtain
\begeqarray %
\frac{\gamma + 1}{2} \left(-\frac{\xi \xstarO \sigma}{\xstarZ} + \frac{\v1}{\xstarZ} - \frac{\vZ \xstarO}{\xstarZ^2} \right) \frac{d\RhoZ}{d \xi} + \left(\frac{\gamma + 1}{2} \right) \Rho1 \sigma  + \left(\frac{\gamma + 1}{2} \right) \frac{\vZ}{\xstarZ} \frac{d \Rho1}{d \xi} & & 
\nonumber \\
 + \left(\frac{\Rho1}{\xstarZ} - \frac{\RhoZ \xstarO}{\xstarZ^2} \right) \frac{d \vZ}{d \xi} + 
\frac{\RhoZ}{\xstarZ} \frac{d \v1}{d \xi} + \left(\frac{\RhoZ \v1 + \Rho1 \vZ}{\xstarZ} \right) \left(\frac{3/2}{\xi + \frac{r_c}{\xstarZ}} + \frac{1}{\xi + \frac{r_c - \rs}{\xstarZ} }\right) & & 
\nonumber \\
- \RhoZ \vZ \left[\frac{3/2 \xi \xstarO}{\xstarZ^2 \left(\xi + \frac{r_c}{\xstarZ} \right)^2} + \frac{\xi \xstarO}{\xstarZ^2 \left(\xi + \frac{r_c - \rs}{\xstarZ} \right)^2}\right] & = & 0  \ , %
\label{eq41} %
\fineqarray
\begeqarray %
\left(-\frac{\xi \xstarO \sigma \RhoZ}{\xstarZ} + \frac{\RhoZ \v1 + \Rho1 \vZ}{\xstarZ} - \frac{\RhoZ \vZ \xstarO}{\xstarZ^2} \right) \frac{d \vZ}{d \xi} + \RhoZ \v1 \sigma + \frac{\RhoZ \vZ}{\xstarZ} \frac{d \v1}{d \xi} & & 
\nonumber \\
+ \frac{1}{\xstarZ} \frac{d \P1}{d \xi} - \frac{\xstarO}{\xstarZ^2} \frac{d \PZ}{d \xi} - \Rho1 \left[\frac{\ell^2}{\xstarZ^3 \left(\xi + \frac{r_c}{\xstarZ} \right)^3} - \frac{G M}{\xstarZ^2 \left(\xi + \frac{r_c - \rs}{\xstarZ} \right)^2} \right] & & 
\nonumber \\
- \RhoZ \left[-\frac{3 \xi \ell^2 \xstarO}{\xstarZ^4 \left(\xi + \frac{r_c}{\xstarZ} \right)^4} + \frac{2 \xi G M \xstarO}{\xstarZ^3 \left(\xi + \frac{r_c - \rs}{\xstarZ} \right)^3} \right] & = & 0 \ , %
\label{eq42} %
\fineqarray
and
\begeqarray %
\left(- \frac{\xi \xstarO \sigma}{\xstarZ} + \frac{\v1}{\xstarZ} \right) \frac{d \PZ}{d \xi} + \P1 \sigma + \frac{\vZ}{\xstarZ} \frac{d \P1}{d \xi}  
- \frac{\gamma \PZ \Rho1 \sigma}{\RhoZ} & & 
\nonumber \\
- \frac{\gamma \vZ \PZ}{\RhoZ \xstarZ} \frac{d \Rho1}{d \xi} + \gamma \left(\frac{\xi \xstarO \PZ \sigma}{\xstarZ \RhoZ} - \frac{\vZ \PZ}{\xstarZ \RhoZ} \left[\frac{\P1}{\PZ} + \frac{\v1}{\vZ} - \frac{\Rho1}{\RhoZ} \right]\right) \frac{d \RhoZ}{d \xi} & = & 0%
\label{eq43} %
\fineqarray
for the continuity, radial momentum, and energy equations, respectively. These perturbed fluid equations (Equations~(\ref{eq41})-(\ref{eq43})) are then simplified by writing them in dimensionless form with the following transforming variables for the perturbed velocity, density, pressure, and the angular eigenvalue, respectively, as
\begeqarray %
\eta & \equiv & \frac{\v1}{\vstar1} \ , %
\label{eq44} \\
\nonumber \\
\zeta & \equiv & \frac{\uMinus \Rho1}{\vstar1 \rhoMinus} \ , %
\label{eq45} \\
\nonumber \\
\pi & \equiv & \frac{\P1}{\uMinus \rhoMinus \vstar1} \ ,%
\label{eq46}
\\
\nonumber \\
\delta & \equiv & \frac{x_{_{*0}} \sigma}{\uMinus} \ , %
\label{eq47} %
\fineqarray
where $\rhoMinus$ is the upstream density. Substituting Equations (\ref{eq44})-(\ref{eq47}) into Equations (\ref{eq41})-(\ref{eq43}), the perturbed fluid equations become
\begeqarray %
\left(\frac{\gamma + 1}{2} \right) \left(-\xi + \eta - \frac{\w0}{\delta}\right)  \frac{d \y0}{d \xi} + \left(\frac{\gamma + 1}{2} \right) \zeta \delta + \left(\frac{\gamma + 1}{2} \right) \w0 \frac{d \zeta}{d \xi} + \left(\zeta - \frac{\y0}{\delta} \right) \frac{d \w0}{d \xi}  & & 
\nonumber \\
+ \y0 \frac{d \eta}{d \xi} + \left(\y0 \eta + \w0 \zeta \right) \left(\frac{3/2}{\xi + \frac{r_c}{\xstarZ}} + \frac{1}{\xi + \frac{r_c - \rs}{\xstarZ}}\right) - \frac{\xi \y0 \w0}{\delta} \left[\frac{3/2}{\left(\xi +\frac{r_c}{\xstarZ} \right)^2} + \frac{1}{\left(\xi + \frac{r_c - \rs}{\xstarZ} \right)^2} \right] & = & 0  \ , %
\label{eq49} %
\fineqarray
\begeqarray %
\left(-\xi \y0 + \y0 \eta + \w0 \zeta - \frac{\y0 \w0}{\delta} \right) \frac{d \w0}{d \xi} + \y0 \eta \delta + \y0 \w0 \frac{d \eta}{d \xi} + \frac{d \pi}{d \xi} & &
\nonumber \\
- \frac{1}{\delta} \frac{d \z0}{d \xi} 
-\frac{\zeta \xstarZ}{\uMinus^2} \left[\frac{\ell^2}{\xstarZ^3 \left(\xi + \frac{r_c}{\xstarZ} \right)^3} - \frac{G M}{\xstarZ^2 \left(\xi + \frac{r_c - \rs}{\xstar} \right)^2} \right] & & 
\nonumber \\
-\frac{\y0 \xstarZ^2}{\delta \uMinus^2} \left[- \frac{3 \xi \ell^2}{\xstarZ^4 \left(\xi + \frac{r_c}{\xstarZ} \right)^4} + \frac{2 \xi G M}{\xstarZ^3 \left(\xi + \frac{r_c - \rs}{\xstarZ} \right)^3} \right]  & = & 0 \ , %
\label{eq50} %
\fineqarray
\begeqarray%
\left(-\xi + \eta \right) \frac{d \z0}{d \xi} + \pi \delta + \w0 \frac{d \pi}{d \xi} - \frac{\gamma \z0 \zeta \delta}{\y0} - \frac{\gamma \w0 \z0}{\y0} \frac{d \zeta}{d \xi} & &
\nonumber \\
+ \gamma \left(\frac{\xi \z0}{\y0} - \frac{\w0 \z0}{\y0} \left[\frac{\pi}{\z0} + \frac{\eta}{\w0} - \frac{\zeta}{\y0} \right] \right) \frac{d \y0}{d \xi} & = & 0 \, %
\label{eq51} %
\fineqarray
for the continuity, radial momentum, and internal energy conservation equations, respectively, and where we have also used Equations (\ref{eq21})-(\ref{eq23}). The quantities $\eta$, $\pi$, and $\zeta$  are complex eigenfunctions of the gas velocity, pressure, and density, and therefore each quantity contains both real and imaginary components. The quantity $\delta$ is also a complex number, where $\delta_{_r}$ and $\delta_{_i}$ are the eigenvalues. The subscript ``$r$'' denotes the real part and ``$i$'' denotes the imaginary part for each of the above quantities.  The quantity $\delta_{_r}$ therefore indicates the growth rate (positive value) or damping rate (negative value) of a mode, and $\delta_{_i}$ is the oscillatory eigenfrequency [in units of ($\uMinus/\xstarZ$)].

From Equations (\ref{eq49})-(\ref{eq51}) we obtain six coupled first-order differential equations
\begeqarray %
\frac{d\zetar}{d\xi} & = & %
\left(\frac{\xi}{\w0} + \frac{\delta_r}{\delta^2} - \frac{\eta_r}{\w0}\right) \frac{d \y0}{d \xi} - \frac{1}{\w0}\left(\zeta_r \delta_r - \zeta_i \delta_i \right) - \left(\frac{2}{\gamma + 1}\right) \left(\zeta_r - \frac{\y0 \delta_r}{\delta^2} \right) \frac{1}{\w0}  \frac{d \w0}{d \xi}  \nonumber \\ 
& & -\left(\frac{2}{\gamma + 1} \right) \frac{\y0}{\w0} \frac{d \eta_r}{d \xi} - \left(\frac{2}{\gamma + 1}\right) \left(\frac{\y0 \eta_r}{\w0} + \zeta_r \right) \left(\frac{3/2}{\xi + \frac{r_c}{\xstarZ}} + \frac{1}{\xi + \frac{r_c -\rs}{\xstarZ}}\right)  \nonumber \\
& & +\left(\frac{2}{\gamma + 1} \right) \frac{\xi \y0 \delta_r}{\delta^2} \left[\frac{3/2}{\left(\xi + \frac{r_c}{\xstarZ} \right)^2} + \frac{1}{\left(\xi + \frac{r_c - \rs}{\xstarZ} \right)^2} \right]\ , %
\label{eq52} \\ 
\nonumber \\  \nonumber \\
\frac{d\zetai}{d\xi} & = & %
- \left( \frac{\delta_i}{\delta^2} + \frac{\eta_i}{\w0}\right) \frac{d \y0}{d \xi} - \frac{1}{\w0}\left(\zeta_r \delta_i + \zeta_i \delta_r \right) - \left(\frac{2}{\gamma + 1} \right) \left(\zeta_i + \frac{\y0 \delta_i}{\delta^2} \right) \frac{1}{\w0}  \frac{d \w0}{d \xi}  \nonumber \\
& & -\left(\frac{2}{\gamma + 1} \right) \frac{\y0}{\w0} \frac{d \eta_i}{d \xi} - \left(\frac{2}{\gamma + 1}\right) \left(\frac{\y0 \eta_i}{\w0} + \zeta_i \right) \left(\frac{3/2}{\xi + \frac{r_c}{\xstarZ}} + \frac{1}{\xi + \frac{r_c -\rs}{\xstarZ}}\right)  \nonumber \\ 
& & - \left(\frac{2}{\gamma + 1} \right) \frac{\xi \y0 \delta_i}{\delta^2} \left[\frac{3/2}{\left(\xi + \frac{r_c}{\xstarZ} \right)^2} + \frac{1}{\left(\xi + \frac{r_c - \rs}{\xstarZ} \right)^2}  \right]\ , %
\label{eq53} \\
\nonumber \\ \nonumber \\
\frac{d\etar}{d\xi} & = &  %
\frac{\zetar \xstarZ}{\y0 \w0 \uMinus^2} \left[\frac{\ell^2}{\xstarZ^3 \left(\xi + \frac{r_c}{\xstarZ} \right)^3} - \frac{G M}{\xstarZ^2 \left(\xi + \frac{r_c - \rs}{\xstarZ} \right)^2} \right] + \frac{\delta_r}{\w0 \y0 \delta^2} \frac{d\z0}{d\xi} - \frac{1}{\w0 \y0} \frac{d\pir}{d\xi} \nonumber \\ %
& & - \frac{1}{\w0} \left(\eta_r \delta_r - \eta_i \delta_i \right) - \left(-\frac{\xi}{\w0} + \frac{\eta_r}{\w0} + \frac{\zeta_r}{\y0} - \frac{\delta_r}{\delta^2} \right) \frac{d\w0}{d\xi} 
\nonumber \\
& & + \frac{\xstarZ^2 \delta_r}{\w0 \delta^2 \uMinus^2} \left[-\frac{3 \xi \ell^2}{\xstarZ^4 \left(\xi + \frac{r_c}{\xstarZ} \right)^4} + \frac{2 \xi G M}{\xstarZ^3 \left(\xi + \frac{r_c - \rs}{\xstarZ} \right)^3}  \right] \ , %
\label{eq54} \\ 
\nonumber \\ \nonumber \\
\frac{d\etai}{d\xi} & = & %
\frac{\zetai \xstarZ}{\y0 \w0 \uMinus^2} \left[\frac{\ell^2}{\xstarZ^3 \left(\xi + \frac{r_c}{\xstarZ} \right)^3} - \frac{G M}{\xstarZ^2 \left(\xi + \frac{r_c - \rs}{\xstarZ} \right)^2} \right]  - \frac{\delta_i}{\w0 \y0 \delta^2} \frac{d\z0}{d\xi} - \frac{1}{\w0 \y0} \frac{d\pii}{d\xi} \nonumber \\ %
& & - \frac{1}{\w0} \left(\etar \delta_i + \etai \delta_r \right) - \left(\frac{\etai}{\w0} + \frac{\zetai}{\y0} + \frac{\delta_i}{\delta^2} \right) \frac{d\w0}{d\xi} 
\nonumber \\
& & - \frac{\xstarZ^2 \delta_i}{\w0 \delta^2 \uMinus^2} \left[-\frac{3 \xi \ell^2}{\xstarZ^4 \left(\xi + \frac{r_c}{\xstarZ} \right)^4} + \frac{2 \xi G M}{\xstarZ^3 \left(\xi + \frac{r_c - \rs}{\xstarZ} \right)^3}  \right]\ , %
\label{eq55} \\
\nonumber \\ \nonumber \\
\frac{d\pir}{d\xi} & = & %
-\gamma \left(\frac{\xi \z0}{\w0 \y0} - \frac{\z0}{\y0} \left[\frac{\pir}{\z0} + \frac{\etar}{\w0} - \frac{\zetar}{\y0} \right] \right) \frac{d\y0}{d\xi} + \frac{\gamma \z0}{\y0} \frac{d\zetar}{d\xi} \nonumber \\ %
& & + \frac{\gamma \z0}{\w0 \y0} \left(\zetar \delta_r - \zetai \delta_i \right) - \frac{1}{\w0} \left(\pir \delta_r - \pii \delta_i \right) - \frac{1}{\w0} \left(-\xi + \etar \right) \frac{d\z0}{d\xi} \ , %
\label{eq56} \\
\nonumber \\ \nonumber \\
\frac{d\pii}{d\xi} & = & %
\frac{\gamma \z0}{\y0} \left(\frac{\pii}{\z0} + \frac{\etai}{\w0} - \frac{\zetai}{\y0} \right) \frac{d\y0}{d\xi} + \frac{\gamma \z0}{\y0} \frac{d\zetai}{d\xi} \nonumber \\ %
& & + \frac{\gamma \z0}{\w0 \y0} \left(\zetar \delta_i + \zetai \delta_r \right) - \frac{1}{\w0} \left(\pir \delta_i + \pii \delta_r \right) - \frac{\etai}{\w0} \frac{d\z0}{d\xi} \ , %
\label{eq57} %
\fineqarray
where $\delta^2 = \delr^2 + \deli^2$. The eigenvalues $\delr$ and $\deli$ are determined by imposing boundary conditions near the inner sonic point such that the real and imaginary parts of the perturbed radial dimensionless velocity $\eta_{_r}$ and $\eta_{_i}$ must vanish. This implies that $|\eta| = (\eta_{_r}^2 + \eta_{_i}^2)^{1/2} \rightarrow 0$ as $\xi \rightarrow 0$. 

\subsection{Dimensionless Initial Boundary Condition for the Perturbed Quantities}

Starting at the shock location ($\xi=1$), integration of the six differential equations begins with the boundary conditions
\begeqarray %
\zetar & = & 2 \gamma {\cal M}_{_-}^2 \; \; , \; \; \zetai = 0 \; , %
\label{eq58} \\ %
\etar & = & 1 + \frac{1}{\gamma {\cal M}_{_-}^2} \; \; , \; \; \etai = 0 \; ,  %
\label{eq59} \\ %
\pir & = & 2 \; \; , \; \; \pii = 0 \; , %
\label{eq60} %
\fineqarray %
and integration proceeds inward toward the inner sonic radius at $\xi=0$. We obtain the initial conditions in Equations (\ref{eq58})-(\ref{eq60}) by utilizing Equations (\ref{eq38}), (\ref{eq39}), (\ref{eq40}), (\ref{eq44}), (\ref{eq45}), and (\ref{eq46}). Using the Runge-Kutta integration technique, we solved the differential equations for trial values of $\delr$ and $\deli$. The eigenvalues are found when the combination of $\delr$ and $\deli$ satisfies the inner boundary condition, which is having $\etar \rightarrow 0$ and $\etai \rightarrow 0$ as $\xi \rightarrow 0$. The method of determining the eigenvalues involves choosing a grid of points in the complex plane consisting of $\delr$ and $\deli$ and integrating the equations from $\xi = 1$ to $\xi \rightarrow 0$ for each point on the grid. The value of $1/|\eta|$ is then plotted, and the eigenvalues show up as narrow regions where the value of $1/|\eta|$ blows up. This is illustrated in Figures~\ref{fig1}(b) for the inner shock profile of model 2 as a contour plot, where the inverse of the dimensionless velocity $1/|\eta|$ is plotted as a function of the angular eigenvalues ($\delta_r$ and $\delta_i$). The narrow white-colored contours indicate that we have obtained the eigenvalues, $\delta_r$ and $\delta_i$, that yield the eigenfunction $|\eta| \rightarrow 0$ as $\xi \rightarrow 0$, while the large blue-colored regions indicate that the function $|\eta|$ has not converged to zero as $\xi \rightarrow 0$.  

The oscillation period of the perturbed shock wave is determined by $\delta_i$ and the growth or damping rate by $\delta_r$. We find that there are solutions with different modes of oscillation, similar to the WD instability problem~\citep[e.g.,][]{ci82}. By analogy with stellar pulsation theory, these modes can be called the fundamental (F), the first overtone (1O), the second overtone (2O), and so forth, where the first mode is the fundamental, the second mode is the first overtone, etc.~\citep{ci82}. The fundamental mode is usually the mode characterized by having no nodes in the perturbed velocity profile. It is also the mode with the smallest oscillation frequency. The first overtone has one node in the perturbed velocity profile and has the next smallest oscillation frequency. Hence, the first mode is usually the fundamental mode and the next higher modes are the overtones. However, in our analysis we also find a mode with zero oscillation frequency, as discussed below. In this work, we refer to the zero oscillation frequency mode as the ``zeroth mode'' (or Z-Mode; see Table~\ref{tbl-2}) to be consistent with the naming convention. Furthermore, a mode with a negative growth rate ($\delta_r < 0$) or a positive growth rate ($\delta_r > 0$) implies a stable or an unstable mode, respectively.

The small white-colored contour point ($\delta_{_{r}} = 0.16, \delta_{_{i}} = 0$) in inset (c) of Figure~\ref{fig1} represents an unstable solution ($\delta_{_{r}}>0$) with zero oscillation frequency ($\delta_{_{i}} = 0$) for the Z-Mode. In Figure~\ref{fig1}(b) we also show the fundamental mode (F-Mode), first overtone (1O), and second overtone (2O) with increasing eigenfrequency $\delta_i > 0$, respectively. For clarity, in Figure~\ref{fig1}(b), the Z-Mode is absent in this plot; this is due to our utilization of a lower-resolution grid in the interest of minimizing calculation time. This trend is also seen in Figures~\ref{fig2}(b)-\ref{fig6}(b) (for model 0 to model 4 of the inner shock profiles) where the Z-mode or F-Mode is either absent or not clearly shown, hence the inclusion of the inset plots; while in Figures~\ref{fig7}(b)-\ref{fig17}(b) (for models 5 to 7 of the inner shock profiles and models 8 to 1 of the outer shock profiles) we do not have the insets to these figures because both the Z-Mode and the F-Mode are clearly seen in the coarse grid plots. Once the eigenvalues have been determined, the next step is to improve accuracy by selecting a smaller grid near the eigenvalues and repeating the above procedure to achieve higher accuracy. If the values $\delta_r$ and $\delta_i$ give a solution, because the equations are symmetrical about the imaginary axis, the values $\delta_r$ and -$\delta_i$ will also yield the same solution as illustrated in Figure~\ref{fig1}(b) for the inner shock of model 2. Hence, in Figures~\ref{fig2}(b)-\ref{fig17}(b) we only show the positive solution of $\delta_i$ with the full range of $\delta_r$.

\section{RESULTS AND DISCUSSION}

\subsection{Disks/Shocks with Outflows-Regions of Instability}

Using linear analysis, we study different regions (regions A, B, C) of the ($\epsilon_{-},\ell$)-parameter space as shown in Figure~\ref{fig1}(a). In region A, we study nine different shock flow profiles. The steady-state shock flow structures are generated from a selected specific accreted energy $\epsilon_{_{-}}$ with increasing specific accreted angular momentum $\ell$ as shown in Figure~\ref{fig1}(a) and Table~\ref{tbl1}. From model 0 through model 8 we examine the nature of shock instability with shock location starting at $\rstarZ = 13.359$ and moving outward to $\rstarZ = 71.558$. As we move through this model sequence, note that the velocity profiles start to show preshock deceleration but change to preshock acceleration when the shock location moves outward (see Figures~\ref{fig2}(a) through \ref{fig17}(a)). Using the perturbed fluid equations and the method of solving the differential equations described in the previous section, we obtain the eigenvalues $\delta_r$ and $\delta_i$ for the first six modes for all 16 velocity profiles (see Table~\ref{tbl-2}) with their associated eigenvalues contour plots (see Figures \ref{fig2}(b) through \ref{fig17}(b)). For illustrative purposes, the corresponding eigenfunctions $\zetar$, $\zetai$, $\etar$, $\etai$, $\pir$, and $\pii$ are discussed below for model 2, and where we only investigate the first six modes of oscillation while modes of arbitrarily high order can be found.
\hskip-1.5truein
\begin{deluxetable}{lccccccccccccr}
\tablecolumns{13} %
\tabletypesize{\scriptsize} %
\tablecaption{Oscillating Periods and the Growth or Damping Rates of the Z-Mode, F-Mode, and Overtones  \label{tbl-2}}
\tablewidth{0pt} %
\tablehead{ %
&\multicolumn{1}{c}{} %
&\multicolumn{2}{c}{Zeroth Mode} %
&\multicolumn{2}{c}{Fundamental} %
&\multicolumn{2}{c}{First Overtone} %
&\multicolumn{2}{c}{Second Overtone} %
&\multicolumn{2}{c}{Third Overtone} %
&\multicolumn{2}{c}{Fourth Overtone} \\
\cline{3-4} \cline{7-8} \cline{11-12}\\
\colhead{Model} %
&\colhead{Shock Branch} %
&\colhead{$\delta_r$} %
&\colhead{$\delta_i$} %
&\colhead{$\delta_r$} %
&\colhead{$\delta_i$} %
&\colhead{$\delta_r$} %
&\colhead{$\delta_i$} %
&\colhead{$\delta_r$} %
&\colhead{$\delta_i$} %
&\colhead{$\delta_r$} %
&\colhead{$\delta_i$} %
&\colhead{$\delta_r$} %
&\colhead{$\delta_i$}} %
\startdata %
0
&Inner shock
&0.1995&0.0%
&-0.2605&0.9295%
&-0.4971&1.5657%
&-0.5414&2.3191%
&-0.5952&3.1191%
&-0.6357&3.9286 \\  %
1
&Inner shock
&0.2051&0.0%
&-0.2571&0.9810%
&-0.5238&1.6500%
&-0.5762&2.4286%
&-0.6286&3.2857%
&-0.6857&4.1429\\ %
2
&Inner shock
&0.2176&0.0%
&-0.2452&1.0971%
&-0.5871&1.8714%
&-0.6571&2.7143%
&-0.7214&3.6429%
&-0.8048&4.5952\\ %
3
&Inner shock
&0.2502&0.0%
&-0.1990&1.4210%
&-0.7614&2.5329%
&-0.9286&3.4286%
&-1.0857&4.5714%
&-1.2143&5.4667\\ %
4
&Inner shock
&0.2769&0.0%
&-0.1583&1.6755%
&-1.0795&2.8113%
&-0.7113&3.7252%
&-0.8113&5.3046%
&-1.0272&6.6457\\ %
5
&Inner shock
&0.2890&0.0%
&-0.1431&1.7762%
&-1.1377&2.7703%
&-0.5618&3.9190%
&-0.6752&5.6081%
&-0.9088&7.1153\\ %
6
&Inner shock
&0.2953&0.0%
&-0.1357&1.8228%
&-1.1436&2.7475%
&-0.4990&4.0079%
&-0.6129&5.7426%
&-0.8485&7.3119\\ %
7
&Inner shock
&0.3179&0.0%
&-0.1112&1.9604%
&-1.1086&2.6765%
&-0.3275&4.2671%
&-0.4408&6.1177%
&-0.6833&7.8314\\ \hline%
8
&Outer shock
&0.3298&0.0%
&-0.0987&2.0149%
&-1.07824&2.6565%
&-0.2616&4.3659%
&-0.3714&6.2619%
&-0.6200&8.0229 \\ %
7
&Outer shock
&0.3400&0.0%
&-0.0865&2.0518%
&-1.0502&2.6482%
&-0.2147&4.4341%
&-0.3196&6.3628%
&-0.5729&8.1559 \\  %
6
&Outer shock
&0.3692&0.0%
&-0.0496&2.1145%
&-0.97618&2.6694%
&-0.1116&4.5712%
&-0.1966&6.5667%
&-0.4425&8.4333 \\  %
5
&Outer shock
&0.3787&0.0%
&-0.0357&2.1226%
&-0.9549&2.6871%
&-0.0857&4.5996%
&-0.1616&6.6106%
&-0.3992&8.4977 \\  %
4
&Outer shock
&0.3986&0.0%
&-0.0057&2.1247%
&-0.9108&2.7408%
&-0.0416&4.6388%
&-0.0943&6.6729%
&-0.3071&8.5949 \\  %
3
&Outer shock
&0.4453&0.0%
&0.0815&2.0565%
&-0.8120&2.9105%
&0.0232&4.6244%
&0.0371&6.6691%
&-0.0933&8.6327 \\  %
2
&Outer shock
&0.4949&0.0%
&0.1938&1.8934%
&-0.6587&3.1051%
&0.0163&4.4271%
&0.1259&6.3978%
&0.0896&8.3091 \\  %
1
&Outer shock
&0.5098&0.0%
&0.2297&1.8213%
&-0.5774&3.1547%
&-0.0126&4.3163%
&0.1379&6.2375%
&0.1297&8.1062 \\ \hline %
\enddata


\hskip-4.5 truein Note. -- All frequencies are in units of ($\uMinus/\xstarZ$).

\end{deluxetable}
\begfig[t] \hskip-0.3in \epsscale{1.15} \plottwo{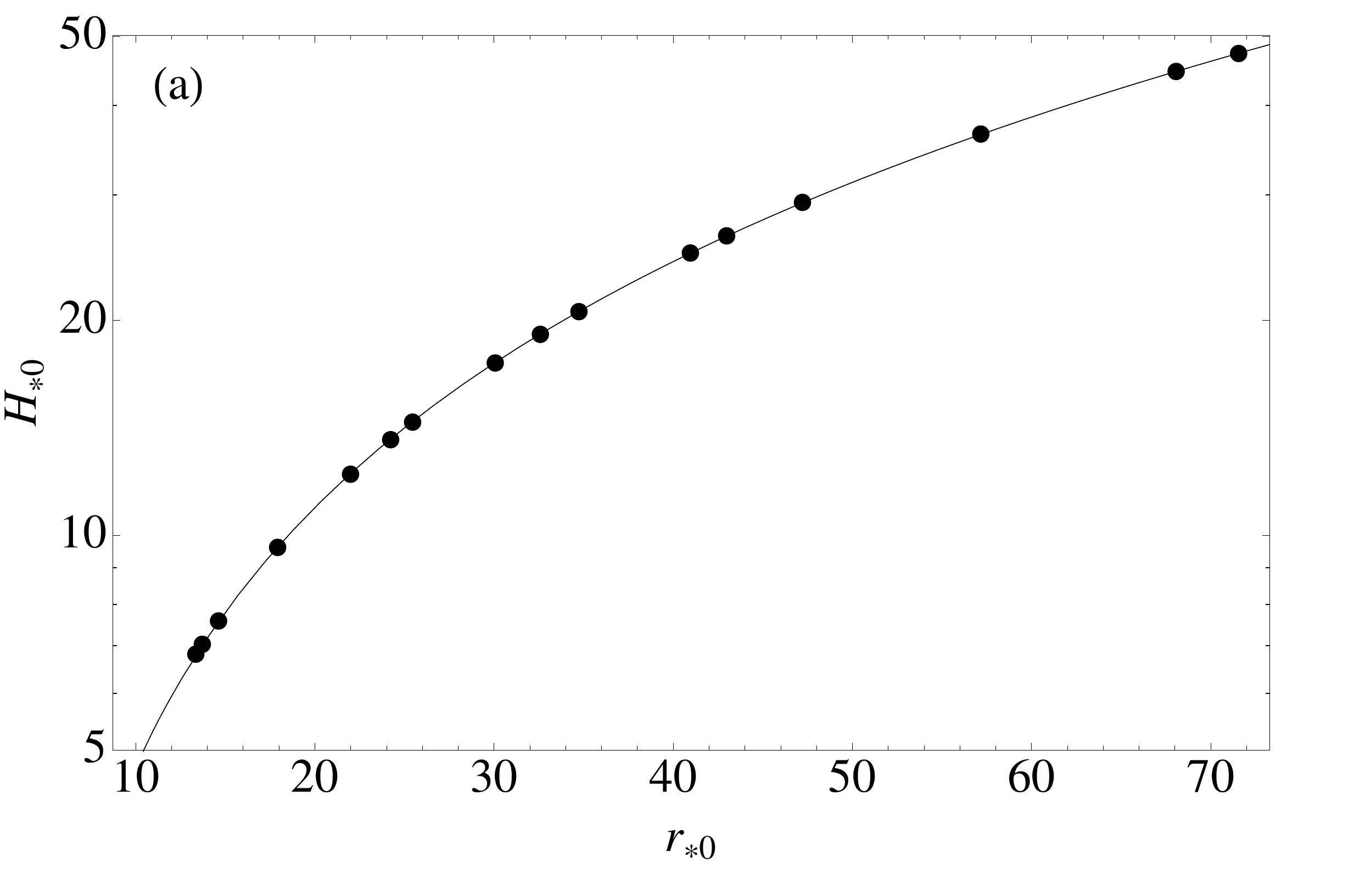}{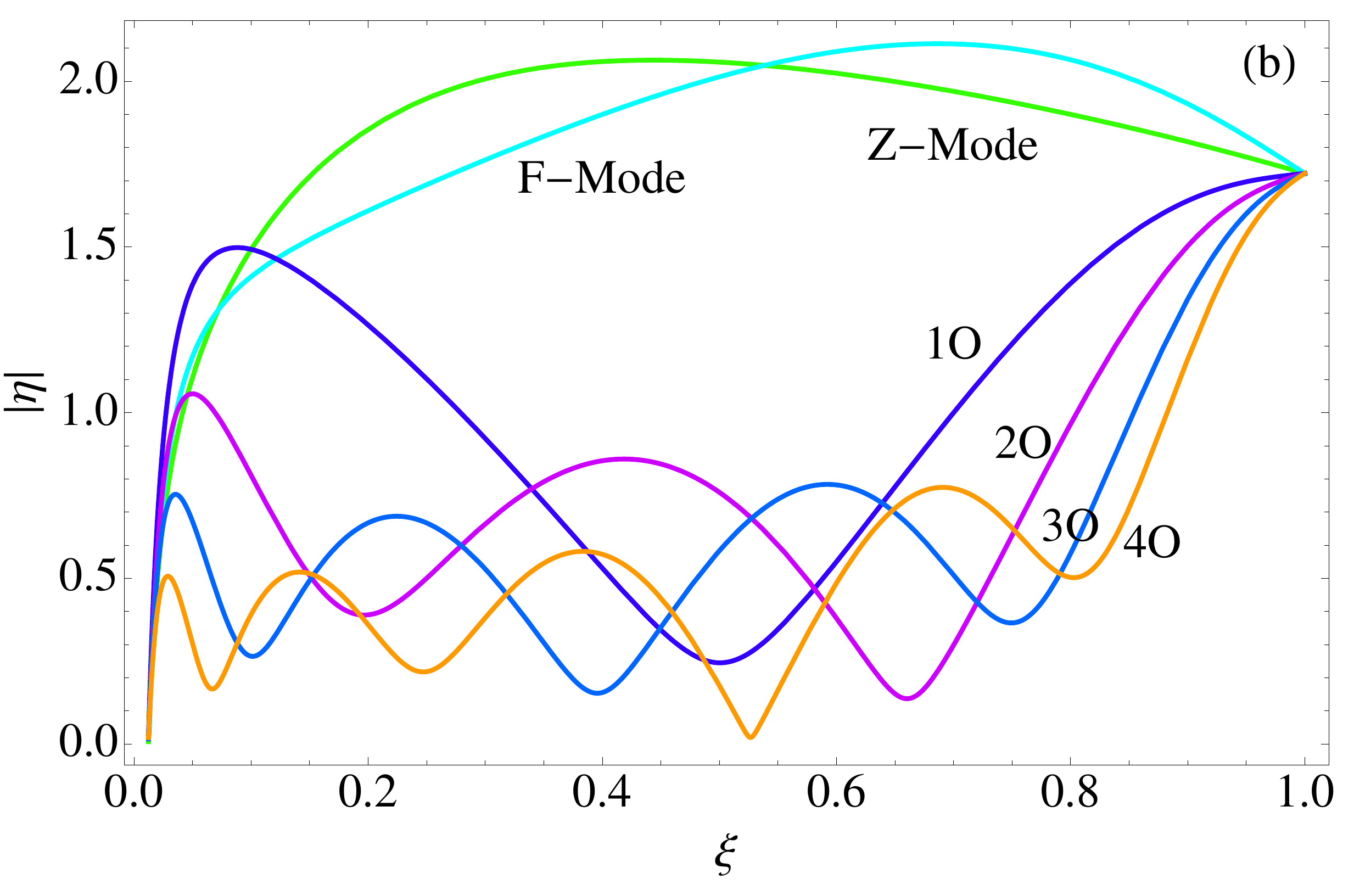}
\caption{\footnotesize (a) This plot illustrates the disk half-height at different shock locations. The plot clearly indicates that the disk half-height for the larger shock location is much larger than for the smaller shock location. The solid circles are the data, while the solid line is the fit through the data. (b) The first six eigenfunctions of the dimensionless perturbed velocity of the inner shock for model 2. These plots demonstrate that the condition $|\eta| \rightarrow 0$ as $\xi \rightarrow 0$ is satisfied.}
\label{fig18} \finfig
\begfig[t] \hskip-0.25in \epsscale{1.15} \plottwo{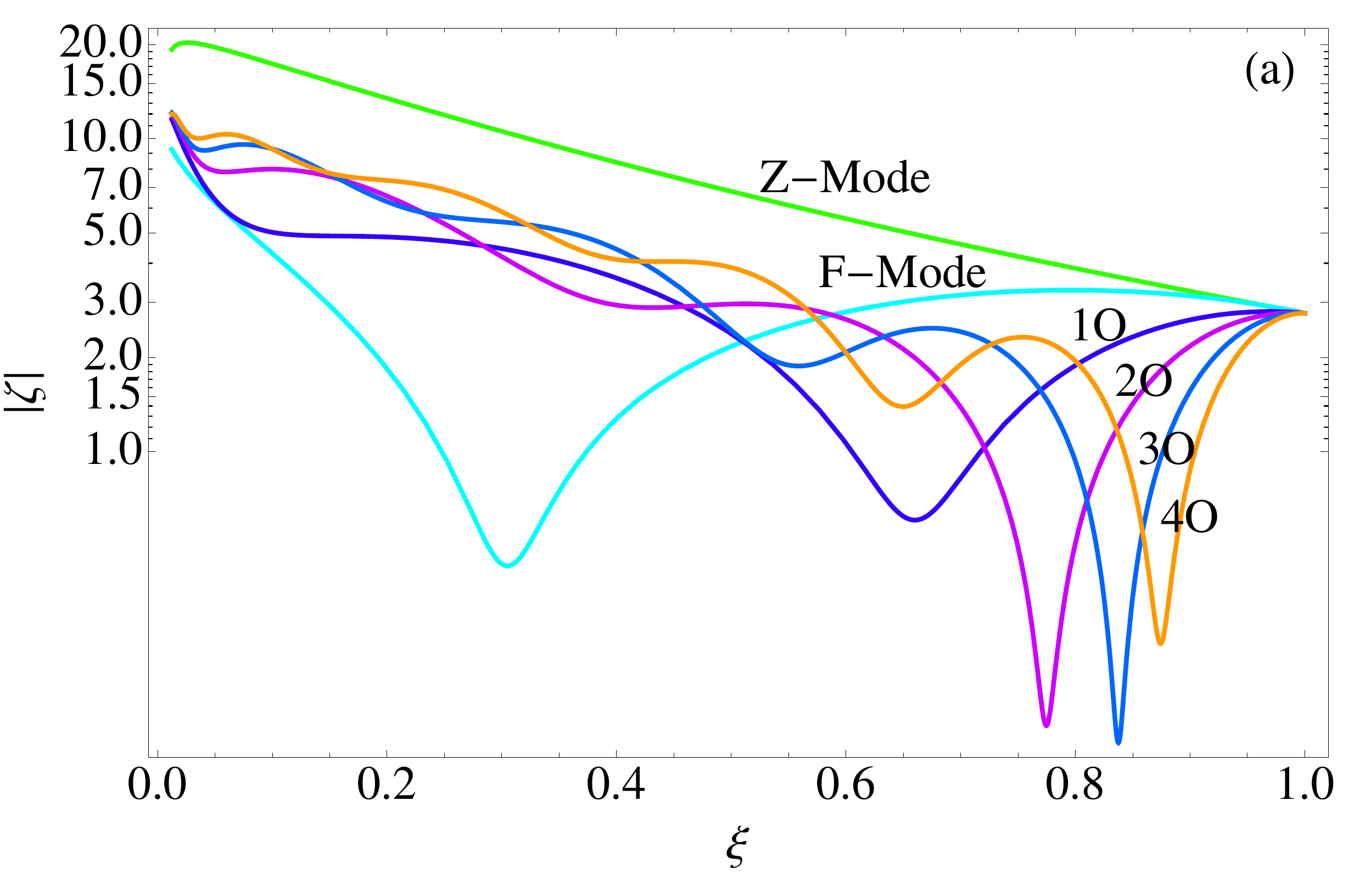}{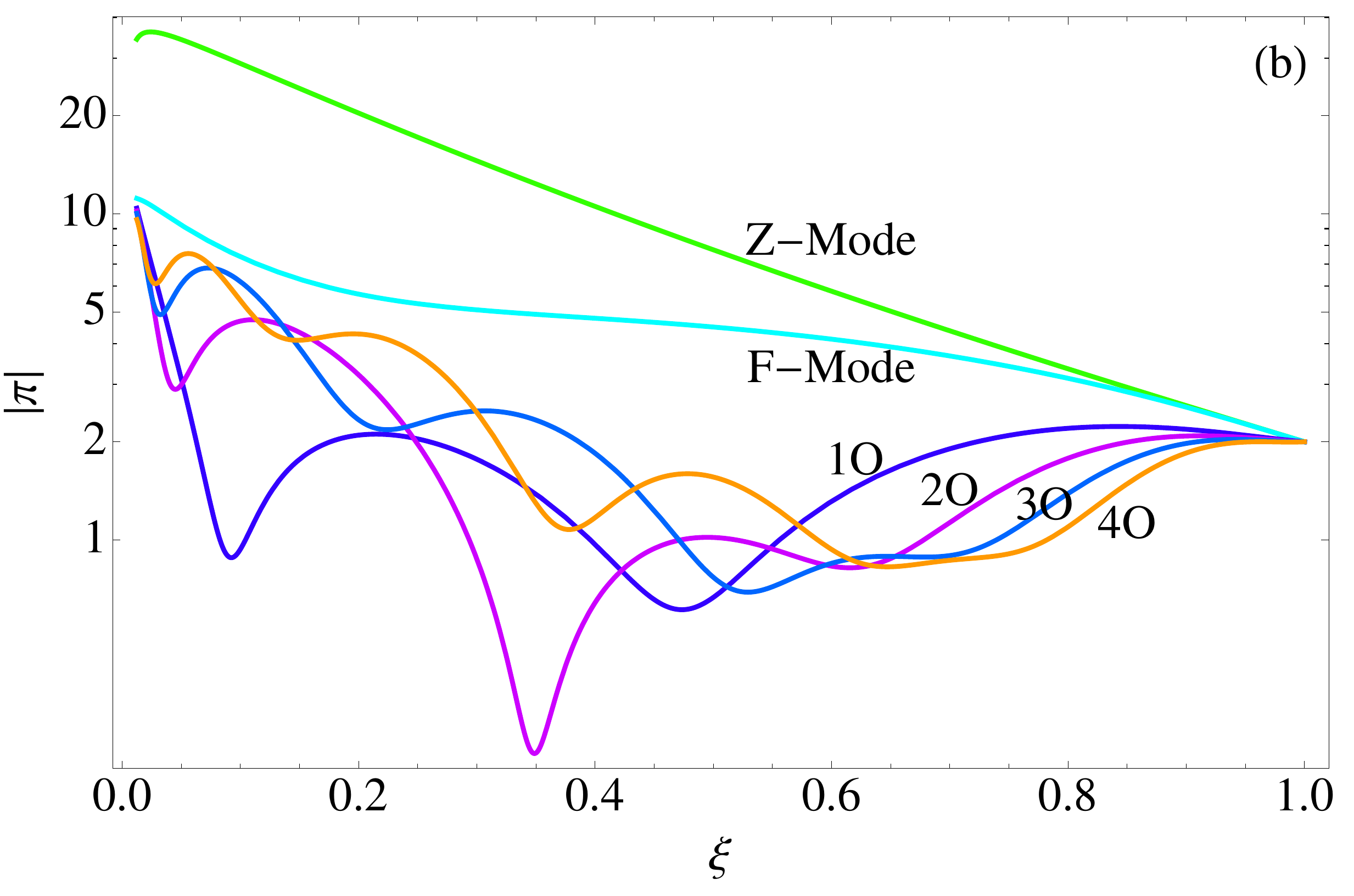}
\caption{\footnotesize (a) These plots illustrate the magnitude of the perturbed gas density and (b) the perturbed gas pressure of the first six modes for model 2. The plots show that the perturbed density and pressure of unstable modes (Z-Mode) have larger magnitudes than the stable ones.}
\label{fig19} \finfig
\begfig[t] \hskip-0.25in \epsscale{1.15} \plottwo{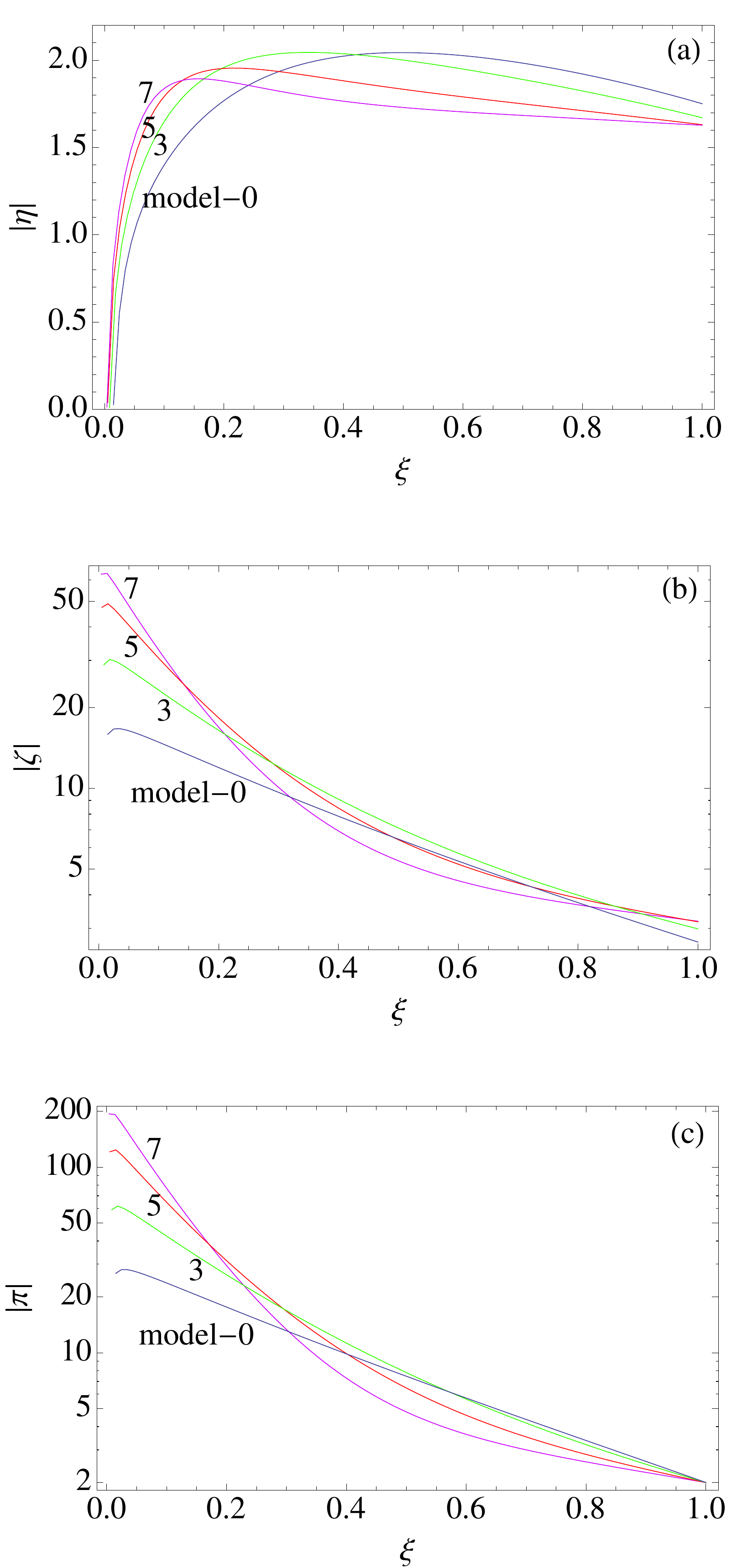}{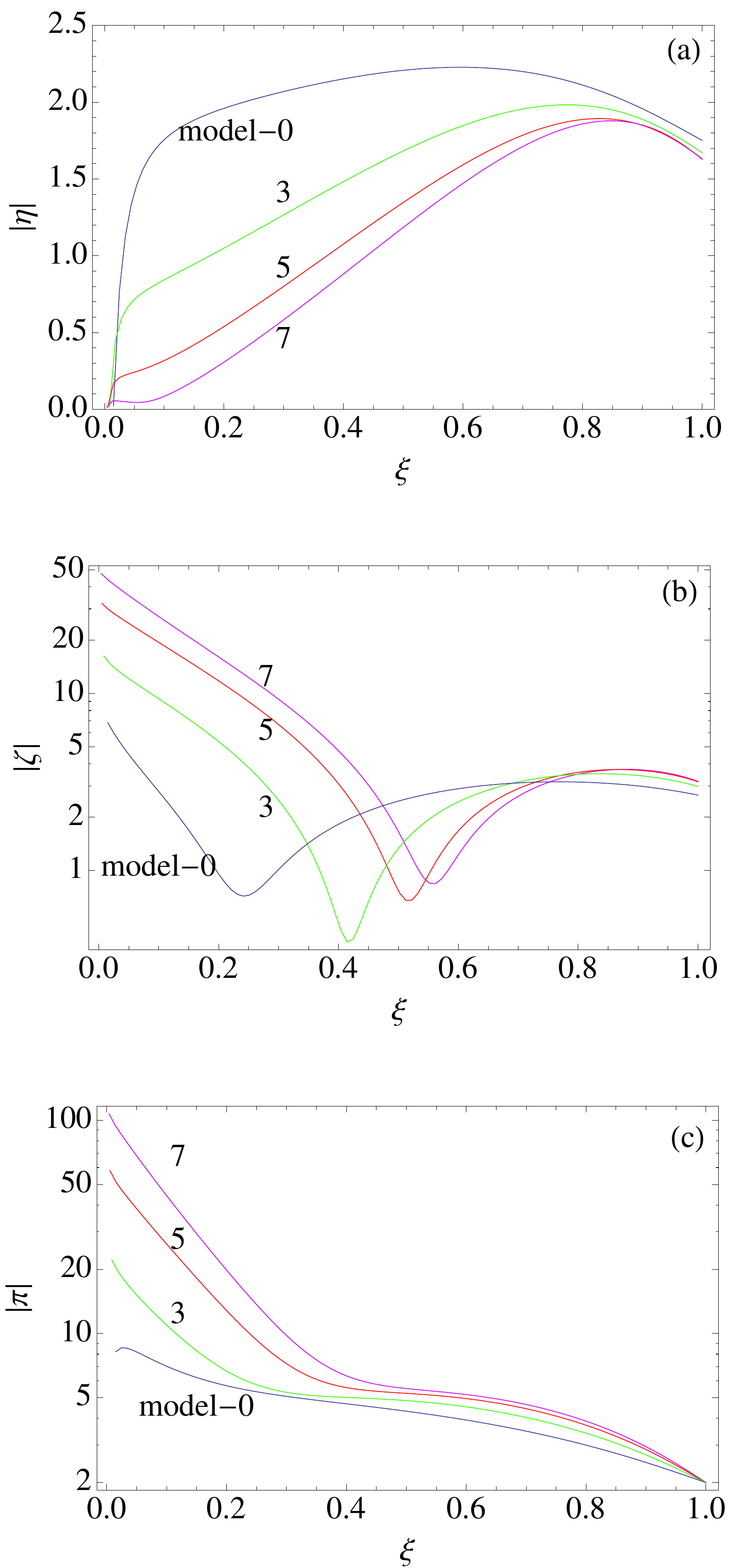}
\caption{\footnotesize The plots in the left column illustrate the eigenfunctions of (a) the perturbed velocity, (b) the perturbed gas density, and (c) the perturbed gas pressure for the Z-Mode mode at four different shock radii that correspond to four different shock models (model 0, 3, 5, and 7). The plots in the right column are the same as the left column but for the F-Mode.}
\label{fig20} \finfig
\begfig[t] \hskip-0.25in \epsscale{1.15} \plottwo{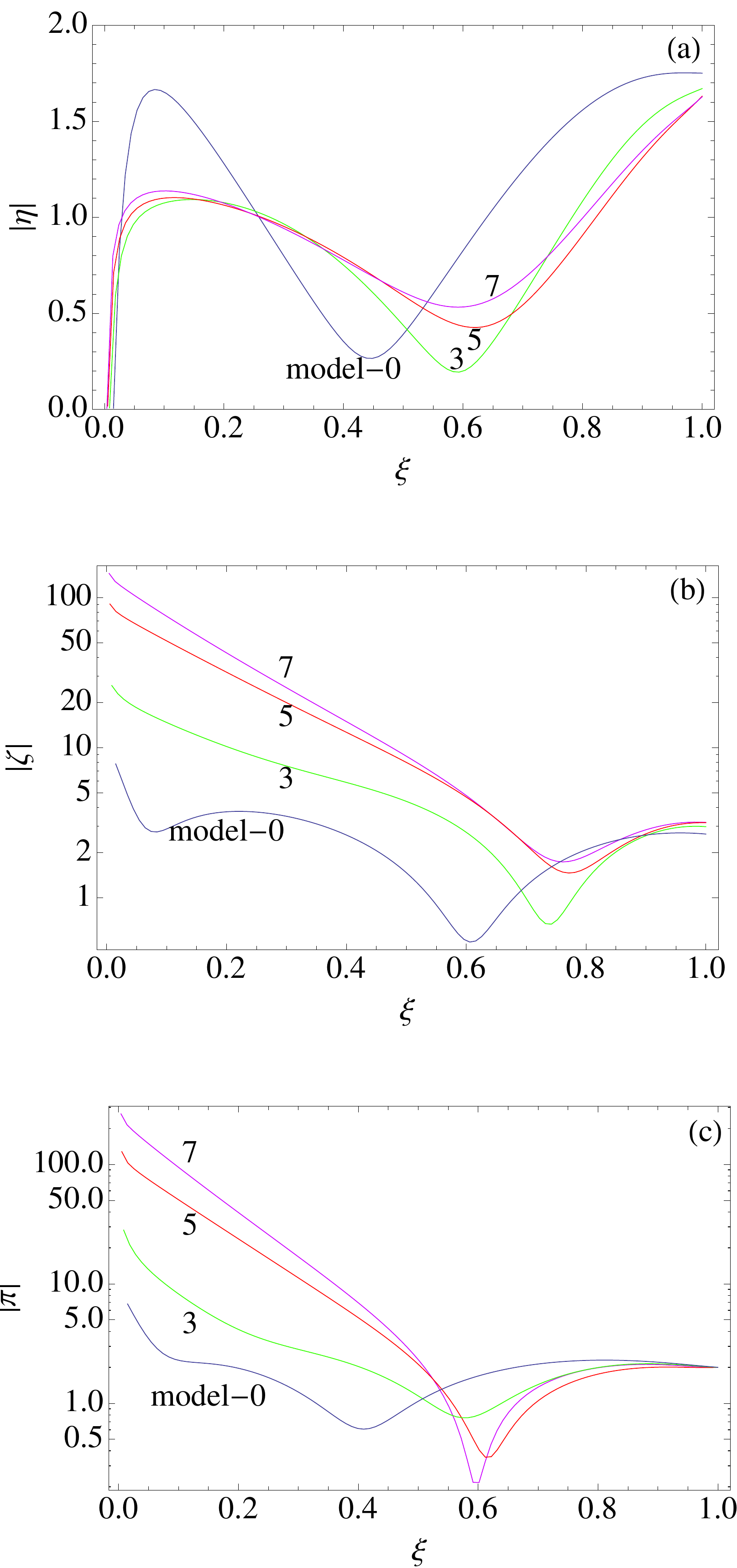}{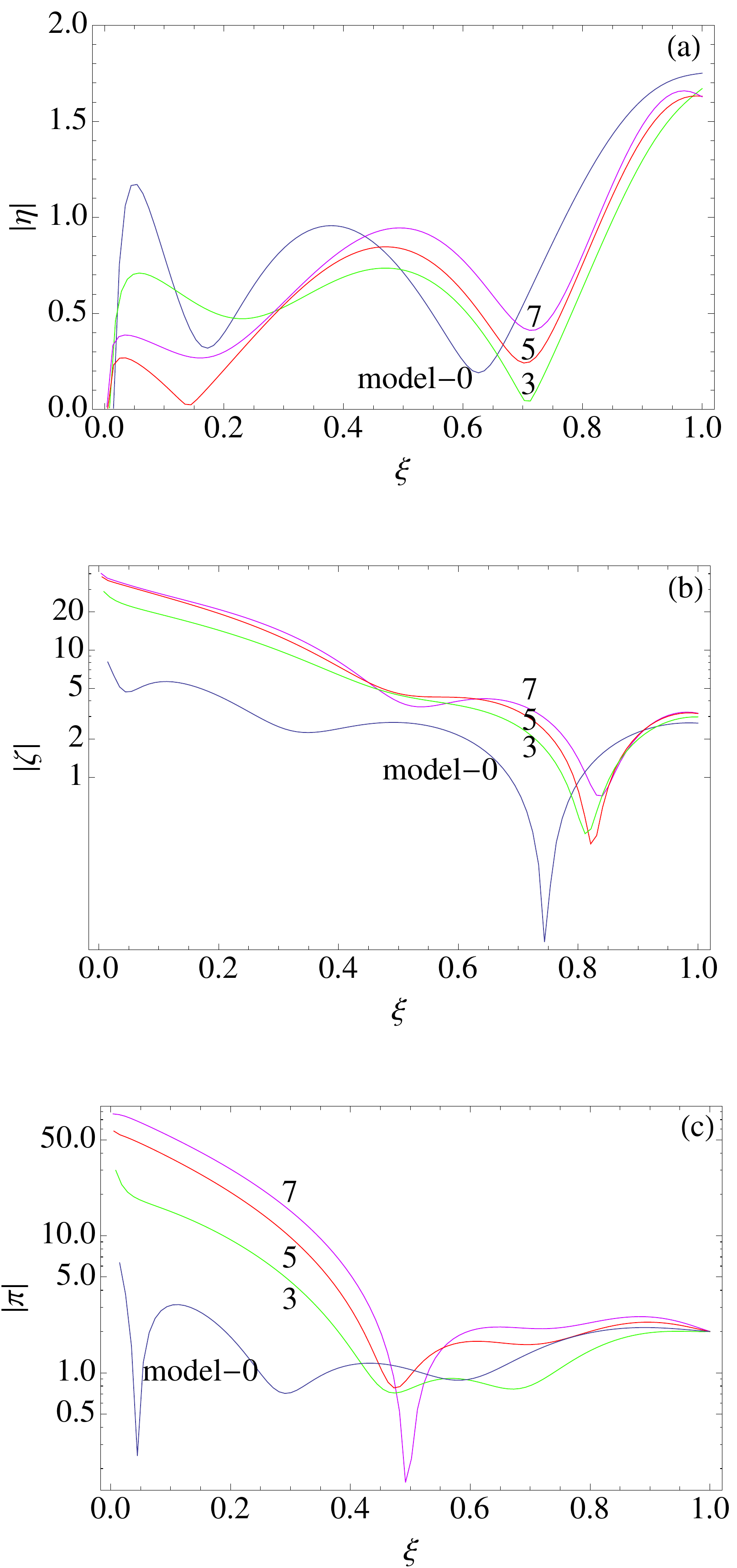}
\caption{\footnotesize The plots in the left and right columns are similar to Figure~\ref{fig20}, but for the 1O and 2O, respectively.}
\label{fig21} \finfig

For model 0 to model 3, the inner shock velocity profiles display preshock deceleration (see Figures~\ref{fig2}(a) through \ref{fig5}(a)), and the fundamental and overtone modes are stable, $\delta_r < 0$, but the zeroth mode with zero oscillation (see Figures~\ref{fig2}(b) through \ref{fig5}(b)) is unstable, $\delta_r > 0$. The shock locations for these models are less than $18$ gravitational radii, and the disk half-height at these shock locations is less than $10$ gravitational radii (see Table~\ref{tbl1}). \citet{nak92,nak94} showed that preshock deceleration causes instability with only one unstable mode, and an unstable mode with zero oscillation is allowed.  It is important to stress here that even though the zeroth mode has a nonzero oscillatory component ($\delta_i = 0$), it can be unstable if the real part of the frequency is positive ($\delta_r > 0$) since it continues to grow with time. Our results, in region A, support Nakayama instability criteria for cases of preshock deceleration. \citet{nh94} also found the mode with zero oscillation frequency for their preshock deceleration profile under a constant disk height assumption; however, it is hard to say conclusively since they only study a single case of preshock deceleration and preshock acceleration.  They also performed a nonlinear evolution perturbation around the steady-state solution for the unstable shock velocity case and showed that the perturbed shock front moves toward and passes the inner sonic point and the flow tends to a shock-free solution. 

Interestingly, \citet{abr81} suggested that disks that are secularly and thermally unstable can be stable with increasing mass loss. Our model contains energy loss from the disk at the shock location due to the presence of an isothermal shock, and the results in all models show that the energy-loss rate $\triangle \epsilon$ increases as the shock location moves inward; hence, we expect all modes to become less unstable as the shock radius decreases in the sequence of our models. From Figures~\ref{fig2}(b)-\ref{fig17}(b) and Table~\ref{tbl-2}, we notice that the growth rate ($\delta_r$) of the zeroth- and higher-order modes decreases (or moving toward a more stable configuration) as the shock radius decreases. However, there is a limitation to this behavior. We find that when a shock occurs below $21$ gravitational radii or when a shock profile exhibits preshock deceleration (below model 4), the disk structure becomes less stable with decreasing shock radius (see Table~\ref{tbl-2}). Furthermore, we expect that these features will be different for different values of $\epsilon_{-}$ parameter.  We have also explored different ($\epsilon_{-}, \ell$)-parameter spaces to obtain maximum energy loss at a shock location, and these results are discussed in Section~5.2, particularly for the behavior of the zeroth mode because we want to know whether the mode can be found to be stable.

For model 4 to model 8 both the inner and outer shock velocity profiles display preshock acceleration (see Figures~\ref{fig6}(a) through \ref{fig17}(a)). The shock locations for these models are greater than $21$ gravitational radii, and the disk half-heights at these shock locations are greater than $12$ gravitational radii (see Table~\ref{tbl1}). Their associated eigenvalues (see Figures~\ref{fig6}(b) through \ref{fig17}(b)) of the Z-Mode are also unstable with zero oscillation frequency. Furthermore, the zeroth mode, the fundamental mode, and the overtones become increasingly unstable with increasing shock radius or increasing disk half-height. These behaviors were not found in~\citet{nh94}. Clearly, our results do not agree with \citet{nak92,nak94}, who has  suggested that any velocity profiles that contain preshock acceleration are stable. Interestingly, \citet{gl06} and \citet{ny08,ny09} used non-axisymmetric and constant disk height assumptions and reported that the results of their outer shock models were unstable. However, it is not clear to us whether their outer shock models also showed preshock acceleration. Nevertheless, \citet{gl06}, \citet{ny08,ny09}, and our results seem to suggest that vertical variation along the disk has the effect of destabilizing a disk structure that exhibits preshock acceleration. This correlates with the disk half-height variation of our model as indicated in  Figure~\ref{fig18}(a), which shows that the disk half-height increases with increasing shock radius as expected according to Equation (\ref{eq7}) (also see Table~\ref{tbl1}). More interestingly, we reach a conclusion similar to \citet{hsw84a,hsw84b}, who have suggested that any standing shock that depends on the scale height and the specific angular momentum of the incoming flow is basically unstable, and that it could move through the system and disappear, and this is true at least in the region of instability of our model.

For illustrative purposes, Figure~\ref{fig18}(b) gives the magnitudes of $\eta$ as a function of $\xi$ for the zeroth mode, the fundamental mode, and the first four overtones for the preshock deceleration solution of model 2. As expected, the eigenfunction of the perturbed velocity $|\eta|$ approaches zero as $\xi$ approaches the inner sonic point. This restriction demands that the perturbed gas density $|\zeta|$ and pressure $|\pi|$ must have the profiles given in Figures~\ref{fig19}(a) and \ref{fig19}(b), respectively. We note that when a mode is unstable, for example, the Z-Mode, both the perturbed density and pressure are relatively large compared to the stable F-Mode or the overtones. According to our results, this feature is a general trend for any unstable modes. In Figures~\ref{fig20} and \ref{fig21}, we examine the eigenfunctions of the Z-Mode and the F-Mode and the 1O and 2O modes, respectively, at four different inner shock flow solutions. Figures~\ref{fig20}(a), (b), and (c) are the eigenfunctions of the perturbed velocity, density, and pressure, respectively, of the Z-Mode and F-Mode for models 0, 3, 5, and 7. We remind the reader that $\xi = 0$ and $\xi =1$ are the locations of the inner sonic point and the shock radius, respectively, as scale according to Equation (\ref{eq27}). As indicated in Table~\ref{tbl1}, model 0 has the smallest inner shock radius, while model 7 has the largest inner shock radius. In the left column of Figure~\ref{fig20}, we notice that the absolute magnitude of the perturbed density and pressure are generally larger with increasing shock radius. This general trend correlates with the first eigenvalue (or the Z-Mode) that we have plotted in Figures~\ref{fig2}(b), \ref{fig5}(b), \ref{fig7}(b), and \ref{fig9}(b) for models 0, 3, 5, and 7, respectively. That is, the white-colored contours of the Z-Mode become increasingly unstable with increasing model number (or increasing shock radius). 

In the right column of figure~\ref{fig20}, we plot the eigenfunctions of the perturbed velocity, density, and pressure of the F-Mode using the same shock models as in the left column. In the F-Mode we notice that the perturbed density and pressure, in general, contain similar patterns to the Z-Mode, i.e. the magnitude of the eigenfunctions is generally higher with increasing shock radius. These similar trends also occur in Figure~\ref{fig21} for the first and second overtones in the left and right columns, respectively.  In Figure~\ref{fig20}(a), the perturbed velocities for the Z-Mode (left column) and F-Mode (right column) have no nodes, while in Figure~\ref{fig21}(a) the perturbed velocities for the IO and 2O show one and two nodes, respectively.  More interestingly, in Figures~\ref{fig20} and \ref{fig21}, the first minimum (nearest to the shock location) of the perturbed velocity, density, and pressure is shifted closer to $\xi = 1$ for all the modes with increasing shock radius, but with the exception of the first overtone. In the first overtone, beyond model 3, the first minimum for the perturbed velocity, density, and pressure remains fixed at about the same $\xi$ or shifts very slowly with increasing shock radius. This feature is consistent with the third eigenvalue (or the 1O) that displays in Figures~\ref{fig7}(b) and \ref{fig9}(b) for models 5 and 7, respectively. In other words, we observe that the eigenvalue of the first overtone remains stable while eigenvalues of higher overtones move from being stable to becoming increasingly unstable as the shock location increases (see Figures~\ref{fig6}(b) to \ref{fig17}(b)). It is important to note here that this behavior only exists in the preshock acceleration solutions. Nevertheless, it is still not clear to us as to why the first minimum of the perturbed velocity, density, and pressure is shifted closer to the shock location for all other modes except the first overtone with increasing shock radius. This problem should be fully investigated using numerical simulations.

\subsection{Disks/Shocks with Outflows-Regions of Stability}

From the above results, we notice that the fundamental mode and the overtones are stable, except the Z-Mode for all cases that exhibit preshock deceleration, and that the Z-Mode is also unstable for cases that exhibit preshock acceleration as well. In region B of Figure~\ref{fig1}(a), we examine 15 different models of ($\epsilon_{-},\ell$) to explore the direction of stability. Table~\ref{tbl3} (models 0B1-2B5) and Figure~\ref{fig22}(a) indicate that the direction of decreasing $\delta_r$ or the direction of stability is in the direction of decreasing $\epsilon_{-}$ and $\ell$. We examine several cases in region C and discover that the Z-Mode in this region exhibits negative growth rate ($\delta_r < 0$), indicating that the Z-mode is stable (see Figure~\ref{fig22}(b)). The velocity profiles and the associated eigenvalues for the Z-mode are shown in Table~\ref{tbl3} (models 5C-9C) and Figures~\ref{fig23}-\ref{fig27}, respectively. One thing to note is that the compression ratios for these models are much larger than the compression ratios for the models in the region of instability (see Tables~\ref{tbl1} and \ref{tbl3}). The results in region C demonstrate the region of stability. We plan to search the boundary between the stable and unstable regions and their implications for QPOs in a subsequent paper.

However, it is interesting to ask whether any of the modes (Z-mode, fundamental mode, or the overtones), in the region of either stability or instability, will still result from a multidimensional analysis. If the shock is both strong and near the horizon, then the high-order modes from our 1D analysis would likely not disappear or dissipate away in the multidimensional case. Because the disk half-height near the horizon is relatively thin compared to the half-height near the outer sonic point (e.g., in our models 0 [inner shock] and 1 [outer shock]), it is possible that the velocity profile near the horizon will not be altered drastically under a 1D+axisymmetric or 2D+axisymmetric analysis. This would suggest that the nature of the shock will be the same near the horizon; thus, any higher modes that exist in a 1D analysis would still exist in a multidimensional analysis.

\begfig[t] \hskip-0.25in \epsscale{1.15} \plottwo{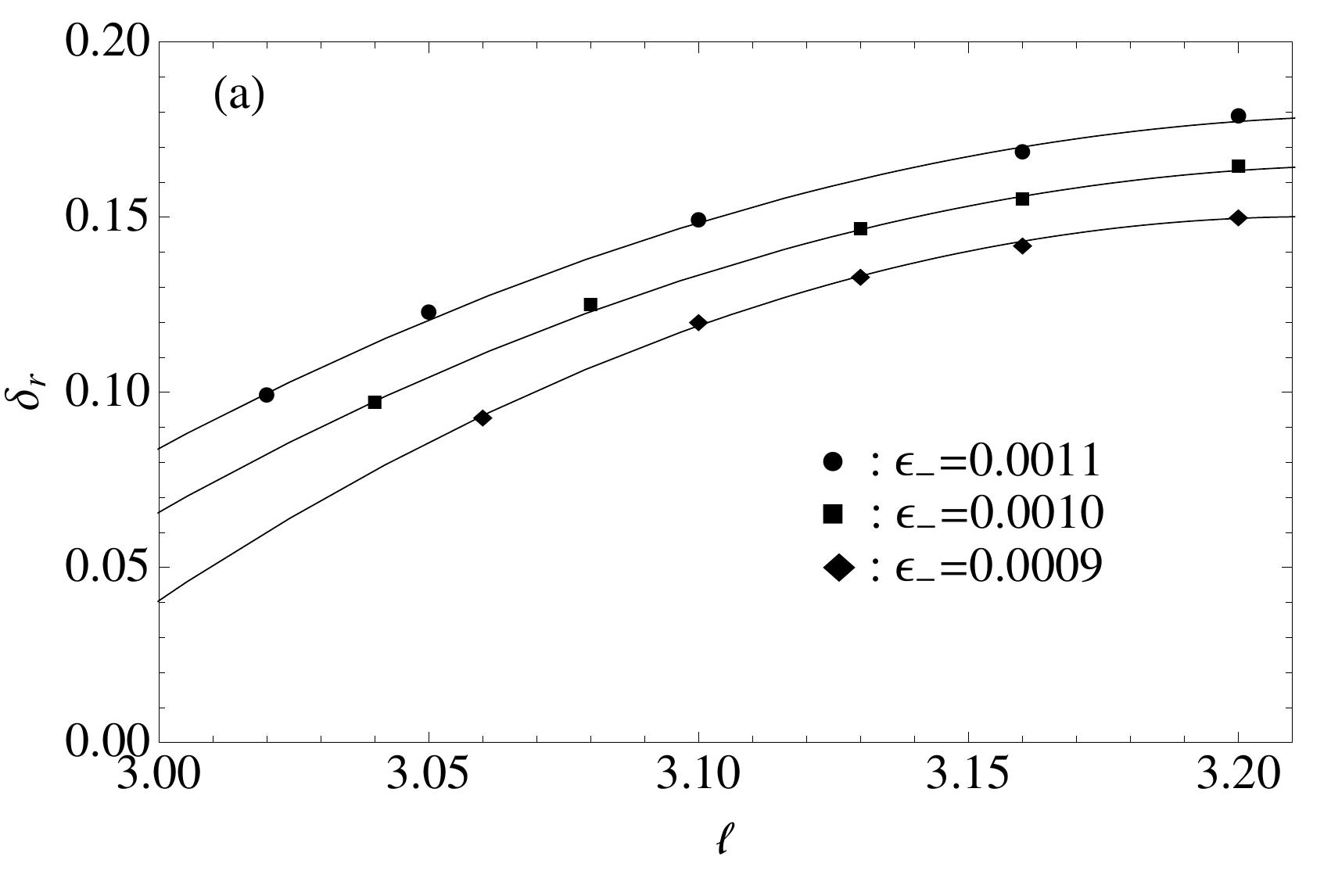}{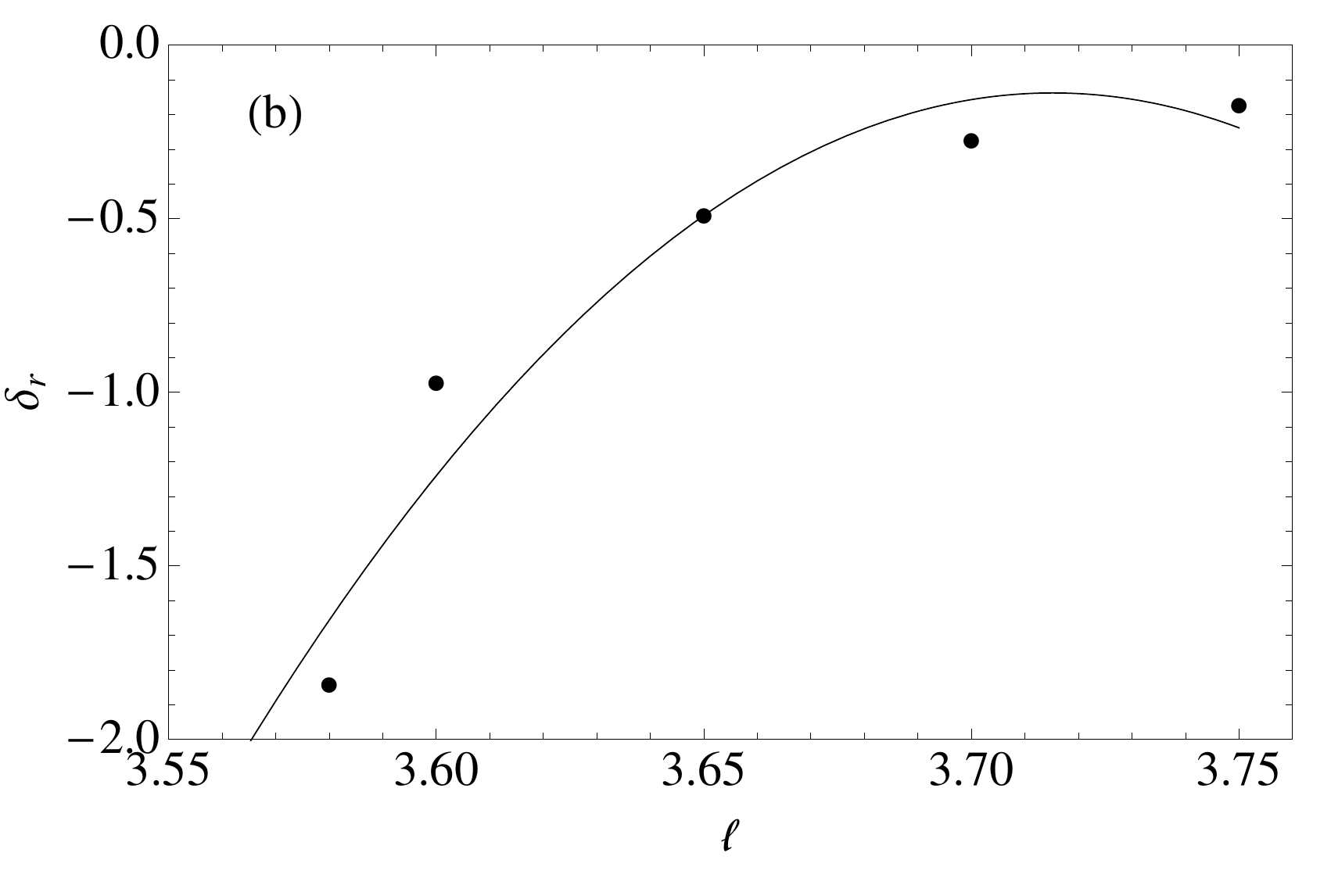}
\caption{\footnotesize (a) These plots show the decreasing of the positive growth rate with decreasing $\epsilon_{-}$ and $\ell$ for three different values of $\epsilon_{-}$ in an unstable region of the ($\epsilon_{-},\ell$)-parameter space. (b) This plot shows the decreasing of the negative growth rate with decreasing $\ell$ for $\epsilon_{-} = 0.00002$ in a stable region of the ($\epsilon_{-},\ell$)-parameter space. The solid lines in both figures are the best-fitted lines through the data points.}
\label{fig22} \finfig
\begfig[t] \hskip-0.25in \epsscale{1.15} \plottwo{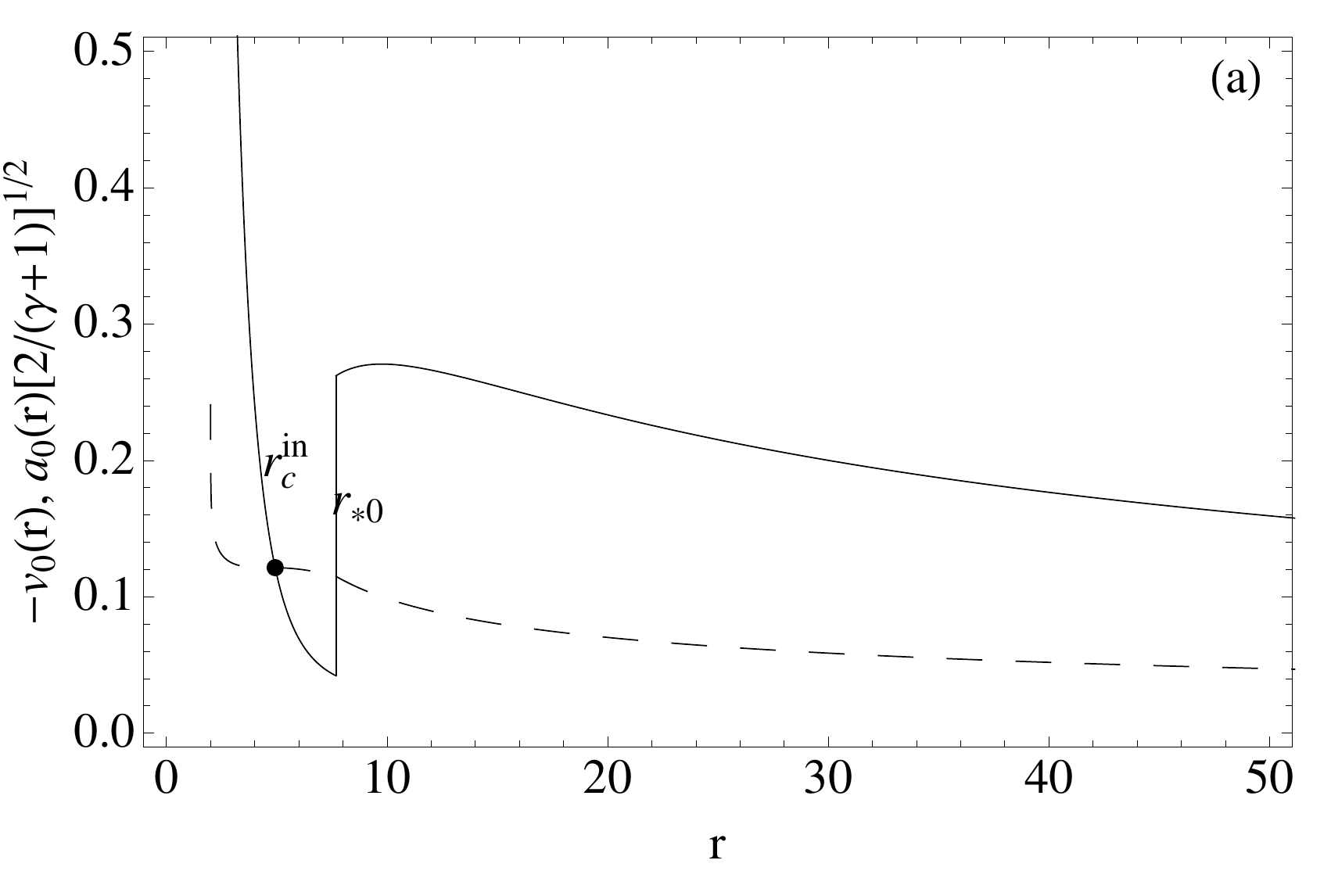}{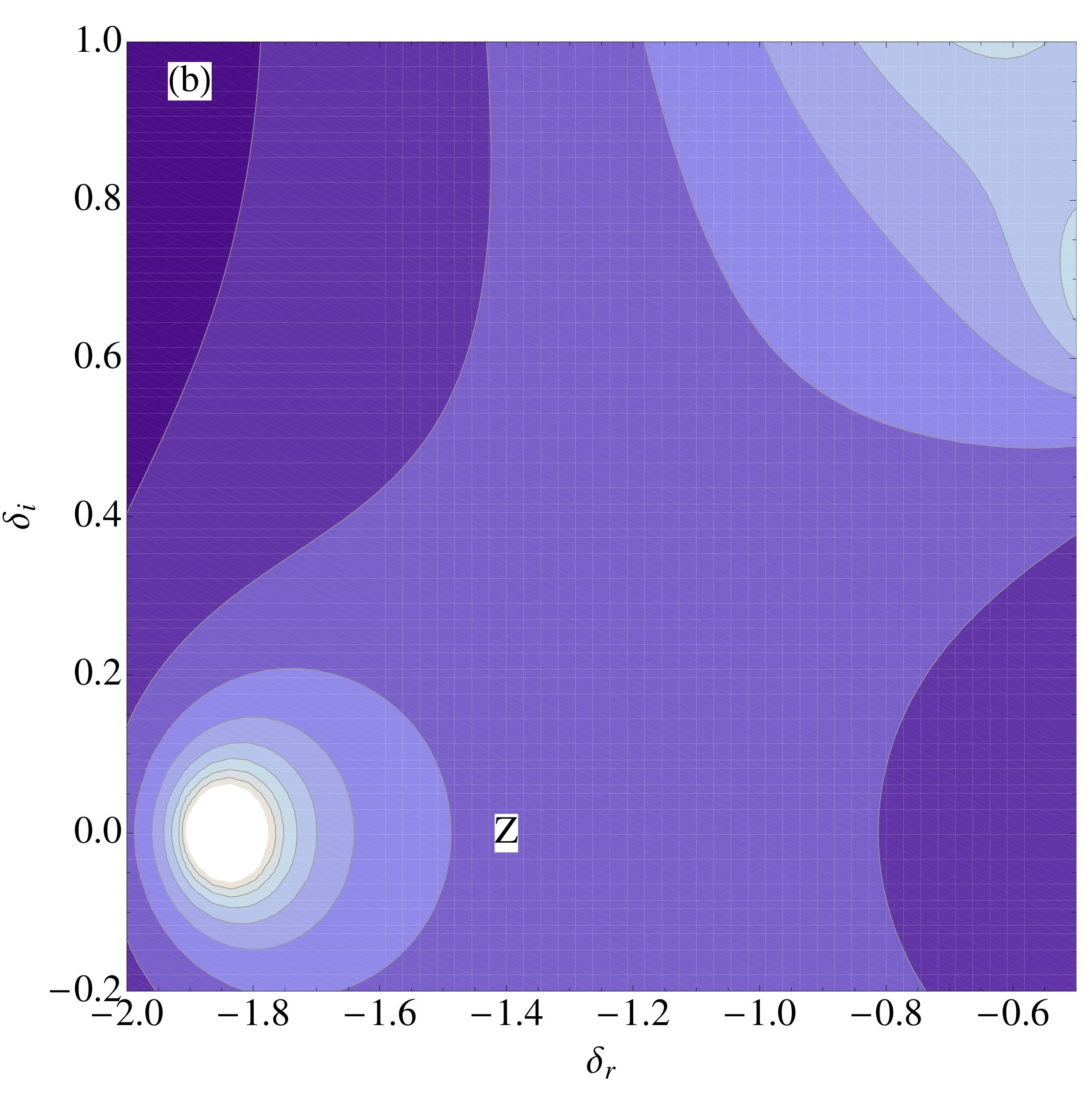}
\caption{\footnotesize Model 5C - inner shock flow: (a) inner shock disk profile with preshock deceleration, and (b) its eigenfrequencies.}
\label{fig23} \finfig
\begfig[t] \hskip-0.25in \epsscale{1.15} \plottwo{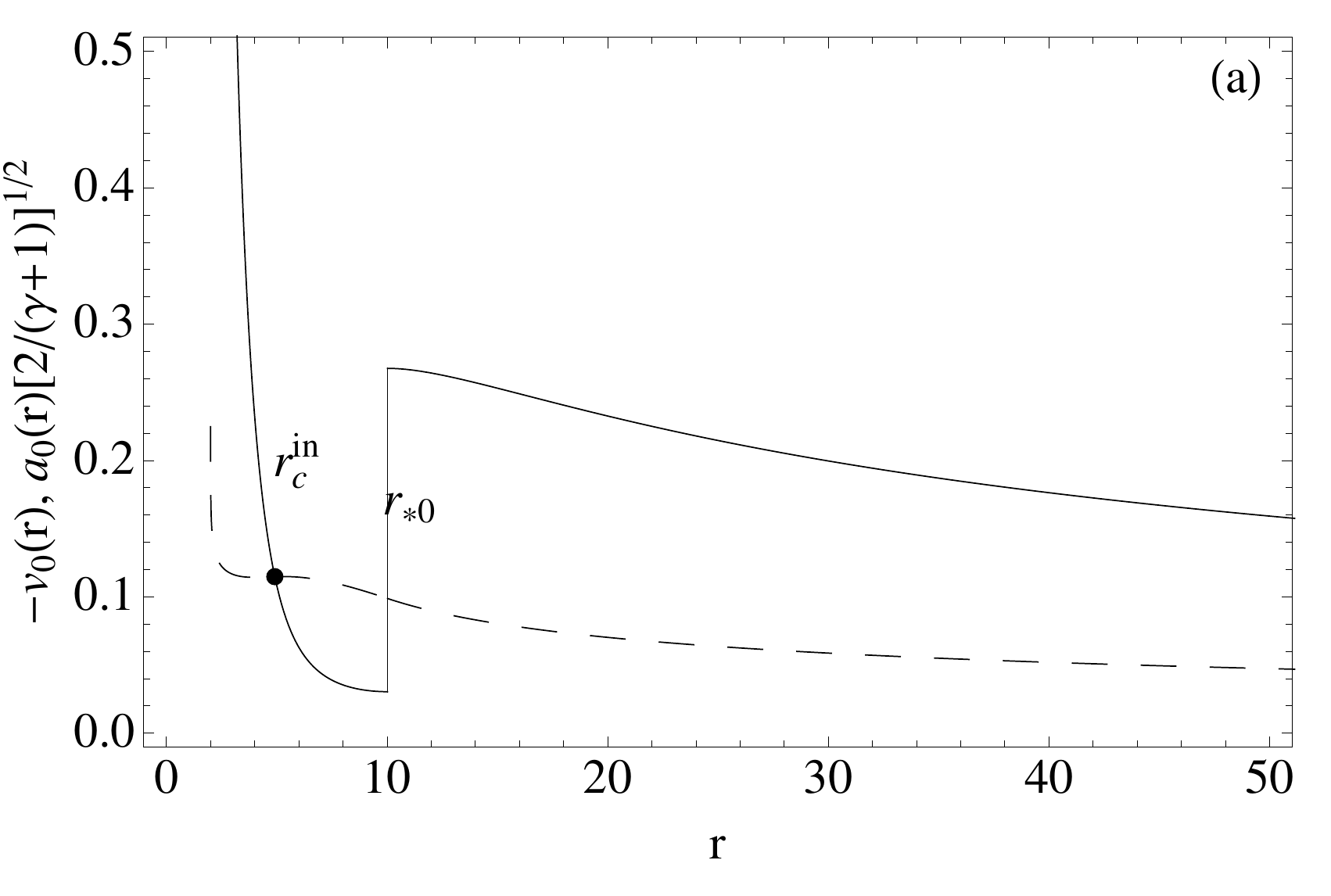}{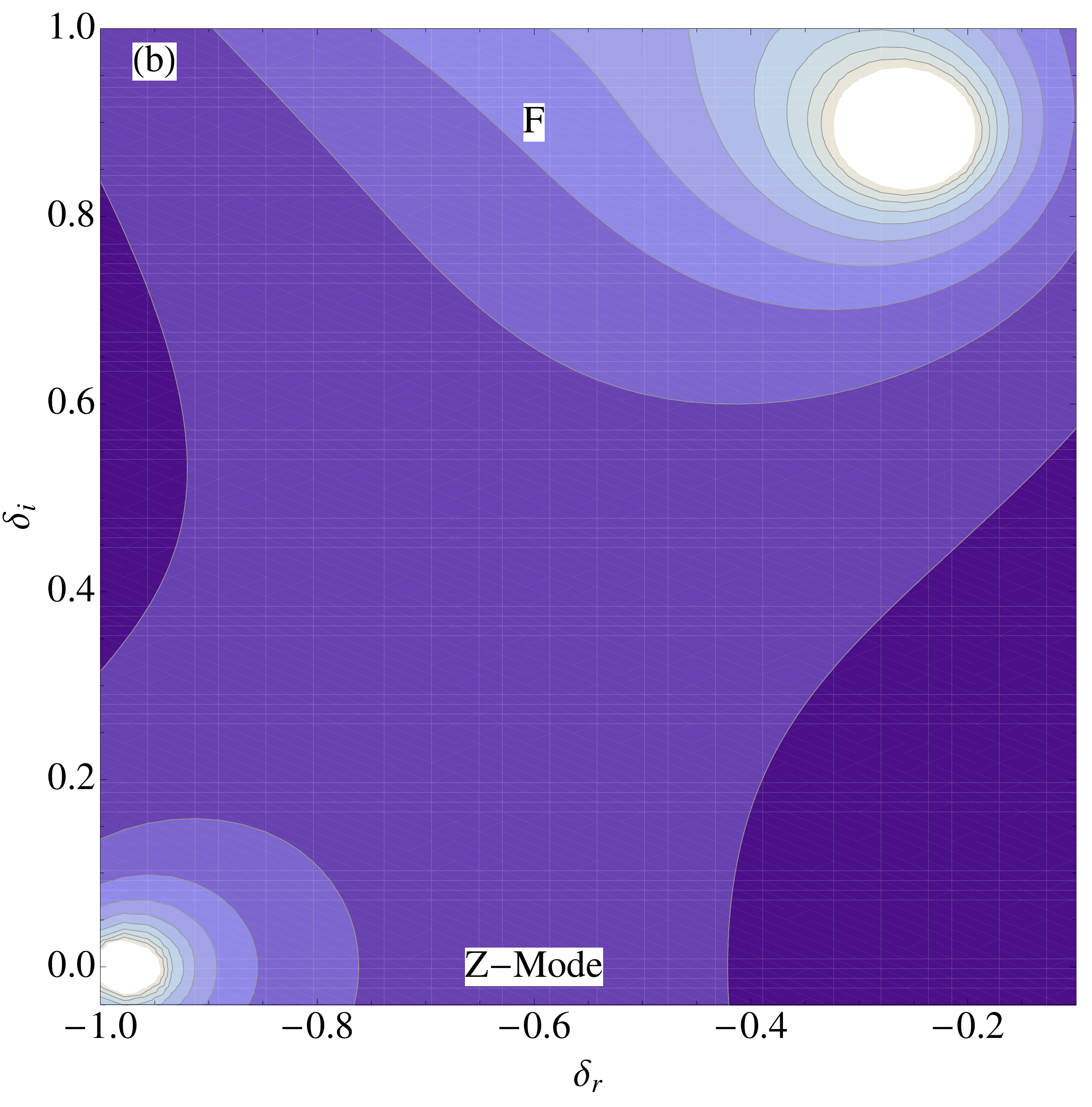}
\caption{\footnotesize Model 6C - inner shock flow: (a) inner shock disk profile with preshock deceleration, and (b) its eigenfrequencies.}
\label{fig24} \finfig
\begfig[t] \hskip-0.25in \epsscale{1.15} \plottwo{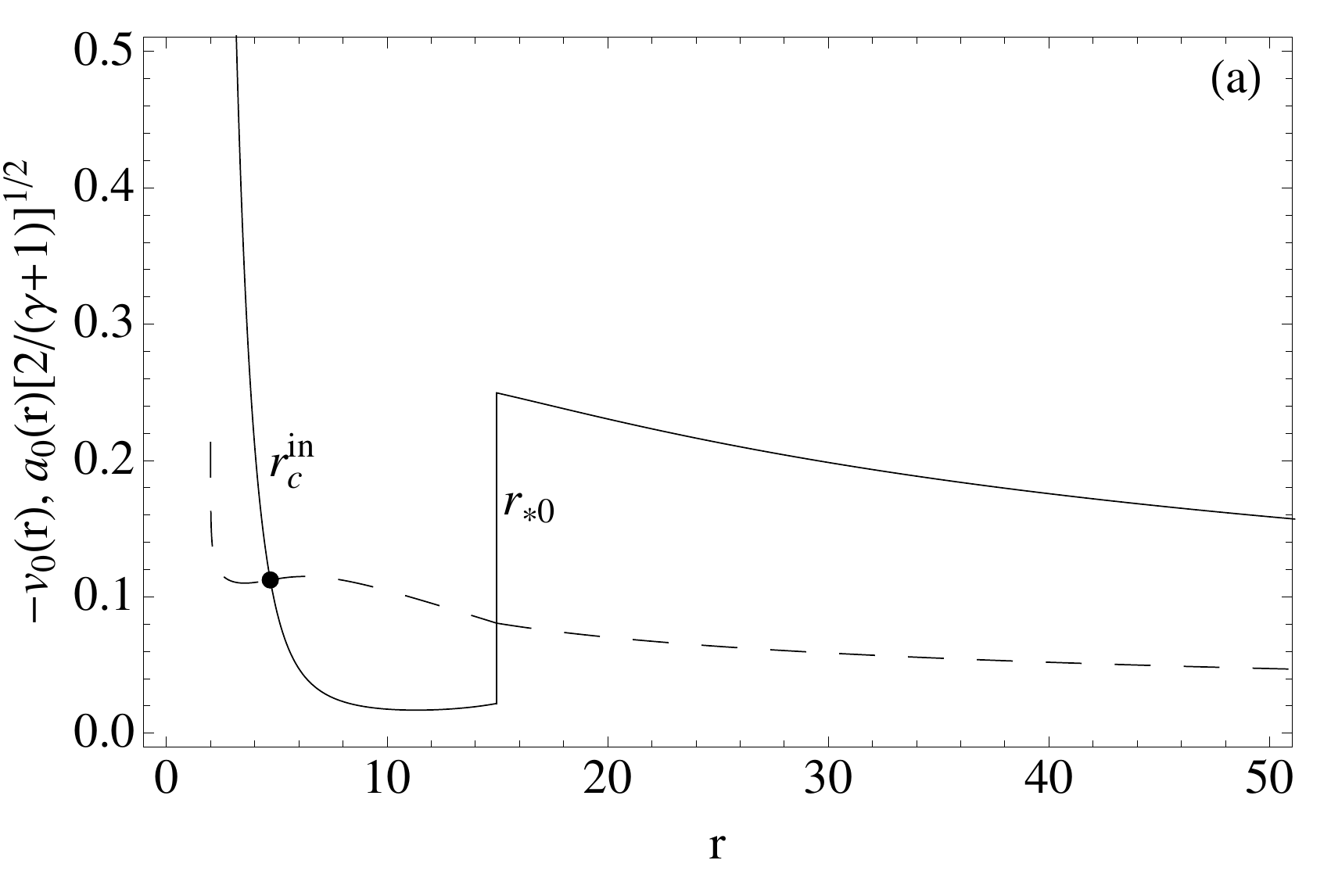}{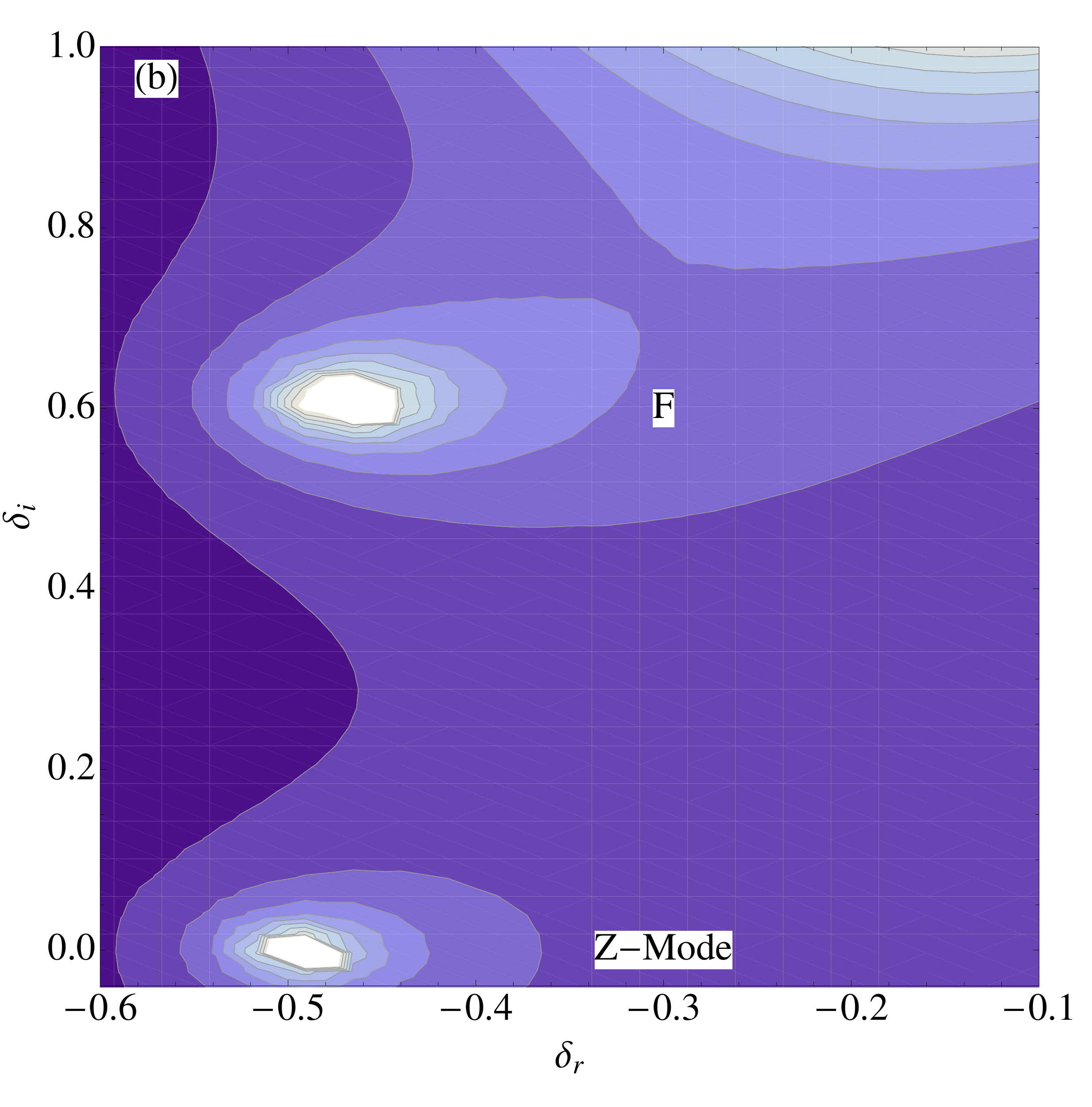}
\caption{\footnotesize Model 7C - inner shock flow: (a) inner shock disk profile with preshock acceleration, and (b) its eigenfrequencies.}
\label{fig25} \finfig
\begfig[t] \hskip-0.25in \epsscale{1.15} \plottwo{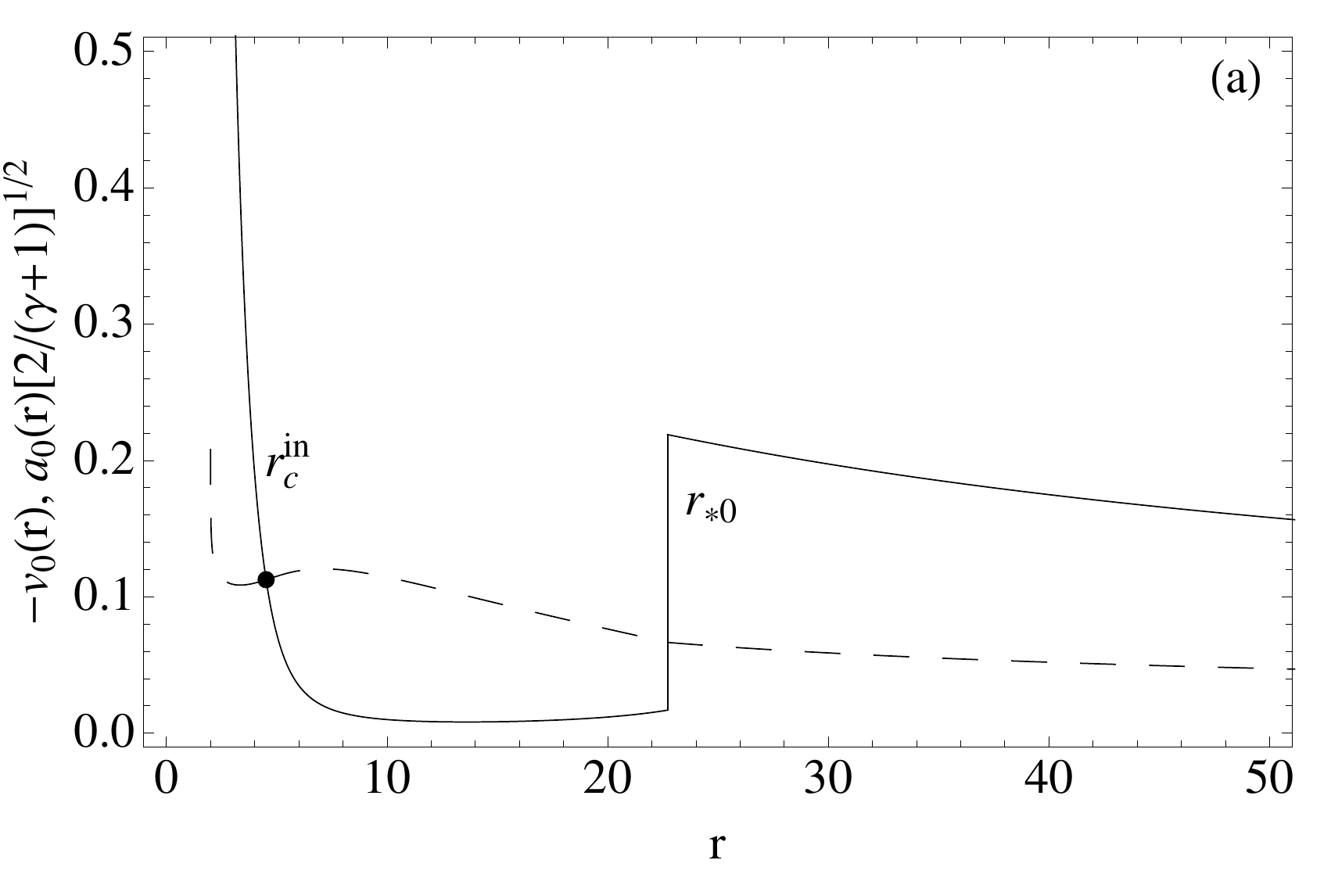}{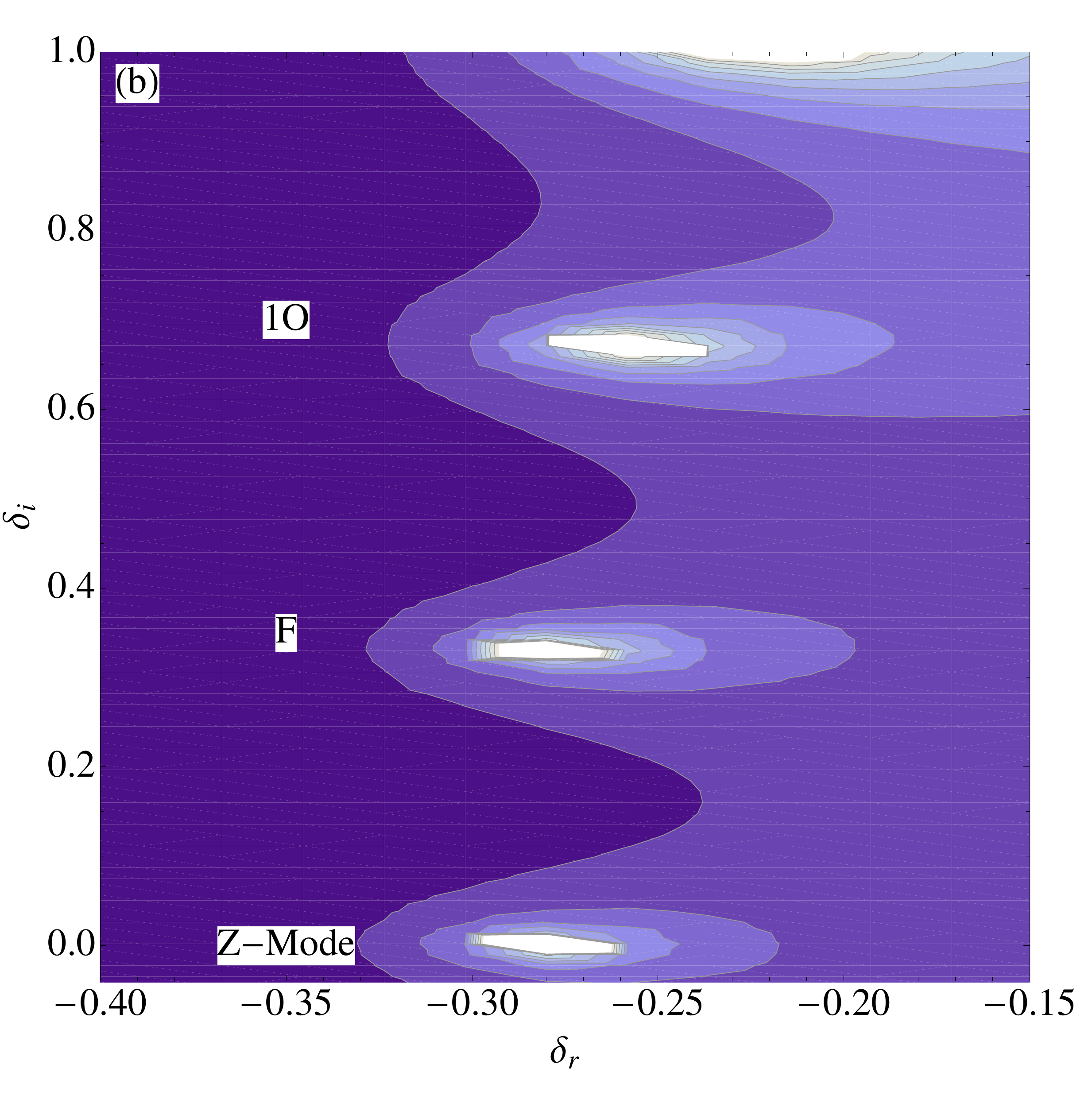}
\caption{\footnotesize Model 8C - inner shock flow: (a) inner shock disk profile with preshock acceleration, and (b) its eigenfrequencies.}
\label{fig26} \finfig
\begfig[t] \hskip-0.25in \epsscale{1.15} \plottwo{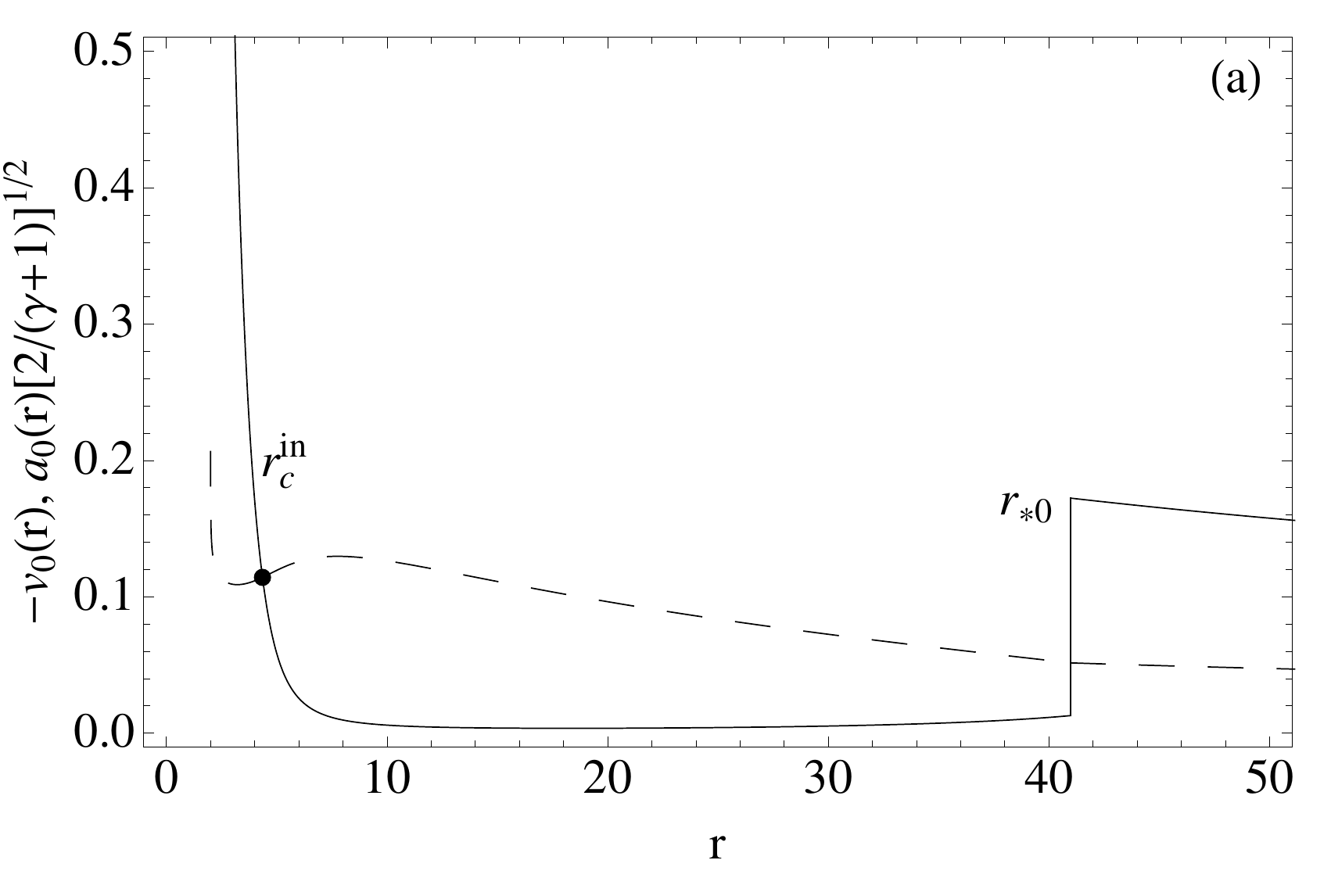}{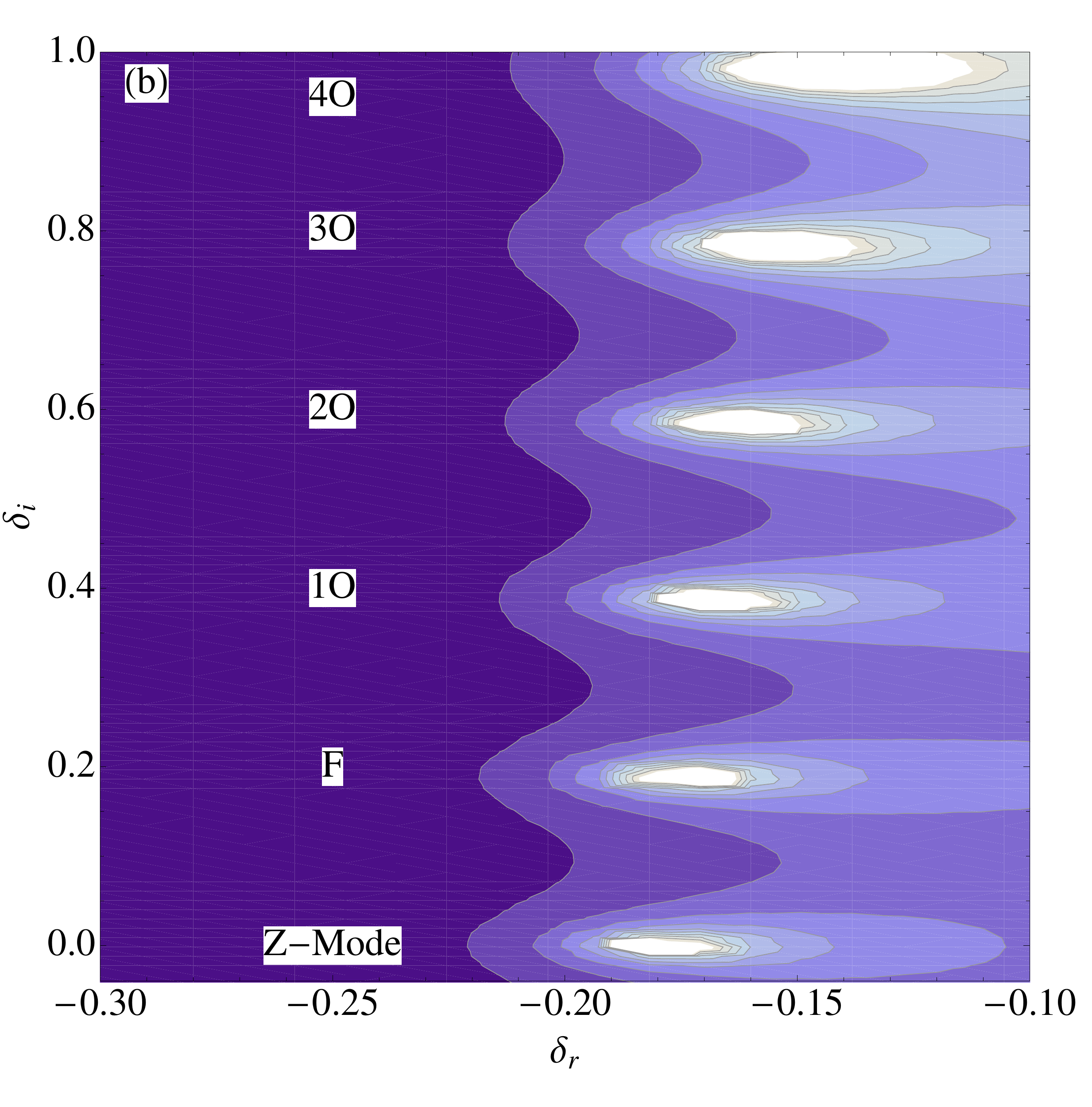}
\caption{\footnotesize Model 9C - inner shock flow: (a) inner shock disk profile with preshock acceleration, and (b) its eigenfrequencies.}
\label{fig27} \finfig

\section{ASTROPHYSICAL IMPLICATIONS OF QPOs IN REGIONS OF INSTABILITY}
Following the \citet{ci82} linearization approach and employing the global spontaneous instability boundary conditions developed by \citet{nak92,nak94}, we have demonstrated that 1D, axisymmetric, inviscid accretion flows exhibiting both preshock deceleration and preshock acceleration in their velocity profiles are unstable to a number of oscillation modes in the region of instability. However, depending on the selected values in the ($\ell,\epsilon_-$)-parameter space, we have also demonstrated that both the preshock deceleration and preshock acceleration velocity profiles can be {\it stable} to the Z-Mode, F-mode, and overtones. Below we discuss the implications of QPOs in the region of instability, while the implications of QPOs in the region of stability will be explored in subsequent papers. Furthermore, it is important to take note that, in the region of instability, the shock velocity profiles that display preshock deceleration (for models 0 to 3 of the inner shocks) and preshock acceleration (models 4 to 7 of the inner shocks and models 4 to 8 of the outer shocks) are unstable because of the zeroth mode (Z-Mode), which has zero oscillation frequency but with positive growth rate.

In order that a particular mode be observable, it is necessary for the configuration that exhibits that mode to last for a sufficiently long time so that the mode can be clearly detected in that time interval. We have found two different types of oscillations, however, and this suggests to us that these two modes will be observable via different manifestations. In the case of the finite oscillation frequency modes X-ray QPOs might be visible for as long as the two-sonic-point-with- shock configuration lasts. However, as time goes on, the zeroth mode may grow to the point that it overwhelms all the other modes and the disk between the two sonic points is disrupted. Or, the shock position may simply move out of the region between the two sonic points without disrupting the disk, but effectively ending the oscillations we predict because of the changed configuration. On the other hand, it may be possible that dissipating processes within the disk cause the growth of the Z-Mode to saturate at some finite amplitude, leaving the disk postshock region to continue to oscillate with one or more of the other finite frequent modes. It might happen that the finite frequency modes manifest themselves first, before the zeroth-frequency mode has had time to grow significantly, and this could constitute a signal to the observer that the disk is nearing an unstable configuration that will ultimately disrupt the postshock disk material, causing it to rain down onto the BH event horizon. The only way to determine the behavior in this case would be nonlinear numerical calculations of the disk dynamics.

In Table~\ref{tbl4} we list the approximate QPO frequency ratios from the fundamental mode to the overtones for models 0 to 7 for the inner shock and models 4 to 8 for the outer shocks computed by taking the ratio between each mode (and to get the ratio into an integer form, we multiply each value in the ratio by a factor of 2). The QPO ratios, in general, are thought to be related to the spin of the compact object~\citep[e.g.][]{rm06,wyh07}. However, if our Z-Mode can be demonstrated to be stable using a nonlinear treatment of the problem, then our model suggests that persistent QPO ratios can be found in our oscillating shock configuration as well. The results from Table~\ref{tbl4} suggest that QPOs ratios 2:3, 3:5, or even higher can be found when the shock is very close to the inner disk at $r_{*0} \sim 13$ gravitational radii. The ratios 2:3 and 3:5 are believed to occur in stellar BH systems such as GRS 1915+105, XTE J1550-564, GRO J1655-40, and SMBHs hole at the galactic center of SgrA*~\citep{asc04,tor05,wag12,ba13}. Furthermore, the ratio 2:3 that we find at $r_{*0} \sim 13$ gravitational radii is consistent with results obtained by~\citet{abr05}, who also found the ratio 2:3 at $r \sim 11 \, GM/c^2$. We emphasize that the ratios 2:3 and 3:5 in our work, for example, are ratios between the fundamental mode and overtones. Different ratios are possible, but any ratios beyond 2:3 or 3:5 might not be observable. 

To see whether  the QPO ratios 2:3 and 3:5 could occur when the shock is located near the horizon, we examine our model 2B1, which has a shock located at $7.499$ gravitational radii, for example. We find that QPO ratios 1:2:3:4:5 could exist at this location. Interestingly, the ratio 1:2 is also consistent with results obtained by~\citet{abr05}, who found similar QPO ratios when the shock was located at $r \sim 8 \,GM/c^2$. The present paper consists of a numerical evaluation of a linearized analysis, rather than a full nonlinear numerical calculation. Thus, if the fundamental mode is suppressed in a full nonlinear treatment, the surviving ratio will be 2:3, and this is not an unrealistic possibility. Barring suppression of mode 4 by some nonlinear physical process, it is difficult to achieve precisely the QPO ratio 3:5 using the model 2B1 parameter values, but the difference between a ratio of 0.67 and a ratio of 0.6 may be difficult to establish observationally. Hence, this demonstrates that our instability model can produce the QPO ratios 2:3 or 3:5 if the shock is located near the horizon and the fundamental mode or the third overtone is suppressed by some nonlinear physical process that is not included here. To determine whether mode suppression is possible in a particular case, a full nonlinear calculation is required, which is beyond the scope of this paper.

From our previous work, for example,~\citet{lb05}  have established a connection between the energy transport rate in the disk and rate of energy loss into the outflows via equation (45), which states that  $\dot{M} c^2 \Delta \epsilon = L_{\rm shock} = L_{\rm outflow}$, where $L_{\rm shock}$ and $L_{\rm outflow}$ are the energy-loss rates from the disk at the shock location and the kinetic luminosity of the outflow, respectively, and $\Delta\epsilon$ is the specific energy transport rate in the disk. Hence, based on the observed outflows, we can see that our model results do depend on the mass accretion rate $\dot{M}$ and therefore the implied value of $\dot{M}$ will either increase or decrease depending on the energy jump $\Delta \epsilon$ at the location of the shock, for a source with a given kinetic luminosity in the outflow. Nevertheless, the relationship between $\dot{M}$ and $L_{\rm outflow}$ is still an open question~\citep[e.g.][]{bs12}, which we will continue to study in future work. We note that \citet{ak01}, \citet{abr05}, and references therein suggested that QPOs ratio could represent oscillations occurring at different radii in the inner disk, which is also consistent with our findings. In future work, we plan to explore the stable region to search for the Z-Mode, the fundamental mode, and any overtones that are potentially associated with QPOs observed in stellar and SMBH systems.

\section{CONCLUSIONS}

\citet{nak92,nak94} demonstrated that preshock deceleration causes instability between a shock point and an inner sonic point of an inviscid, axisymmetric, isothermal standing shock wave and  concluded that a velocity profile that displays preshock acceleration or preshock deceleration is stable or unstable, respectively. Following the~\citet{ci82} linearization method and employing the~\citet{nak92,nak94} instability boundary conditions, we perform a global linear stability analysis of an inviscid axisymmetric ADAF disk with an isothermal standing shock wave and a standard hydrostatic disk half-height. We find that disks/shocks with outflows can be stable or unstable depending on the values of ($\epsilon_{-}, \ell$)-parameter space. 

In the region of instability, for demonstrative purposes we select a value of $\epsilon_{-} = 0.002$ with different values of $\ell$, and our results show that velocity profiles that display either preshock deceleration or acceleration are {\it unstable} to the zeroth mode with zero oscillation frequency, but are {\it stable} to the fundamental mode and overtones when the shock location is at less than~48 gravitational radii. However, velocity profiles that display preshock acceleration are unstable to the zeroth mode, as well as to the fundamental mode and some of the overtones with increasing disk height, when the shock location is farther out than~$57$ gravitational radii. The disk half-heights for the preshock deceleration profiles are less than $10$ gravitational radii, while greater than $12$ gravitational radii for the preshock acceleration profiles. The solutions that display preshock deceleration agree with~\citet{nak92,nak94} {\it global spontaneous instability criteria} but not for the preshock acceleration cases. Furthermore, we have also shown that if the Z-Mode can be demonstrated to be stable using a nonlinear treatment, then our model suggests the possible existence of QPOs with ratios 2:3 and 3:5 very close to the inner disk in BH systems, as observed in the microquasar GRS 1915+105 and the SMBH SgrA*.

In the region of stability, we see that the Z-Mode, in general, is stable for both preshock deceleration and acceleration velocity profiles, with the exception that the Z-Mode becomes less stable for the preshock acceleration when the compression ratio increases; we will explore this behavior in more details in subsequent papers.  From these results, we have therefore demonstrated, for the first time, that disks with standing shocks powering outflows can be linearly stable.
%
\begin{deluxetable}{lcccccccccr}
\tabletypesize{\scriptsize} \tablecaption{Growth Rates of the Z-Mode \label{tbl3}} \tablewidth{0pt} \tablehead{
\colhead{Model}  %
&\colhead{$\ell$} %
&\colhead{$\epsilon_{_{-}}$} %
&\colhead{$\epsilon_{_{+}}$} %
&\colhead{$r_{_{out}}$} %
&\colhead{$\rstarZ$} %
&\colhead{$r_{_{in}}$} %
&\colhead{$H_{_{*0}}$} %
&\colhead{${\cal R}_{_{*0}}$} %
&\colhead{$\delta_{r}$} %
&\colhead{$\delta_{i}$}} %
\startdata %
0B1
& 3.02%
&0.0011%
&-0.004985%
&155.846%
&7.980  %
&6.258 %
&3.377 %
&1.254 %
&0.0986    %
&0.0 \\ %
0B2
& 3.05%
&0.0011%
&-0.007104%
&154.959%
&8.987%
&6.233%
&3.922 %
&1.411 %
&0.1223%
&0.0 \\ %
0B3
& 3.1%
&0.0011%
&-0.009100%
&153.434%
&10.827%
&6.048%
&4.923 %
&1.666 %
&0.1488%
&0.0 \\ %
0B4
& 3.16%
&0.0011%
&-0.009920 %
&151.520%
&13.798    %
&5.748     %
&6.562 %
&1.980 %
&0.1682    %
&0.0 \\ %
0B5
&3.20%
&0.0011%
&-0.009572 %
&150.189%
&16.847    %
&5.542     %
&8.281 %
&2.190 %
&0.1783    %
&0.0 \\ %
1B1
& 3.04%
&0.001%
&-0.005342%
&173.850%
&7.761 %
&6.149%
&3.227 %
&1.263 %
&0.0964%
&0.0 \\ %
1B2
& 3.08%
&0.001%
&-0.008285%
&172.694%
&9.066%
&6.119%
&3.921 %
&1.491 %
&0.1246%
&0.0 \\ %
1B3
& 3.13%
&0.001%
&-0.010177%
&171.201%
&10.912%
&5.934%
&4.911 %
&1.770 %
&0.1462%
&0.0 \\ %
1B4
& 3.16%
&0.001%
&-0.010709 %
&170.278%
&12.278    %
&5.793     %
&5.651 %
&1.942 %
&0.1547    %
&0.0 \\ %
1B5
&3.20%
&0.001%
&-0.010782 %
&169.015%
&14.663    %
&5.596     %
&7.026 %
&2.175 %
&0.1640    %
&0.0 \\ %
2B1
&3.06%
&0.0009%
&-0.005618 %
&195.903%
&7.499    %
&6.041     %
&3.052 %
&1.266 %
&0.0927%
&0.0 \\ %
2B2
&3.10%
&0.0009%
&-0.008894 %
&194.776%
&8.759    %
&6.032     %
&3.711 %
&1.514 %
&0.1201    %
&0.0 \\ %
2B3
&3.13%
&0.0009%
&-0.010363 %
&193.912%
&9.779    %
&5.947     %
&4.247 %
&1.697 %
&0.1331%
&0.0 \\ %
2B4
& 3.16%
&0.0009%
&-0.011287 %
&193.029%
&10.943    %
&5.828     %
&4.864 %
&1.884 %
&0.1421    %
&0.0 \\ %
2B5
&3.20%
&0.0009%
&-0.011824 %
&191.825%
&12.884    %
&5.647     %
&5.902 %
&2.140 %
&0.1501    %
&0.0 \\ \hline%
5C
& 3.58%
&0.00002%
&-0.033442 %
&9964.22 %
&7.688%
&4.931    %
&2.028 %
&6.233     %
&-1.8476    %
&0.0 \\ %
6C
&3.6%
&0.00002%
&-0.035305 %
&9963.72 %
&10.026 %
&4.915%
&2.805    %
&8.812     %
&-0.9786    %
&0.0 \\ %
7C
&3.65%
&0.00002%
&-0.030888 %
&9962.44 %
&14.967 %
&4.709%
&4.523    %
&11.494     %
&-0.4970    %
&0.0 \\ %
8C
&3.7%
&0.00002%
&-0.023803 %
&9961.14%
&22.720 %
&4.520    %
&7.343 %
&13.004     %
&-0.2813    %
&0.0 \\ %
9C
&3.75%
&0.00002%
&-0.014752 %
&9959.83%
&40.982 %
&4.356    %
&14.371 %
&13.437     %
&-0.1792    %
&0.0 \\ \hline%
\enddata


\hskip-0.0 truein Note. -- All quantities are expressed in
gravitational units ($GM=c=1$).

\end{deluxetable}
\hskip-1.5truein
\begin{deluxetable}{lccr}
\tabletypesize{\scriptsize} \tablecaption{QPO Ratios \label{tbl4}} \tablewidth{0pt} \tablehead{
\colhead{Model}  %
&\colhead{Shock Branch} %
&\colhead{$\rstarZ$} %
&\colhead{QPO Ratio}} %
\startdata %
0
&Inner shock %
&13.359    %
&2:3:5:6:8\\ %
1
&Inner shock %
&13.722    %
&2:3:5:7:8 \\ %
2
&Inner shock %
&14.627    %
&2:4:5:7:9 \\ %
3
&Inner shock %
&17.925    %
&3:5:7:9:11\\ %
4
&Inner shock %
&21.993    %
&3:6:7:11:13\\ %
5
&Inner shock %
&24.232    %
&4:6:8:11:14\\ %
6
&Inner shock %
&25.457    %
&4:5:8:11:14\\ %
7
&Inner shock %
&30.069    %
&4:5:9:12:16\\ %
8
&Outer shock %
&32.588    %
&4:5:9:13:16\\ %
7
&Outer shock %
&34.740    %
&4:5:9:13:16\\ %
6
&Outer shock %
&40.959    %
&4:5:9:13:17\\ %
5
&Outer shock %
&42.983    %
&4:5:9:13:17\\ %
4
&Outer shock %
&47.214    %
&4:5:9:13:17\\ \hline%
\enddata


\hskip-0.0 truein Note. -- $\rstarZ$ is expressed in gravitational units ($GM=c=1$).

\end{deluxetable}

\acknowledgements T.L. wishes to acknowledge Lev Titarchuk, Charles D. Dermer, and Martin Laming for discussions. The authors also wish to acknowledge the anonymous referee for several comments and insightful suggestions that significantly improved the paper. Much of this work was completed while T.L. was a National Research Council research associate at the Naval Research Laboratory. K.S.W. and M.T.W. are supported by the Chief of Naval Research.

\newpage

\section*{APPENDIX}

\appendix

\section{Boundary Conditions at the Shock Radius}

Let the velocity of the shock be $v_{_*} = \vstar1 e^{\sigma t}$ (see Equation (\ref{eq28})) in the observer frame, and then in the shock frame the upstream velocity changes to
\begeq \vIn = - u_{_{-}} - v_{_*} \ , \label{A1} \fineq
where $\vIn$ is the incoming gas velocity and $\uMinus > 0$ is the preshock velocity. At the shock location, note that the Mach number $\cal{M}_{_-}$ will modify since the shock location is oscillating. Hence, we write 
\begeq
M_{_{\rm n}} = {\cal M}_{_-} + M_{_1} \ , 
\label{A2}
\fineq
where $M_{_{\rm n}}$ and $M_{_1}$ denote the new and perturbed Mach
number, respectively, with $M_{_1} \ll \cal{M}_{_-}$. If $v$ is
the velocity of the postshock gas in the observer frame, then
$v-v_{_*}$ will be the velocity as seen in the shock frame; hence,
using the velocity jump condition in Equation (\ref{eq15}) for isothermal
shock condition, we have
\begeq 
v-v_{_*} = \gamma^{-1} M^{-2}_{_{\rm n}} \vIn \; , 
\label{A3}
\fineq
where we change $v_{_0} \rightarrow v-v_{_*}$, $\uMinus \rightarrow v_{_{in}}$, and ${\cal M}_{_-} \rightarrow M_{_{\rm n}}$ in Equation (\ref{eq15}). Substituting Equations (\ref{A1}) and (\ref{A2}) into Equation (\ref{A3}) for $\vIn$ and $M_{_{\rm n}}$, and solving for $v$ where only linear in first order quantities of $M_{_1}$ and $v_{_*}$ are retained, we obtain
\begeq 
v = -\gamma^{-1} {\cal M}^{-2}_{_-} \uMinus + \left(1 +
\frac{1}{\gamma {\cal M}^{2}_{_-}}\right) v_{_*} \; . 
\label{A4}
\fineq
Comparing Equation (\ref{A4}) with Equations (\ref{eq19}) and (\ref{eq35}), we have
\begeq 
\v1 = \left(1 + \frac{1}{\gamma {\cal M}^{2}_{_-}}\right)
\vstar1 \; . 
\label{A5} 
\fineq
For the perturbation in density, we use the isothermal shock
condition, which expresses the conservation of mass across the
shock, namely,
\begeq 
\rhoIn \vIn = \rho (v-v_{_*}) \; . 
\label{A6} 
\fineq
Utilizing Equations (\ref{A2}) and (\ref{A3}) and solving for $\rho$ from Equation (\ref{A6}), we obtain
\begeq 
\rho = \gamma \rhoMinus M_n^2 = \gamma \rhoMinus {\cal
M}^{2}_{_-} + 2 \gamma \rhoMinus \frac{v_{_*}}{\uMinus} {\cal
M}_{_-}^2 \; , 
\label{A7} 
\fineq
where we have taken $\rhoIn = \rhoMinus$. Comparing Equation (\ref{A7}) with Equations (\ref{eq18}) and (\ref{eq34}), we have
\begeq 
\Rho1 =  2 \gamma \vstar1 \frac{\rhoMinus}{\uMinus} {\cal
M}_{_-}^2 \; . 
\label{A8} 
\fineq
Finally, for the pressure perturbation, we use the isothermal jump
condition for the conservation of momentum flux, namely,
\begeq 
P - P_{_{\rm in}} = \rhoIn \vIn^2 - \rho (v-v_{_*})^2 \; .
\label{A9} 
\fineq
Upon substituting $\vIn$, $v-v_{_*}$, and $\rho$ using Equations (\ref{A1}), (\ref{A3}), and (\ref{A7}), respectively, and utilizing Equation (\ref{A2}) for $M_{_{\rm n}}$ and Equation (\ref{eq8}) for $\rhoIn = \gamma P_{_{\rm in}}/a^2_{_-}$, we find
\begeq 
P = \gamma {\cal M}^2_{_{-}} P_{_{-}}  + 2 \uMinus \rhoMinus v_{_*}
 \; , 
 \label{A10} 
 \fineq
where we have taken $P_{_{\rm in}} = P_{_{-}}$. Comparing Equation (\ref{A10}) with Equations (\ref{eq20}) and (\ref{eq36}), we obtain
\begeq 
\P1 = 2 \uMinus \rhoMinus \vstar1  \; . 
\label{A11} \fineq

\end{document}